\documentclass[a4paper,11pt]{article}

\usepackage{jcappub} 
\usepackage[export]{adjustbox} 
\usepackage[utf8]{inputenc}
\usepackage{color}
\usepackage{graphicx}
\usepackage{amsmath,amsfonts,amssymb}
\usepackage{acronym}
\usepackage{xspace}
\usepackage{multirow}
\usepackage{tabularx}
\usepackage{orcidlink}
\usepackage{ragged2e}
\usepackage{enumitem}
\usepackage{setspace}
\usepackage{caption}
\usepackage{subcaption}
\usepackage{booktabs}
\usepackage{comment}

\usepackage{listings}
\usepackage[ruled,vlined]{algorithm2e}

\SetCommentSty{mycommfont}

\newcommand{\msun}{M_{\odot}}

\newcommand{\fpbh}{f_{\rm PBH}}

\newcommand{\td}{{\rm d}}

\newcommand{\pd}{{\partial}}

\definecolor{rossocorsa}{rgb}{0.83, 0.0, 0.0}

\newcommand{\MPBH}{M_{{\text{\tiny PBH}}}}

\def\lsim{\mathrel{\rlap{\lower4pt\hbox{\hskip0.5pt$\sim$}}
 \raise1pt\hbox{$<$}}}         
\def\gsim{\mathrel{\rlap{\lower4pt\hbox{\hskip0.5pt$\sim$}}
 \raise1pt\hbox{$>$}}}         

\newcommand{\MPl}{M_{\textrm{\tiny{Pl}}}}

\newcommand{\vx}{\mathbf{x}}

\newcommand{\Pz}{\mathcal P_\zeta}

\newcommand{\calP}{\mathcal{P}}

 \def\be   {\begin{equation}}   \def\ee   {\end{equation}}
 \def\ba   {\begin{array}}      \def\ea   {\end{array}}
 \def\bea  {\begin{equation}\begin{aligned}}   \def\eea  {\end{aligned}\end{equation}}
 \def\bean {\begin{eqnarray*}}  \def\eean {\end{eqnarray*}}

\definecolor{TardisBlueAcceso}{RGB}{0, 80, 200}

\definecolor{darkgreen}{rgb}{0,0.5,0}
\definecolor{oucrimsonred}{rgb}{0.6, 0.0, 0.0}
\definecolor{persianblue}{rgb}{0.11, 0.22, 0.73}
\definecolor{forestgreen}{rgb}{0.13,0.35,0.13}
\definecolor{lightgray}{rgb}{0.83, 0.83, 0.83}
\definecolor{blunotte}{HTML}{006778}
\definecolor{darkredcustom}{rgb}{0.435, 0.039, 0.094}
\definecolor{tardisblue}{RGB}{0,100,180}

\definecolor{oucrimsonred}{rgb}{0.6, 0.0, 0.0}

\newcommand{\Msun}{M_\odot}

\definecolor{verdes}{rgb}{0.1, 0.5, 0.1}%
\definecolor{cornellred}{rgb}{0.7, 0.11, 0.11}

\definecolor{VioletRed4}{rgb}{0.55, 0.13, .32}
\definecolor{rossocorsa}{rgb}{0.83, 0.0, 0.0}

\title{Memoirs of the curvaton: non-perturbative non-Gaussianity and supermassive primordial black holes}

\author[a,b]{S.~Allegrini\orcidlink{0009-0004-2664-7440}}
\emailAdd{sasha.allegrini@kbfi.ee}
\author[c]{A.~J.~Iovino\orcidlink{0000-0002-8531-5962}}
\emailAdd{a.iovino@nyu.edu}
\author[a]{G.~Perna\orcidlink{0000-0002-7364-1904}}
\emailAdd{gabriele.perna@phd.unipd.it}
\author[a]{H.~Veerm\"ae\orcidlink{0000-0003-1845-1355}}
\emailAdd{hardi.veermae@cern.ch}

\affiliation[a]{Keemilise ja Bioloogilise F\"u\"usika Instituut, R\"avala pst. 10, 10143 Tallinn, Estonia}
\affiliation[b]{Tallinn University of Technology, Akadeemia tee 23, Tallinn, 12618, Estonia}
\affiliation[c]{Center for Astrophysics and Space Science (CASS), New York University Abu Dhabi, PO Box 129188, Abu Dhabi, UAE}

\abstract{
The curvaton provides a simple mechanism for generating strongly non-Gaussian curvature perturbations after inflation, with potentially important consequences on small scales. 
We study curvaton dynamics beyond the standard quadratic potential and construct the local non-Gaussian map $\zeta=F(\zeta_{\rm G})$ relating the curvature perturbation to an auxiliary Gaussian field $\zeta_{\rm G}$. 
Curvaton self-interactions make the onset of oscillations field dependent and modify the effective equation of state once the curvaton enters the adiabatic regime. 
We incorporate these effects using the abbreviated action, which provides a compact way to connect the frozen and oscillatory regimes and exposes sources of non-Gaussianity absent in the purely quadratic case. 
We apply the formalism to quadratic, monomial, quartic, and cosine potentials, for which we derive the mapping $F(\zeta_{\rm G})$ and show that self-interactions can either enhance or suppress the resulting non-Gaussianity depending on the potential and initial conditions. We consider non-perturbative aspects in the strongly non-Gaussian regime, and show how strong non-Gaussianity can suppress the power spectrum.
As an application, we provide a bottom-up scenario in which strongly positive curvaton non-Gaussianity allows primordial supermassive black hole seeds at peak amplitudes $\mathcal{A}_{\rm pk}\sim10^{-3}$, which are compatible with the COBE/FIRAS $\mu$-distortion bounds. This opens a new primordial scenario for the Little Red Dots observed by the JWST.
The axion-like curvaton provides a particularly natural setting for this mechanism.
}

\begin{document}
\maketitle
\flushbottom

\section{Introduction}
\label{sec:intro}

Observations of the cosmic microwave background (CMB) and the Lyman-$\alpha$ forest imply nearly Gaussian primordial curvature perturbations and an almost scale-invariant power spectrum, with amplitude $ A_s\sim 2.1 \cdot10^{-9}$~\cite{Planck:2018jri,Planck:2018vyg,Planck:2019kim,Chabanier:2019eai,Cyr:2023pgw,DESI:2024mwx}. 
However, at scales $k \gg {\rm Mpc}^{-1}$,  primordial power spectrum is still largely unconstrained. The strongest bounds for $ {\rm Mpc}^{-1} \ll  k \lesssim 10^5{\rm Mpc}^{-1}$ arise from CMB spectral distortions~\cite{Chluba:2012we,Chluba:2013dna}, and, at the smallest scales $k \gg 10^{5}\,{\rm Mpc}^{-1}$ from the non-observation of scalar-induced gravitational waves (SIGWs)~\cite{Tomita:1975kj,Matarrese:1993zf,Acquaviva:2002ud,Mollerach:2003nq,Ananda:2006af,Baumann:2007zm,Domenech:2021ztg} by pulsar timing arrays (PTA)~\cite{You:2023rmn, Iovino:2024tyg}, and from the overproduction of primordial black holes (PBHs) via the critical collapse of large curvature perturbations~\cite{Zeldovich:1967lct,Hawking:1971ei,Carr:1974nx,Carr:1975qj}, which sets a limit of about $\mathcal{P}_{\zeta} \lesssim \mathcal{O} (10^{-2})$~\cite{Inomata:2018epa,Byrnes:2018txb,Karam:2022nym,Kushwaha:2026msi}.

The primordial Universe at scales much smaller than those accessible through recombination can be probed by a plethora of phenomena that are directly related to primordial curvature perturbations. These include phenomena sensitive to changes in the equation of state of the Universe~\cite{Jedamzik:1996mr,Jedamzik:1998hc,Musco:2023dak,Domenech:2019quo,Domenech:2021ztg,Blas:2026xws,Perna:2026szd} and to the statistical properties of small-scale curvature perturbations, including primordial non-Gaussianity (NG). In particular, due to the limited observational information available on these scales, there is currently no empirical reason to assume that primordial curvature perturbations are even approximately Gaussian. NG has been shown to play a prominent role in PBH formation and abundance~\cite{Bugaev:2013vba,Young:2014ana,Nakama:2016gzw,Byrnes:2012yx,Franciolini:2018vbk,Yoo:2018kvb,Kawasaki:2019mbl,Atal:2018neu,Atal:2019cdz,Kehagias:2019eil,Figueroa:2020jkf,Palma:2019lpt,Riccardi:2021rlf,Taoso:2021uvl,Biagetti:2021eep,Kitajima:2021fpq,Hooshangi:2021ubn,Meng:2022ixx,Young:2022phe,Escriva:2022pnz,Hooshangi:2023kss,Ferrante:2022mui,Gow:2022jfb,Ianniccari:2024bkh,Raatikainen:2025gpd,Iovino:2025cdy,Zhang:2026hqg} and in determining the shape of SIGW spectra~\cite{Kohri:2018awv,Cai:2018dig,Cai:2019elf,Unal:2018yaa,Hajkarim:2019nbx,Yuan:2020iwf,Adshead:2021hnm,Atal:2021jyo,Domenech:2021and,Garcia-Saenz:2022tzu,Abe:2022xur,Liu:2023ymk,Yuan:2023ofl,Li:2023xtl,Perna:2024ehx,Zeng:2025cer,Iovino:2024sgs,Perna:2026szd}.

For instance, strongly non-Gaussian small-scale perturbations have been invoked in the context of avoiding spectral distortion constraints on PBHs~\cite{Unal:2020mts,Nakama:2017xvq,Hooper:2023nnl,Byrnes:2024vjt, Pritchard:2025yda,Iovino:2024tyg}, and they have been shown to be relevant when probing PBHs with SIGWs by future GW experiments such as LISA~\cite{Garcia-Bellido:2016dkw,Garcia-Bellido:2017aan,Bartolo:2018evs,Bartolo:2018rku,Iovino:2025cdy,Hong:2026rcl} or PTA experiments~\cite{Franciolini:2023pbf,Iovino:2024tyg,Gouttenoire:2025jxe}. A thorough understanding of the theoretical mechanisms capable of generating strongly non-Gaussian curvature perturbations is therefore of central importance.

During the last two decades, several studies have highlighted how a light spectator field, known as the \emph{curvaton}, can generate large-amplitude perturbations after inflation~\cite{Lyth:2001nq,Enqvist:2001zp,Moroi:2001ct,Lyth:2002my,Wands:2002bn,Gordon:2002gv,Enqvist:2003mr,Dimopoulos:2003ss,Dimopoulos:2003az,Bartolo:2003jx,Bartolo:2002vf,Enqvist:2004gz,Allahverdi:2006dr,Enqvist:2008be,Enqvist:2012tc,Enqvist:2013qba,Enqvist:2011jf}, which, under suitable conditions, are capable of producing PBHs~\cite{Kohri:2012yw,Kawasaki:2012wr,Bugaev:2013vba,Ando:2017veq,Ando:2018nge,Chen:2019zza,Inomata:2020xad,Kawasaki:2021ycf,Pi:2021dft,Ferrante:2023bgz,Chen:2023lou,Hooper:2023nnl,Gow:2023zzp,Chen:2024pge,Kuroda:2025coa,Kasai:2026yna} and SIGWs. In the curvaton framework, non-Gaussian curvature perturbations $\zeta$ arise naturally from the dynamics of the curvaton field~\cite{Bartolo:2003jx,Sasaki:2006kq,Kawasaki:2011pd,Mukaida:2014wma,Liu:2020zlr,Ferrante:2023bgz} and can be altered by self-interactions~\cite{Enqvist:2005pg,Enqvist:2008gk,Enqvist:2009zf,Enqvist:2009ww,Enqvist:2009eq,Enqvist:2010dt,Byrnes:2010xd,Byrnes:2011gh}. 

The main focus of this work is on NGs of local type, characterized by the non-Gaussian curvature perturbation $\zeta({\bf x})$, which, at a given spatial position, depends only on some auxiliary Gaussian field $\zeta_{\rm G}({\bf x})$ at that point. The non-Gaussian field can thus be written as a function of the auxiliary Gaussian field in the form
\be\label{eq:FullNGs}
    \zeta({\bf x}) = F\left(\zeta_{\rm G}({\bf x})\right)\,,
\ee
with $F$ a generic non-linear function. A common assumption in the literature is that deviations from Gaussianity are mild, allowing us to rewrite Eq.~\eqref{eq:FullNGs} via the perturbative series
\be\label{eq:FirstExpansion}
    \zeta({\bf x}) 
    = \zeta_{\rm G}({\bf x}) + \frac{3}{5}f_{\rm NL}\,\zeta_{\rm G}^2({\bf x}) + \frac{9}{25}g_{\rm NL}\,\zeta_{\rm G}^3({\bf x}) + \dots\,.
\ee
In most applications, the series is truncated at the quadratic order, under the assumption that NG is weak enough and its effects can be adequately captured by the $f_{\rm NL}$ term only. However, in the curvaton scenario the coefficients $f_{\rm NL},\, g_{\rm NL}, \dots$ are not independent parameters but follow from a single underlying dynamical relation.

In this work, we go beyond the standard quadratic curvaton scenario and derive the non-Gaussian map, Eq.~\eqref{eq:FullNGs}, for a class of self-interacting potentials in which the onset of the oscillatory phase is itself field dependent. This dependence, which is absent in the purely quadratic case, introduces a novel source of non-linearity that propagates into $F(\zeta_{\rm G})$ and, consequently, into the statistics of $\zeta$. Crucially, rather than truncating the expansion~\eqref{eq:FirstExpansion} at finite order or postulating an \emph{ad hoc} ansatz for the distribution of $\zeta$, we derive $F(\zeta_{\rm G})$ directly from the curvaton dynamics, thereby obtaining a fully non-perturbative description of the resulting non-Gaussian statistics.

A well-motivated application of this framework is the formation of PBH seeds for SMBHs. In typical critical collapse scenarios, the power spectrum at the relevant scales is strongly constrained by CMB spectral distortions, which requires extremely large NGs in order to maintain a non-negligible PBH abundance. Many earlier studies have relied on perturbative expansions in $f_{\rm NL}$, which break down precisely in the strongly non-Gaussian regime of interest, or on ad hoc ans\"{a}tze for the distribution of $\zeta$~\cite{Byrnes:2024vjt, Pritchard:2025yda,Hooper:2023nnl}. Strong NGs derived from first principles in the curvaton model were considered in Ref.~\cite{Hooper:2023nnl}; however, that work was restricted to the quadratic potential and employed a simplified treatment of the PBH abundance that did not account for the full non-linear compaction function, nor for the non-perturbative contribution of large NGs to the curvature power spectrum entering the spectral distortion constraints.

As pointed out in Ref.~\cite{Ferrante:2022mui}, whenever such a closed-form expression is available, as is the case for the curvaton, the truncated expansion~\eqref{eq:FirstExpansion} fails to reproduce the correct PBH abundance, especially in the far tail of the PDF where PBH formation occurs. Determining the full non-Gaussian relation, Eq.~\eqref{eq:FullNGs}, for different curvaton potentials is, therefore, of pivotal importance for obtaining reliable estimates of the PBH abundance in these models.
Moreover, taking into account NGs when computing the shape of the power spectrum~\cite{Veermae:2026yzz} is crucial for determining whether the formation of PBH seeds is compatible with the constraints arising from CMB spectral distortions.

This work is organised as follows: Sec.~\ref{sec:deltaN} reviews the $\delta N$ formalism and outlines the curvaton dynamics for a generic potential, and Sec.~\ref{sec:NonGaussianities} derives the non-Gaussian relation~\eqref{eq:FullNGs} for several benchmark potentials and  provides a simple approximate analytic ansatz for $F(\zeta_{\rm G})$ for quadratic and cosine potentials. Sec.~\ref{sec:NGPS} examines the impact of these NGs on the full curvature power spectrum. How large primordial NGs can sustain the formation of supermassive PBHs while evading the CMB $\mu$-distortion constraints is shown in Sec.~\ref{sec:SMBH}. Specifically, how this scenario can be realised in the case of an axion-like curvaton. We conclude in Sec.~\ref{sec:conc}.

Reduced Planck units $\hbar=c=\MPl = 1$ are used throughout this work.

\section{$\delta N$ formalism: from curvaton dynamics to curvature perturbations} 
\label{sec:deltaN}

It is well known that even in the simplest quadratic curvaton scenario, significant NGs can be generated due to the non-linear relation between the curvature perturbation and the curvaton density contrast~\cite{Bartolo:2003jx,Lyth:2005fi,Sasaki:2006kq,Pi:2022ysn}. When included, curvaton self-interactions can play a crucial role in either suppressing or enhancing the resulting NGs~\cite{Enqvist:2009eq, Enqvist:2009ww, Enqvist:2009zf, Enqvist:2010dt, Byrnes:2010xd, Byrnes:2011gh,Kawasaki:2011pd, Mukaida:2014wma}, as shown below, depending on the shape of the potential and the choice of its parameters.

The conversion of the initial fluctuations in the curvaton field into adiabatic curvature perturbations can be evaluated with the $\delta N$ formalism~\cite{Starobinsky:1982ee, Salopek:1990jq, Sasaki:1995aw, Lyth:2004gb, Lyth:2005fi}, which provides a powerful framework for computing the non-linear evolution of cosmological perturbations. In particular, it focuses on super-horizon fluctuations, that is, fluctuations at scales $k^{-1}$ much larger than the horizon scale $(aH)^{-1}$. Its applicability relies on the conservation of curvature perturbations at the non-linear level~\cite{Lyth:2004gb}, which can be demonstrated using the gradient expansion, where $k/(aH)$, rather than the amplitude of the perturbation, is taken as the expansion parameter~\cite{Salopek:1990jq, Shibata:1999zs}. In this case, each patch of the Universe can be evolved locally as a separate FLRW universe with a spatial metric
\be \label{eq:PFLRW}
    g_{ij} = \tilde a^2 \delta_{ij}\,,
    \qquad \qquad
    \tilde a(t,\mathbf{x}) = a(t) e^{\zeta(t, \mathbf{x})}\,,
\ee
where $\tilde a(t,\mathbf{x})$ and $a(t)$ denote respectively the local and global scale factors and $\zeta$ corresponds to the primordial scalar curvature perturbation in the constant density gauge. Consequently, scalar curvature perturbations are directly related to the amount of local expansion within this framework.

In detail, super-horizon curvature fluctuations can be estimated from the number of $e$-folds it takes to pass from an initial flat hypersurface at $t_{\rm in}$ ($\zeta (t_{\rm in}) = 0$) to a uniform-density hypersurface at $t$ as~\cite{Starobinsky:1982ee,Salopek:1990jq,Sasaki:1995aw},
\be\label{eq:deltaN_general}
    \zeta(t, \mathbf{x})
    = \delta N(t, \mathbf{x}) 
    \equiv N(t, \mathbf{x}) - \bar N (t)\,,
\ee
where $N(t, \mathbf{x}) = \ln (\tilde a(t,\mathbf{x})/a(t_{\rm in}) )$ and $\bar N (t) \equiv \ln( a(t)/a(t_{\rm in}) )$ describe the number of $e$-folds in the perturbed and global background. Since non-linear super-horizon fluctuations are considered, the definition of the latter suffers from potential ambiguities: the global background reveals itself in an empirically meaningful way only after the causally isolated patches have come into causal contact at a later time due to the expansion of the Universe (e.g., today). When we demand that the curvature perturbations vanish on average, we must stipulate that
\be\label{eq:zeta_mean}
    \langle \zeta(t, \mathbf{x}) \rangle = 0 
    \qquad\Leftrightarrow \qquad
    \bar N(t) = \langle N(t, \mathbf{x})\rangle\,,
\ee
where the average is taken over all Hubble patches in the constant-density hypersurface at time $t$.

The $\delta N$ formalism is particularly well suited to the curvaton scenario since the curvaton field generates isocurvature perturbations $\zeta_\phi$ during inflation, which are converted into adiabatic perturbations $\zeta$ after its decay. 

Once the parameters of the curvaton potential $V(\phi)$ and the decay rate $\Gamma_{\phi}$ are fixed, the spatial dependence of the curvature perturbation is entirely encoded by the initial field in each patch at the end of inflation, $\zeta(t,\,{\bf x}) = \delta N(t,\phi_\star)$, where $\phi_\star \equiv \phi(t_\star,\,{\bf x})$ denotes the initial (potentially perturbed) field in a given Hubble patch. The curvature perturbation can therefore be obtained by evolving the curvaton field from the end of inflation until it has decayed and computing the number of $e$-folds in each patch, $N(t,\phi(t_\star,\,{\bf x}))$, for some given $H(t) \ll \Gamma_{\phi}$. After the curvaton has completely decayed, the Universe enters a radiation-dominated phase and evolves adiabatically. As a result, the curvature perturbation $\zeta$ remains conserved on super-horizon scales until horizon re-entry, $\dot{\zeta} = 0$. Therefore, evaluating $\delta N$ briefly after the curvaton decay is sufficient to determine the curvature perturbations generated by the spectator field.

\begin{figure}[t]
	\begin{center}
	\includegraphics[width=1.\textwidth]{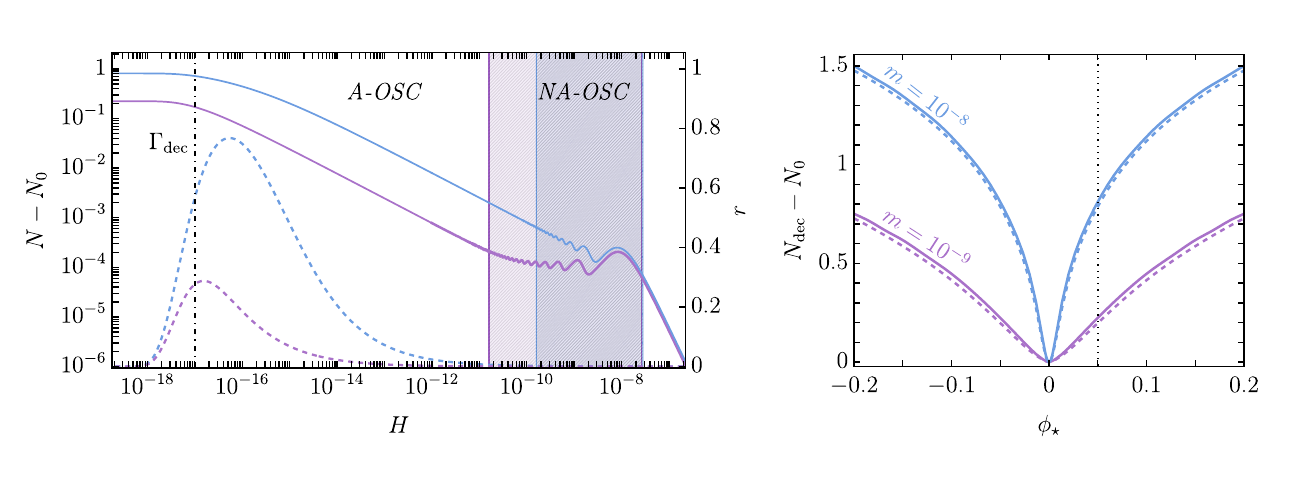}
\caption{ {\it Left panel}: $N-N_0$ (solid) and the energy density parameter $r\equiv 3\rho_{\phi}/(4\rho_{r} + 3\rho_{\phi})$ (dashed) as functions of $H$ (with time evolving from right to left) for a non-instantaneous decay scenario and quartic self-interactions (potential in Eq.~\eqref{eq:quarticV}).
In both panels, we fix $\phi_\star = 0.05$, $H_\star = 10^{-6.5}$, $\lambda = 10^{-13}$, and $\Gamma_{\phi} = 10^{-17}$, with two benchmark cases, $m=10^{-8}$ (light blue) and $m=10^{-9}$ (light purple). $N-N_0$ is computed numerically from Eq.~\eqref{eq:deltaN_general}, while the density parameter $r$ is computed by evolving the energy densities using the system in Eq.~\eqref{eq:EOM}, supplemented by Eq.~\eqref{eq:fluidEOM} once the evolution enters the adiabatic regime. The coloured regions highlight the non-adiabatic oscillation phase (NA-OSC) between $H^{-1}=T$ and $H^{-1}=10\,T$, after which the system enters the adiabatic oscillation regime (A-OSC). {\it Right panel}: $N_\textrm{dec}-N_0$ as a function of the initial field $\phi_\star$. The dotted line marks the initial value corresponding to the curves shown in the left panel. The dashed lines show the results obtained assuming instantaneous decay at $H=\Gamma_{\phi}$.
}
    \label{fig:zeta}
	\end{center}
\end{figure}

An illustrative example is given in Fig.~\ref{fig:zeta} for two benchmark scenarios of a quartic potential (cf. Eq.~\eqref{eq:quarticV} in Sec.~\ref{sec:quartic_V}). The left panel shows the time evolution of $N - N_0$ for a fixed initial $\phi_\star$ in the different models, together with the evolution of the energy fraction $r \equiv 3\rho_{\phi}/(4\rho_{r} + 3\rho_{\phi})$. The right panel, instead, shows how curvaton's initial value $\phi_\star$ affects the number of $e$-folds $N_{\rm dec} \equiv N(t_{\rm dec})$ until its decay on a constant density hypersurface characterised by some $H(t) \ll \Gamma_{\phi}$. $N_0$ corresponds to the $e$-foldings in the absence of the curvaton, i.e., when $\phi_\star = 0$. This case produces a minimal delay in the number of $e$-folds, and, expectedly, the difference $N_{\rm dec}-N_0$ is seen to vanish. The dashed lines assume that the curvaton decays instantaneously when $H(t_{\rm dec}) = \Gamma_{\phi}$ (cf. Sec.~\ref{sec:NonGaussianities}). 

Importantly, Fig.~\ref{fig:zeta} demonstrates that the outlined approach is independent of the distribution of the curvaton field at the end of inflation. Once $N_{\rm dec}(\phi_\star)$ has been computed, one can map the distribution of $\phi_\star$ onto the distribution of curvature perturbations via Eq.~\eqref{eq:deltaN_general}, obtaining
\be\label{eq:zeta_phi}
    \zeta = F(\phi_\star)\,.
\ee
Since the initial fluctuations of $\phi_\star$ are Gaussian, this relation essentially has the form \eqref{eq:FullNGs} defining a locally non-Gaussian field. However, in order to agree with \eqref{eq:FirstExpansion} when the perturbations are small, we will define the auxiliary Gaussian field through the curvaton perturbation $\delta \phi \equiv \phi_\star - \langle\phi_\star\rangle$ as
\be\label{def:zeta_G}
    \zeta_{\rm G} 
    \equiv {\pd_{\phi_\star}F(\langle\phi_\star\rangle)} \delta \phi \,,
\ee
so that $\pd_{\zeta_{\rm G}} F|_{\zeta_G=0} = 1$. This definition is especially suitable when $\zeta$ is approximately Gaussian, because then $\zeta \approx \zeta_{\rm G}$ at the leading order. When the corrections are small, then $\zeta_{\rm G}$ will qualitatively capture the behaviour of $\zeta$. For strong NGs, this definition may fail, especially when ${\pd_{\phi_\star}F(\langle\phi_\star\rangle)} = 0$. In this case, it is simpler to consider a standardised auxiliary Gaussian random field with unit variance
\be\label{def:hatzeta_G}
    \hat{\zeta}_{\rm G} 
    \equiv \frac{\delta \phi}{\sigma_{\phi}}\,,
\ee
where 
$
    \sigma_{\phi}^2 
    \equiv \langle \phi_\star^2\rangle - \langle\phi_\star\rangle^2
    \equiv \langle \delta \phi^2 \rangle  =  \int \frac{\td k}{k} \mathcal{P}_{\delta \phi}(k)\,
$
is the variance of the curvaton fluctuation. Here, $\calP_{\delta \phi}$ denotes the dimensionless perturbation spectrum defined through
\be 
\langle \delta\phi_{\bf k} \delta \phi_{\bf k'}\rangle  = (2 \pi)^3 \delta({\bf k}+{\bf k'}) \frac{2 \pi^2}{k^3} \calP_{\delta \phi} (k)\,.
\ee

Finally, the constant term is typically omitted in the local expansion \eqref{eq:FirstExpansion}, so that $F(\zeta_{\rm G} = 0) = 0$. However, a vanishing average \eqref{eq:zeta_mean} implies that we must impose
\be\label{eq:mean_zeta_cond}
\langle F(\zeta_{\rm G})\rangle = 0\,,
\ee
but, since $F$ can be constructed from the curvaton dynamics only up to an additive constant and without any reference to the distribution of $\phi_\star$, we will keep this constant implicit. That is, we will report $F(\zeta_{\rm G})$ (or $F(\hat\zeta_{\rm G})$), with the understanding that the average $\zeta = F(\zeta_{\rm G}) - \langle F(\zeta_{\rm G})\rangle$ must be subtracted whenever it does not vanish. Note that, by Eq.~\eqref{eq:PFLRW}, constant shifts in $\zeta$ can always be absorbed by the global scale factor, and therefore this procedure does not affect cosmological perturbations in any way. 

Based on the choices made in the local expansion, we will often report our results fixing the additive constant so that $F(\zeta_{\rm G} = 0) = 0$. Perturbatively, this is imposed by subtracting the contributions to the mean at each order in perturbation theory,
\bea\label{eq:FirstExpansion_mean}
    \zeta 
    = \sum_{n\geq1} F_{{\rm NL},n} \left( \zeta_{\rm G}^n - \langle\zeta_{\rm G}^n\rangle \right)\,
    = \zeta_{\rm G} + \frac{3}{5}f_{\rm NL}\,(\zeta_{\rm G}^2 - \langle\zeta_{\rm G}^2\rangle)  + \frac{9}{25}g_{\rm NL}\,\zeta_{\rm G}^3 + \dots\,,
\eea
where the first equality introduces the indexed non-linearity parameters $F_{{\rm NL},1} = 1$, $F_{{\rm NL},2} = (3/5)f_{\rm NL}$ and $F_{{\rm NL},3} = (9/25)g_{\rm NL}$, etc.

\subsection{Timeline of curvaton dynamics for general potentials}
\label{sec:NGsPot}

A generic curvaton model is characterised by a spectator field $\phi$ evolving in a general potential $V(\phi)$, whose energy density remains subdominant throughout the inflationary epoch. Following the standard curvaton scenario, after the end of inflation, the spectator field can become dynamically relevant and contribute significantly to the total energy density of the Universe. During this stage, the isocurvature perturbations generated by the spectator field are eventually converted into adiabatic curvature perturbations once the field decays~\cite{Mollerach:1989hu,Lyth:2004gb}. 

In this picture, a generic curvaton will proceed through the following stages:
\begin{enumerate}[leftmargin=*]
\renewcommand{\labelenumi}
{\textbf{\roman{enumi}.}}
    \item \textbf{Inflationary epoch.}
   The curvaton is subdominant, with
$\bar{\rho}_\phi \ll H_\star^2$,
where $H_\star$ is the approximately constant Hubble rate at the end of
inflation, $t_\star$. Hereafter, the subscript ``$\star$'' denotes quantities
evaluated at $t_\star$. Inflation erases initial inhomogeneities of a spectator field, leaving a nearly homogeneous background centred around $\langle \phi_\star \rangle \equiv \langle \phi(t_\star,\,{\bf x}) \rangle$. During this stage, we assume a negligible initial curvaton velocity,
so that its kinetic energy can be neglected, $K_\phi \simeq 0$,  and $\bar{\rho}_\phi \simeq \langle V(\phi_\star)\rangle$. For a light spectator field, quantum fluctuations are generated during inflation, such that at horizon exit the field acquires Gaussian and nearly scale-invariant perturbations with a typical amplitude $\delta\phi_\star \sim H_\star/2\pi$~\cite{Bardeen:1983qw}.

    \item \textbf{Frozen spectator.}
    After the end of inflation, the inflaton reheats the Universe by decaying into radiation. During this stage, the curvaton is frozen, meaning that the Hubble friction term prevents the field from rolling, since $H^{2} \gg |V''|$. At this point, the curvaton carries minuscule isocurvature perturbations generated by the primordial density contrast field $\zeta_\phi \sim \delta\rho_{\phi}/\bar{\rho}_{\phi}$.
    To prevent the curvaton from driving a secondary phase of inflation by dominating the energy density before the onset of oscillations, we require its initial energy density to remain subdominant throughout the frozen phase, namely $\bar{\rho}_\phi \ll H^2$ during this period. As seen in Fig.~\ref{fig:zeta}, the first curvature perturbations are generated in this phase, although their contribution is negligible when compared to the perturbations generated in the subsequent evolution.

    \item \textbf{Thawing and non-adiabatic oscillations.}
    In the radiation-dominated Universe, the Hubble rate decreases as $H^2 \propto a^{-4}$ and eventually becomes comparable to the effective frequency of the spectator field oscillations, defining the time $H_{\mathrm{osc}}^2 \sim |V''|$.
   The field therefore starts to accelerate as it rolls down the potential, and its kinetic energy becomes relevant to the curvaton energy budget.
    As we will discuss in Sec.~\ref{sec:oscillations}, this stage can be dominated by self-interactions in the spectator potential. This will determine the onset of oscillations and can be affected by the initial field perturbation. This phase is indicated by NA-OSC in Fig.~\ref{fig:zeta}, and is characterised by the oscillations in $N-N_0$. The duration of this phase is seen to differ for the two benchmark scenarios.

     \item \textbf{Adiabatic oscillations.} After a few oscillations about the minimum of the potential, the Hubble time $H^{-1}$ becomes much longer than the period of oscillations $T$, and the oscillations enter an adiabatic regime. Although both the kinetic and potential components contribute to the curvaton energy budget, in this regime, their time averages are related by the virial relation $2 \langle K_\phi\rangle \simeq \langle \phi V'(\phi)\rangle$~\cite{Turner:1983he,Cembranos:2015oya}. Therefore, the oscillating field can be described as an effective perfect fluid with equation of state $P_\phi=P_\phi(\rho_\phi)$. Here, $P_\phi$ denotes the time-averaged pressure of the field, while the energy density is determined by the oscillation amplitude $\Phi$, such that $\rho_\phi \simeq V(\Phi)$. This description avoids tracking the kinetic and potential energies separately, simplifying the computation. If the equation of state parameter drops below $P/\rho < 1/3$ as the curvaton evolves, its energy density will be less diluted than that of radiation. This will enhance the curvaton's contribution to the energy budget of the Universe. This is the case when the curvaton has a non-vanishing mass, such that its energy density will start scaling as $\rho_\phi \propto a^{-3}$ at late times. This phase is labelled as A-OSC in Fig.~\ref{fig:zeta} and displays the dominant growth of the curvature perturbation in both benchmark cases.

    \item \textbf{Decay.}
    Once $H \simeq \Gamma_{\phi}$, the curvaton will decay into radiation with a characteristic timescale $\Gamma_{\phi}^{-1}$. After that stage, only adiabatic curvature perturbations remain and the curvature perturbation can be seen to be frozen in Fig.~\ref{fig:zeta}. We stress that this picture is valid only for perturbations that remain superhorizon until that point.
\end{enumerate}

Accurately solving the curvaton dynamics is essential to determine the curvature perturbation generated after its decay. In the next subsections, we outline how to treat the different dynamical regimes of the curvaton in a general fashion, i.e. without specifying the potential $V(\phi)$.

\subsection{Thawing and post-inflationary dynamics} 
\label{sec:dynamics}

For the purpose of this work, we assume that the inflaton instantaneously decays into radiation after inflation ends, so that $\rho(t_\star) = 3 H_\star^2 = \rho_r(t_\star)$. Given this, a generic patch of the Universe can be described by a background radiation fluid with energy density $\rho_r = \rho_r(t)$ and pressure $P_r = \rho_r(t)/3$, together with a curvaton field that has a local initial value $\phi_\star \equiv\phi(\mathbf{x}, t_\star) $. Adapting the $\delta N$ formalism, we can treat the field in each Hubble patch as homogeneous with an initial value $\phi_\star$. The energy density and pressure of the curvaton are given by
\be
    \rho_\phi = \frac{1}{2} H^2 \left(\pd_{N}\phi\right)^2 + V(\phi)\,, 
    \qquad
    P_\phi = \frac{1}{2} H^2 \left(\pd_{N}\phi\right)^2 - V(\phi)\,,
\ee
where we have introduced the number of $e$-folds as the time variable, $\td N = H \td t = \td \log a$, with $N_\star = 0$ at the end of inflation.

The system then evolves according to the following equations: 
\bea\label{eq:EOM}
    \pd_{N}^2\phi
    + \left(3 + \frac{\pd_{N}H}{H} + \frac{\Gamma_{\phi}}{H} \right)
    \pd_{N}\phi
    + \frac{V'(\phi)}{H^2}
    &= 0\,, \\ 
    \pd_{N}\rho_r + 4 \rho_r - \Gamma_{\phi} H \left(\pd_{N}\phi\right)^2 
    &= 0\,,
\eea
and the Hubble rate is determined by the Friedmann equation
\be \label{eq:Friedmann1}
    \left[3 - \frac{1}{2} \left(\pd_{N}\phi\right)^2 \right]H^2 
    = V(\phi) + \rho_r\,.
\ee
As initial conditions, we set $\rho_r|_{N=0} = 3 H_\star^2$ and $\phi|_{N=0} = \phi_\star$, with vanishing initial velocity $\left.\pd_{N}\phi\right|_{N=0} = 0$. Immediately after inflation, the energy exchange between the curvaton and radiation can be neglected in Eq.~\eqref{eq:EOM}, as long as $\Gamma_{\phi} \ll H$. We stress that, in the context of the gradient expansion, Eq.~\eqref{eq:EOM} should be thought of as an equation of motion for the \emph{inhomogeneous} scalar field $\phi({\bf x},t)$, where all spatial gradients have been omitted as they are suppressed by $k/(aH)$ and therefore do not alter the field evolution at the leading order.

After inflation, the curvaton remains frozen. During this stage, it satisfies the slow-roll conditions $(V')^2/V \ll H^2$ and $|V''| \ll H^2$, so that the field velocity is negligible $\dot{\phi} \simeq 0$, and therefore behaves as a fluid with an equation of state $P_\phi \simeq -\rho_\phi$. Consequently, the spectator's energy density can be approximated as $\rho_\phi \simeq V(\phi_\star)$ up to the onset of oscillations at $H = H_{\mathrm{osc}}$. To avoid triggering a secondary phase of inflation, the curvaton must remain subdominant with respect to radiation throughout the interval $H_{\mathrm{osc}} \leq H \leq H_\star$.

In the purely quadratic case, a good estimate for the onset of oscillations is given by the condition that the Hubble rate becomes comparable to the mass, $H_{\mathrm{osc}} \simeq m$. For a general potential, curvaton dynamics is not determined by the mass and the criterion can be generalized as the condition where the Hubble rate becomes comparable to the angular frequency,
\be\label{eq:Hosc_cond}
    H_{\mathrm{osc}}\, T(\phi_{\star}) \lesssim 2\pi \, ,
\ee
where $T=T(\rho_{\phi,\star}) = 2\pi/\omega_{\phi,\star}$ is the initial period of oscillations at amplitude $\phi_{\star}$ estimated by neglecting Hubble friction (see e.g. Eq.~\eqref{eq:perioOsc}). Since $T = 2\pi/m$ for quadratic potentials, the standard case is trivially recovered. For a few $e$-folds around the onset of oscillations, the period $T$ is comparable to the Hubble timescale and the regime is non-adiabatic causing the fluid description to break down. For the purpose of our work, this intermediate non-adiabatic stage can either be determined numerically or be bypassed by matching the frozen regime directly onto the subsequent adiabatic oscillatory phase described in the next subsection.

\subsection{Adiabatic oscillations}
\label{sec:oscillations}

Expansion will quickly drive the Hubble timescale to become much longer than the period of oscillations. In this case, the Hubble friction can be treated adiabatically in the field equation, and it is possible to define an effective time-averaged equation of state $P_\phi(\rho_\phi)$ for the oscillating spectator field. This equation of state depends on the shape of the potential.\footnote{Such an effective description can break down if the spectator field fragments due to self-interactions. In that case, it will subsequently behave as radiation or dust~\cite{Lozanov:2019jxc}. Curvaton fragmentation is not considered in this work.}

In the regime of coherent oscillations, potentials of the form $V \propto \phi^n$ yield a barotropic index $w = (n-2)/(n+2)$~\cite{Turner:1983he}. More generally, in the adiabatic regime, the effective time-averaged equation of state can be expressed in terms of the abbreviated action~\cite{Tomberg:2021bll} (see appendix~\ref{app:potentials})
\be\label{eq:EoS_W}
    P(\rho_{\phi}) = \frac{W}{\pd_{\rho_{\phi}} W} - \rho_{\phi}\,, \qquad
    W(\rho_{\phi}) \equiv \int^{\Phi_1(\rho_{\phi})}_{\Phi_2(\rho_{\phi})} \td \phi \sqrt{2(\rho_{\phi} - V(\phi))}\,,
\ee
with the turning points $\Phi_i$ defined by $\rho_{\phi} = V(\Phi_i)$. This description is valid as long as the effect of Hubble friction is minimal during a single oscillation, requiring that $T H \ll 1$. The non-adiabatic behaviour at the onset of oscillations arises because the condition is not yet fully satisfied. The period of oscillations in the adiabatic regime is given by the abbreviated action as
\be \label{eq:period}
    T \equiv 2\pd_{\rho_{\phi}} W \,,
\ee
providing a simple way of tracking its evolution as the curvaton density is diluted by expansion. 

Once the field enters the oscillatory regime, each oscillation cycle requires several integration steps, and solving the full equations of motion~\eqref{eq:EOM} becomes increasingly slow as the oscillation frequency grows. Since this typically occurs deep in the adiabatic regime, we can instead stop the integration of~\eqref{eq:EOM} at some $H=\Lambda \ll H_\star$. The subsequent evolution is then accurately described in terms of interacting fluids
\bea \label{eq:fluidEOM}
    3 H^2 & = \rho_r + \rho_\phi \equiv \rho \, , \\
    \dot{\rho}_r + 4H \rho_r &= +\Gamma_{\phi} \rho_\phi \, ,\\
    \dot{\rho}_\phi + 3H(\rho_\phi + P_\phi) &= -\Gamma_{\phi} \rho_\phi \, ,
\eea
with initial conditions set by the solution of Eq.~\eqref{eq:EOM} at $N=N_\Lambda$. Interestingly, the continuity equation in Eq.~\eqref{eq:fluidEOM} can be recast in terms of $W$ as
\be \label{eq:Wcontinuity}
    \dot{W} + 3H W = - \Gamma \frac{\pd W}{\pd \ln \rho_\phi} \,,
\ee
where we have combined the equation of state $P(\rho_\phi)$ from Eq.~\eqref{eq:EoS_W} with the continuity equation. The abbreviated action $W$ thus plays a role analogous to the entropy density for fluids in local thermal equilibrium. In particular, when the decay is negligible ($\Gamma_{\phi} \ll  H$), $W(\rho_\phi)$ is an adiabatic invariant and is conserved in a comoving volume,
\be \label{eq:Wconserved}
      W(\rho) \propto a^{-3}.
\ee
Therefore, it determines how the expansion dilutes the energy density of a coherently oscillating curvaton $\rho_\phi$. The curvaton's energy density is related to the amplitude of the oscillation $\Phi$ as $\rho_\phi \simeq  V(\Phi)$,\footnote{We use symmetric potentials in all of the worked examples in this paper, so the turning points are $\Phi_i \equiv \pm \Phi$.} so when compared to the scalar field EoM~\eqref{eq:EOM}, Eq.~\eqref{eq:fluidEOM} tracks the envelope of the $\phi$ while disregarding the phase.

\begin{figure}
	\begin{center}
		\includegraphics[width=0.7\textwidth]{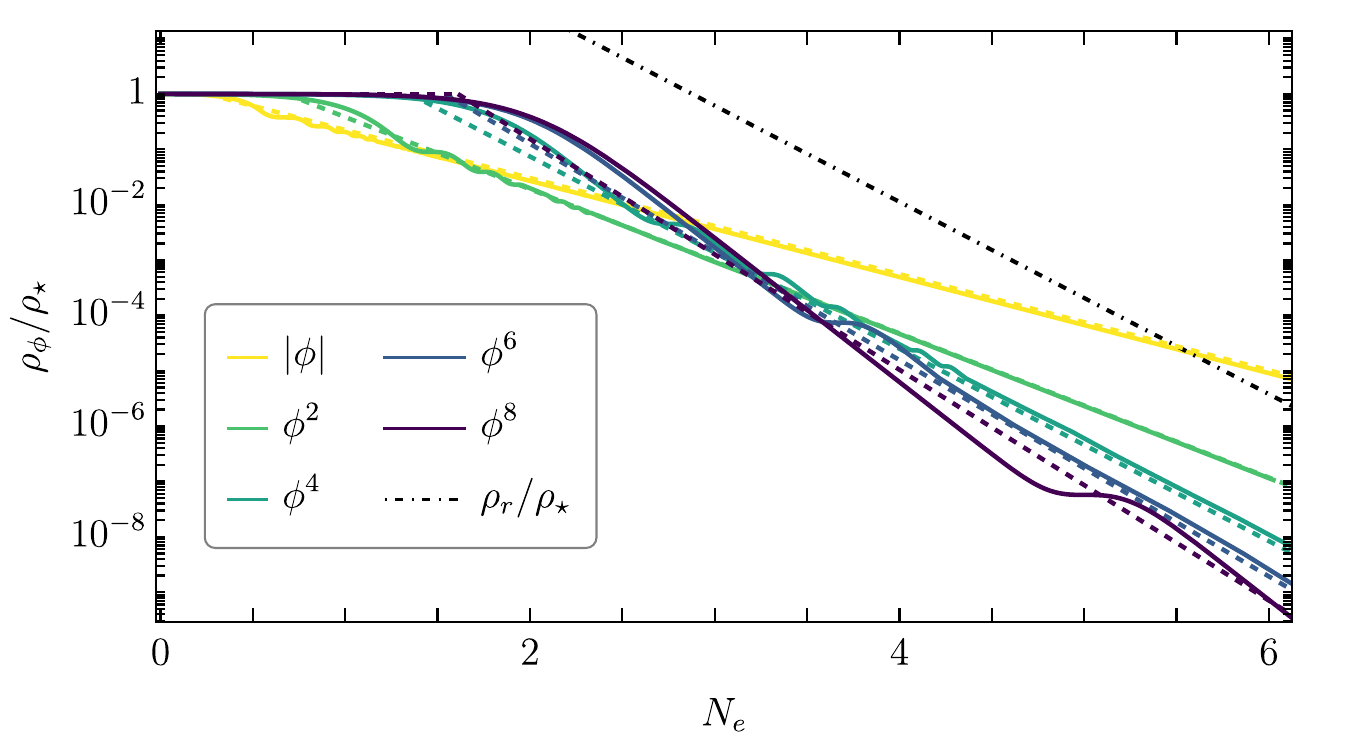}
		\caption{Comparison between the numerical evolution (solid lines) of the energy density and its approximate scaling given by Eq.~\eqref{eq:Wevol} (dashed lines) for monomial potentials $V = \lambda \left|\phi\right|^n$. The onset of the oscillatory regime for even $n$ is determined using Eq.~\eqref{eq:Hosc}. For the linear case, instead, we use the analytic result outlined in Section~\ref{sec:linear_V}. In all cases, we fix $H_\star = 10^{-7}$, $\rho_{\phi,\star}/\rho_\star = 10^{-5}$, and $\lambda = 10^{-15}$.}
	\label{fig:density}  
	\end{center}
\end{figure}

\subsection{Instantaneous thawing approximation}
\label{sec:insta_thawing}

If the onset of oscillations can be determined analytically, one can bypass the need to explicitly model the intermediate non-adiabatic regime. In this case, the frozen phase can be directly glued to the subsequent adiabatic evolution in such a way that the transition from a frozen to an adiabatically oscillating spectator is instantaneous when $H = H_{\rm ad}$, so that the time evolution can be described by
\be\label{eq:Wad}
    W = 
    \begin{cases}
        W_\star\,,        \quad & H > H_{\rm ad}\,,\\
        W_\star \left( a/a_{\rm ad}\right)^{-3} \approx W_\star \left( H/H_{\rm ad}\right)^{3/2} \,, \quad &  H \leq H_{\rm ad}\,,
    \end{cases}
\ee
where we defined $W_{\star} \equiv  W\left( V(\phi_{\star})\right)$ and used the $W \propto a^{-3}$ scaling law~\eqref{eq:Wconserved} and, up to the onset of the adiabatic regime, the Universe is radiation-dominated, so that $H \propto a^{-2}$.
Although $H_{\rm ad}$ is approximated roughly with $H_{\rm osc}$ as given by Eq.~\eqref{eq:Hosc_cond}, it must be estimated numerically in general cases by first evolving the full system~\eqref{eq:EOM} until $HT \ll 1$ is satisfied at some $a_{\rm ref}$ and then matching the solution to~\eqref{eq:Wad} so that $W(\rho_{\phi}(a_{\rm ref})) = W_\star \left( a_{\rm ref}/a_{\rm ad}\right)^{-3}$.
In some cases, $H_{\rm ad}$ can be estimated analytically. For instance, for monomial potentials 
\be
    V = \lambda |\phi|^n\,
\ee
for which
\be\label{eq:Wevol}
    W(\rho_\phi) \simeq \frac{4 \sqrt{2}}{2+n} \lambda^{-1/n} \rho_\phi^{\frac{n+2}{2n}} \,,
\ee
we can estimate that (for the derivation, see App.~\ref{app:mono})
\be\label{eq:Hosc}
    H_{\rm ad} 
    = A(n)\, \sqrt{\frac{\pi}{2}}\,
    \frac{\Gamma\!\left(\tfrac{1}{2} + \tfrac{1}{n}\right)}{\Gamma\!\left(1+\tfrac{1}{n}\right)}\,
    \sqrt{\lambda} \phi_\star^{\frac{n}{2}-1}\,,
\ee
where $A(n)$ is an $\mathcal{O}(1)$ coefficient. For an even $n \geq 2$, we find that $A(n) = \sqrt{(n-1)/3}$ works very well. A comparison between solutions of~\eqref{eq:EOM} and the instantaneous thawing approximation is shown in Fig.~\ref{fig:density} for a few benchmark monomial potentials. It shows that apart from cases in which the non-adiabatic oscillatory phase persists for a relatively long time (such as $n=6, 8$), this intermediate stage can be safely neglected by performing a direct matching. We also note that $H_{\rm ad}  \propto \sqrt{V''(\phi_\star)}$ (when $n \neq 1$), in agreement with the naive estimate $H_{\rm ad} \approx m$.

\subsection{Instantaneous decay approximation}
\label{sec:insta_dec}

The instantaneous thawing approximation is most useful when combined with the approximation in which the curvaton decays instantaneously into radiation when
\be\label{eq:insta_dec}
     H_{\rm dec} = \Gamma_{\phi}
\ee
as it can reduce reliance on numerical techniques. In that case, the energy density at the decay is determined by
\be\label{eq:W_dec}
    W(\rho_{\phi, \rm dec}) = W_\star \left(a_{\rm dec}/a_{\rm ad}\right)^{-3}\,.
\ee
In this case, $\rho_{\phi, \rm dec}$ can be fixed by the decay criterion \eqref{eq:insta_dec} and the decay parameter $r_{\rm dec}$. This relates the number of $e$-folds and allows us to derive an analytical prescription, valid for an arbitrary curvaton potential, to determine the non-Gaussian curvature perturbation $\zeta$.

The above reasoning can be outlined in terms of the standard derivation of curvaton NGs from the instantaneous decay approximation~\cite{Sasaki:2006kq}. To this end, we begin by noting that for a barotropic fluid, such as radiation or the curvaton during its adiabatic regime, the corresponding curvature perturbation is conserved on super-horizon scales when energy transfer between the fluids is negligible~\cite{Lyth:2004gb}. For each isolated component $i$, we therefore have
\bea \label{eq:cons_zeta_i}
    \zeta_i({\bf x}) 
    &= \delta N({\bf x}, t) + \frac{1}{3}\ln \frac{W_i(\rho_i({\bf x}, t))}{W_i(\bar{\rho}_i(t))} \,,
    \qquad
    W_i(\rho) = \exp \left( \int \frac{\td \rho}{\rho + P_i(\rho)} \right)\,,
\eea
where $\bar{\rho}_i$ denotes a background density. In this formulation, a uniformly oscillating scalar in its adiabatic regime is treated identically to a thermal bath of particles. Notably, the conservation of $\zeta_i$ is directly related to the scaling property $W \propto a^{-3}$ and implies that
\be \label{eq:cons_zeta_i_integrated}
    W_i(\rho_i({\bf x}, t)) = W_i(\bar\rho_i(t)) e^{3(\zeta_i({\bf x}) - \delta N({\bf x},t))}\,.
\ee
Since we only have two components -- radiation and the curvaton -- we will drop the index and reserve $W$ for the curvaton only while treating radiation in the usual way.\footnote{We would have $W_r \propto \rho_r^{3/4}$ for radiation and $W_m \propto \rho_m$ for non-relativistic matter.}

A uniform density hypersurface at a time $t$ is described by
\be \label{eq:decaysurface}
    \rho_\phi({\bf x}, t) + \rho_r({\bf x}, t) = \bar{\rho}(t)\,.
\ee
Since a uniform density slicing implies that $\zeta = \delta N$, Eq.~\eqref{eq:cons_zeta_i_integrated} allows us to recast inhomogeneities in the fields as
\be\label{eq:decayW}
    \frac{1}{\bar \rho(t)} W^{-1}\left(W(\bar \rho_{\phi}(t)) e^{3(\zeta_{\phi} ({\bf x}, t)-\zeta({\bf x}, t))} \right)
    + (1-\bar\Omega_{\phi}) e^{4(\zeta_{r} ({\bf x}, t)-\zeta({\bf x}, t))} = 1\,,
\ee
where $\Omega_\phi \equiv \bar{\rho}_\phi/\bar{\rho}|_{t}$ is defined in terms of background energy densities, $W^{-1}$ denotes the inverse function of $W$, i.e., $W^{-1}(W(\rho_{\phi}))= \rho_{\phi}$. For notational simplicity, the explicit time dependence of the background quantities has been suppressed. This relation holds for an arbitrary curvaton potential and at any time during the adiabatic regime. From now on, the explicit dependence on space and time coordinates will be omitted whenever no confusion arises, with $\zeta$ being the final curvature perturbation after curvaton decay and $\zeta_\phi$ the conserved curvature perturbation associated with the curvaton during the adiabatic regime. 

Assuming that the curvaton decays instantaneously into radiation at $H=\Gamma_{\phi}$, no energy transfer between the scalar field and radiation takes place before reaching the decay hypersurface, so that $\zeta_\phi$ remains conserved throughout the adiabatic regime up to the decay time. Identifying the uniform total density slicing with the decay hypersurface at time $t_{\rm dec}$, defined by $H=\Gamma_{\phi}$, gives the relation between $\zeta_{\phi}$ and $\zeta$. In the standard curvaton paradigm, the pre-existing radiation component carries only a subdominant and nearly Gaussian fluctuation inherited from the inflaton sector, while the dominant super-horizon curvature perturbation is generated by the curvaton. In this case, the subdominant contribution $\zeta_{r}$ can be neglected, and it remains only to relate $\zeta_{\phi}$ to the initial fluctuation in the curvaton field $\phi_\star$.

Given the conservation of curvature perturbations~\eqref{eq:cons_zeta_i} and an initially flat slicing ($\delta N=0$), together with the definition~\eqref{eq:Wad}, we have that during the whole adiabatic regime
\be \label{eq:zetaphi}
    e^{3\zeta_{\phi}} \big |_{t_{\rm dec}}
    = e^{3\zeta_{\phi}} \big |_{t_{\rm ad}}
    = \frac{W(\rho_\phi)}{W(\bar{\rho}_\phi)}\,, 
\ee
where $t_{\rm ad}$ denotes the beginning of the adiabatic regime. 

Combining Eq.~\eqref{eq:Wad} with Eq.~\eqref{eq:zetaphi}, the curvature perturbation generated by the curvaton at decay is given by
\be \label{eq:zetaW1}
    e^{3\zeta_{\phi}} 
    =\frac{W(V(\phi_\star))}{W(V(\bar\phi_\star))} 
    \left( \frac{H_{\rm ad}(\phi_\star)}{H_{\rm ad}(\bar\phi_\star)}\right)^{-3/2} \,.
\ee
Plugging this result into Eq.~\eqref{eq:decayW}, once a background $\bar{\phi}_\star$ is fixed, allows one to determine the full non-Gaussian curvature perturbation as a function of the initial field value, namely $\zeta=\zeta(\phi_\star)$. Explicitly, Eq.~\eqref{eq:decayW} gives
\bea\label{eq:zeta_phistar}
    e^{4\zeta} - \frac{e^{4\zeta}}{\bar \rho} W^{-1}
    \left[ W(\bar\Omega_{\phi} \bar \rho) \frac{W(V(\phi_\star))}{W(V(\bar \phi_\star))}
    \left(\frac{H_{\rm ad}(\phi_\star)}{H_{\rm ad}(\bar\phi_\star)}\right)^{-3/2} e^{-3\zeta} \right]
    + \bar\Omega_{\phi} - 1  = 0\,,
\eea
where we remind that $\bar\Omega_{\phi} \equiv \bar \rho_{\phi}/\bar \rho$ corresponds to the initial background field $\bar\phi_\star$, while the final background field $\bar\phi$ is independent of it as it is fixed by $H=\Gamma_{\phi}$.

Eqs.~\eqref{eq:zetaW1} and \eqref{eq:zeta_phistar} show that the dependence on $\phi_\star$ arises from two sources: \emph{a)} $W$ depends on $\phi_\star$ as it sets the \emph{initial energy density} $\rho_{\phi,\star} = V(\phi_\star)$ in the instantaneous thawing approximation~\eqref{eq:insta_dec}, and \emph{b)} the field dependence in $H_{\rm ad} = H_{\rm ad}(\phi_\star)$ sets the \emph{timing} at which the frozen curvaton transitions into an adiabatically oscillating field. Importantly, it accounts for the non-adiabatic evolution during the first oscillations. We stress that the instantaneous thawing approximation is therefore accurate as long as the decay takes place in the adiabatic regime, $HT \ll1$.

\subsubsection{Choosing the background}
\label{sec:background}

At this stage, it is important to clarify the meaning of background quantities entering the definition of $\zeta$. The curvaton perturbation $\zeta_\phi$ is defined with respect to a background energy density $\bar \rho$, which implicitly determines the \emph{initial} background field $\bar \phi_\star$ so that $\bar \rho_\phi = \rho_\phi(\bar \phi_\star)$ at the decay surface.

We emphasize freedom in choosing $\bar\phi_\star$ since different choices for fixing $\bar\phi_\star$ are related by a global shift in $\zeta$ and thus will not affect the NGs of curvature perturbations. This is most intuitive in the context of the $\delta N$ formalism, where a choice of $\bar \rho_\phi$ corresponds to a choice of a reference scenario for the number of $e$-folds
\be
    \bar N \equiv N(\bar\phi_\star)
\ee
that in turn is used to define 
\be \label{eq:deltaNDef}
    \delta N \equiv N(\phi_\star) - \bar N = \zeta
\ee 
so it can only affect the curvature perturbations up to an irrelevant global additive constant, as stated above. 

An important consequence of this argument is that the background $\phi_\star = \bar\phi_\star$ corresponds to the absence of curvature perturbations $\zeta = 0$.\footnote{This also follows from Eq.~\eqref{eq:zeta_phistar}, which gives
\begin{equation*}
    \zeta = 0 \,: 
    \qquad
    W^{-1}
    \left[ W(\bar\Omega_{\phi} \bar \rho) \frac{W(V(\phi_\star))}{W(V(\bar \phi_\star))}
    \left(\frac{H_{\rm ad}(\phi_\star)}{H_{\rm ad}(\bar\phi_\star)}\right)^{-3/2} \right]
    = \bar\Omega_{\phi}\bar \rho\,.
\end{equation*}
This equation is solved by $\phi_\star = \bar\phi_\star$.} Thus, a typical choice is $\bar\phi_\star = \langle \phi_\star \rangle$, so that the curvature fluctuation vanishes for the mean Gaussian fluctuation $F(\langle \phi_\star \rangle) = 0$. For instance, this choice was left implicit in the seminal study~\cite{Sasaki:2006kq}. This choice can be problematic when $V(\langle \phi_\star \rangle) = 0$, as it implies that the curvaton's background energy density vanishes. 

This highlights that $\bar\Omega_{\phi} \equiv \rho_\phi(\bar \phi_\star)/\rho$ will generally not coincide with the cosmological average over all Hubble patches,
\be
    \langle \Omega_{\phi} \rangle 
    \equiv \langle \rho_{\phi}(\phi_\star) \rangle / \rho\,.
\ee
Demanding that $\bar\Omega_{\phi} = \langle\Omega_{\phi}\rangle$ leads to a subtle but important point that is often overlooked in the literature: in general, the ensemble average of the field, $\langle \phi_{\star} \rangle$, does not coincide with the background $\bar \phi_\star$. This mismatch has a purely statistical origin. While $\langle \phi_{\star} \rangle$ characterises the mean of the field distribution, $\bar \phi_\star$ is instead defined through the mean energy density, which depends non-linearly on the field.

This distinction is particularly relevant when field fluctuations are sizeable, namely, when the initial variance $\sigma_{\phi\star}^2$ of the curvaton perturbations, $\delta \phi \equiv \phi_\star - \langle\phi_\star\rangle$, is such that $\sigma_{\phi\star} \gg \bar\phi_\star$. In this regime, the curvaton's energy density is dominated by fluctuations rather than by the homogeneous field, and the condition determining the background $\bar \phi_\star$ generally depends on the primordial power spectrum of curvaton fluctuations in addition to the mean $\langle \phi_{\star} \rangle$. By contrast, in the opposite regime, $\bar\phi_\star \gg \sigma_{\phi\star}$, field fluctuations become negligible, and one finds $\langle \phi_\star \rangle \simeq \bar\phi_\star$ to excellent accuracy. In this limit, the standard treatment commonly adopted in the literature is fully justified.

Overall, the shifting argument above shows that the choice of the background will not affect the NGs of curvature perturbations but rather the interpretation of $\bar\Omega_{\phi}$ (or the parameter $r$). For this reason, we will consider different conditions for the background. For instance, in Fig.~\ref{fig:zeta}, the background was effectively fixed at $\bar \phi_\star = 0$. A reasonable condition for fixing $\bar\phi_{\star}$ that approximately preserves the average is\footnote{Assuming the curvaton field is Gaussian, the averages are given by
\begin{equation*}
\langle V(\phi_\star)\rangle 
    \equiv \int \frac{\td \phi}{\sqrt{2 \pi} \sigma_{\phi}}
    V(\phi)\,
    \exp\left(- \frac{(\phi - \langle\phi_\star\rangle)^2}{2 \sigma_\phi^2}\right)\,.
\end{equation*}}
\be
    V(\bar\phi_\star ) 
    = \langle V(\phi_\star)\rangle 
\ee
This condition implies that the energy density of the background after inflation agrees with the cosmological average. However, it does not account for the non-linear evolution after thawing, which can result in some disagreement between the average curvaton density and the background density at the decay hypersurface.

Finally, we emphasize that, in order to interpret $\zeta$ as describing fluctuations about a vanishing mean, one may always perform an appropriate shift \emph{after} the relation $\zeta = F(\zeta_{\rm G})$ has been derived to enforce the condition $\langle \zeta \rangle = 0$ in Eq.~\eqref{eq:mean_zeta_cond}. Consequently, the specific choice of a background is inconsequential from the point of view of characterizing NGs.

\subsubsection{Decays in the quadratic regime}
\label{sec:quadratic_dec}

As Hubble friction gradually damps the amplitude of oscillations, the curvaton will eventually be driven to a  minimum of its potential. In the case of smooth potentials, the behaviour around the minimum is expected to be quadratic. Although reaching a quadratic regime before decay is not necessary, as shown in Sec.~\ref{sec:linear_V}, it naturally increases the energy fraction of the curvaton, which is a necessary condition to induce a non-negligible curvature perturbation at its decay.

According to Eq.~\eqref{eq:Wevol}, one has $W(\rho_\phi)\propto \rho_\phi$ in the quadratic regime so that, with $\zeta_R \simeq 0$, Eq.~\eqref{eq:decayW} simplifies to
\be \label{eq:sasaki}
    e^{4\zeta}
    - \bar\Omega_{\phi,\rm dec}e^{3\zeta_{\phi}+\zeta} 
    + \bar\Omega_{\phi,\rm dec} - 1 = 0\,,
\ee
recovering the well-known relation~\cite{Sasaki:2006kq}.
It is a quadratic equation in $e^{\zeta}$ and can be solved analytically
\be \label{eq:MasterX}
    \zeta = 
    \ln\left[X\left(\frac{r_{\rm dec}}{3+r_{\rm dec}} e^{3\zeta_\phi},\frac{1-r_{\rm dec}}{3+r_{\rm dec}}\right)\right]\,,
\ee
where we defined the auxiliary functions
\bea
    X(A,B) &= \frac{1}{\sqrt 2} \left[Q(A,B) + \sqrt{\sqrt{8}A/Q(A,B)-Q(A,B)^2}\right]\,,
    \\
    Q(A,B) &= \sqrt{\left(A^2 + \sqrt{A^4+B^3}\right)^{1/3} - B \left(A^2+\sqrt{A^4+B^3}\right)^{-1/3}}\,
\eea
that give the relevant branch of the solution of the quartic equation $X^4-4A X-3 B=0$.

The leading-order perturbative solution for small $\zeta_{\phi}$ reads
\be\label{eq:zetaR}
    \zeta = r_{\rm dec}\zeta_{\phi} + \mathcal{O}(\zeta_{\phi}^2)\, ,
\ee
where
\be \label{eq:rdec}
    r_{\rm dec} \equiv \frac{3 \bar\Omega_{\phi,\rm dec}}{4 - \bar\Omega_{\phi,\rm dec}}
\ee
The parameter $r_{\rm dec}$ thus quantifies the efficiency of converting the curvaton perturbation into the final curvature perturbation. In particular, it encodes the entire background evolution of the curvaton relative to radiation between the onset of oscillations and the decay time. As a consequence, Eq.~\eqref{eq:sasaki} depends on the curvaton dynamics only through the single parameter $r_{\rm dec}$, without explicit dependence on the decay rate $\Gamma_{\phi}$ or on other parameters of the curvaton potential, which arise instead when relating the fluctuations in $\phi_\star$ to $\zeta_\phi$. This is not entirely surprising since Eq.~\eqref{eq:sasaki} assumes that curvaton dynamics is completely dominated by the quadratic minimum at the time of decay.

\section{Non-Gaussianities in various curvaton models}
\label{sec:NonGaussianities}

\subsection{The free curvaton: quadratic potentials}
\label{sec:quadratic_V}

The standard case that has been extensively studied in the literature is the free, non-interacting curvaton
\be \label{eq:quadraticV}
    V(\phi) = \frac{1}{2} m^2 \phi^2 \,.
\ee
It is important not only for its simplicity but also because, in most curvaton models, the quadratic regime is reached relatively quickly after the onset of oscillations. Thus, as shown in Sec.~\ref{sec:quadratic_dec}, the difference in the amount of NG between such models arises primarily from differences in the thawing conditions of the field.

To make direct contact with the $\delta N$ formalism introduced in Sec.~\ref{sec:deltaN}, we consider the evolution of two patches of the universe: one populated by a free curvaton field with an initial value $\phi_\star$, and another consisting of a pure radiation fluid. Under the instantaneous decay approximation, the decay rate $\Gamma_{\phi}$ can be neglected in Eqs.~\eqref{eq:fluidEOM}, so that the curvaton and radiation evolve independently up to the decay hypersurface. Fixing $N=0$ at the instantaneous thawing condition, $H=H_{\rm ad}$, the universe in the two patches expands by different numbers of $e$-folds before reaching the decay hypersurface
\bea \label{eq:quadEvol}
    3 \Gamma_\phi^2 & = 3 H_{\rm ad}^2 e^{-4N_0} \, , \\
    3 \Gamma_\phi^2 &= 3 H_{\rm ad}^2 e^{-4N_{\rm dec}} + V(\phi_\star) e^{-3N_{\rm dec}} \,.
\eea
Imposing $H_{\rm ad}=m/\sqrt{3}$ from Eq.~\eqref{eq:Hosc} and combining the two equations above, we obtain
\be \label{eq:UniversalQuad}
    e^{4\delta N} - \frac{3^{3/4}}{6}\sqrt{\frac{m}{\Gamma_\phi}}\,\phi_\star^2\,e^{\delta N}-1=0 \,,
\ee
where $\delta N \equiv N_{\rm dec}-N_0$. By equating $\zeta = \delta N$ one can show that this equation is identical to Eq.~\eqref{eq:sasaki}.\footnote{%
Choosing a background with $\bar{\phi}_\star \neq 0$, $\bar{N}$ can be defined through
$3 \Gamma_\phi^2 = 3 H_{\rm ad}^2 e^{-4\bar N} + V(\bar \phi_\star) e^{-3\bar N}$.
Combining this with the second line of Eq.~\eqref{eq:quadEvol} we obtain
\be \label{eq:UniversalQuad2}
    e^{4 \zeta}
    - \left( \frac{m^2}{6 \Gamma_\phi^2}\,\bar \phi_\star^2\, e^{-3 \bar N} \right)
      \left( \frac{\phi_\star}{\bar \phi_\star} \right)^2 e^{\zeta}
    + \left( \frac{m^2}{6 \Gamma_\phi^2}\,\bar \phi_\star^2\, e^{-3 \bar N} \right) - 1
    = 0 \,,
\ee
where we used the definition~\eqref{eq:deltaNDef} for the curvature perturbation $\zeta$.
Since, for a quadratic potential, Eqs.~\eqref{eq:Wevol} and~\eqref{eq:Hosc} imply
\be \label{eq:W,Had_quad}
    W \propto \rho_{\phi} \propto \phi_{\star}^2\,,
    \quad
    H_{\rm ad} = \mathrm{const.}
    \qquad \stackrel{\mbox{\eqref{eq:zetaW1}}}{\Rightarrow} \qquad
    e^{3\zeta_{\phi}} = (\phi_{\star}/\bar\phi_{\star})^2\,,
\ee
we obtain Eq.~\eqref{eq:sasaki}, with $\bar \Omega_{\phi,{\rm dec}} = (m^2/6 \Gamma_\phi^2)\,\bar \phi_\star^2\, e^{-3 \bar N}$.}
Importantly, Eq.~\eqref{eq:UniversalQuad} illustrates the statement made in Sec.~\ref{sec:background}: the NGs of the curvature perturbation depend solely on the non-linear relation between $\phi_\star$ and $\zeta$ and are independent of the choice of the background. Moreover, the free curvaton exhibits a universal profile that depends on the combination $(m/\Gamma_\phi)^{1/4}\phi_\star$.

\begin{figure*}[t!]
\centering
\begin{minipage}{.5\textwidth}
  \centering
  \includegraphics[width=\linewidth]{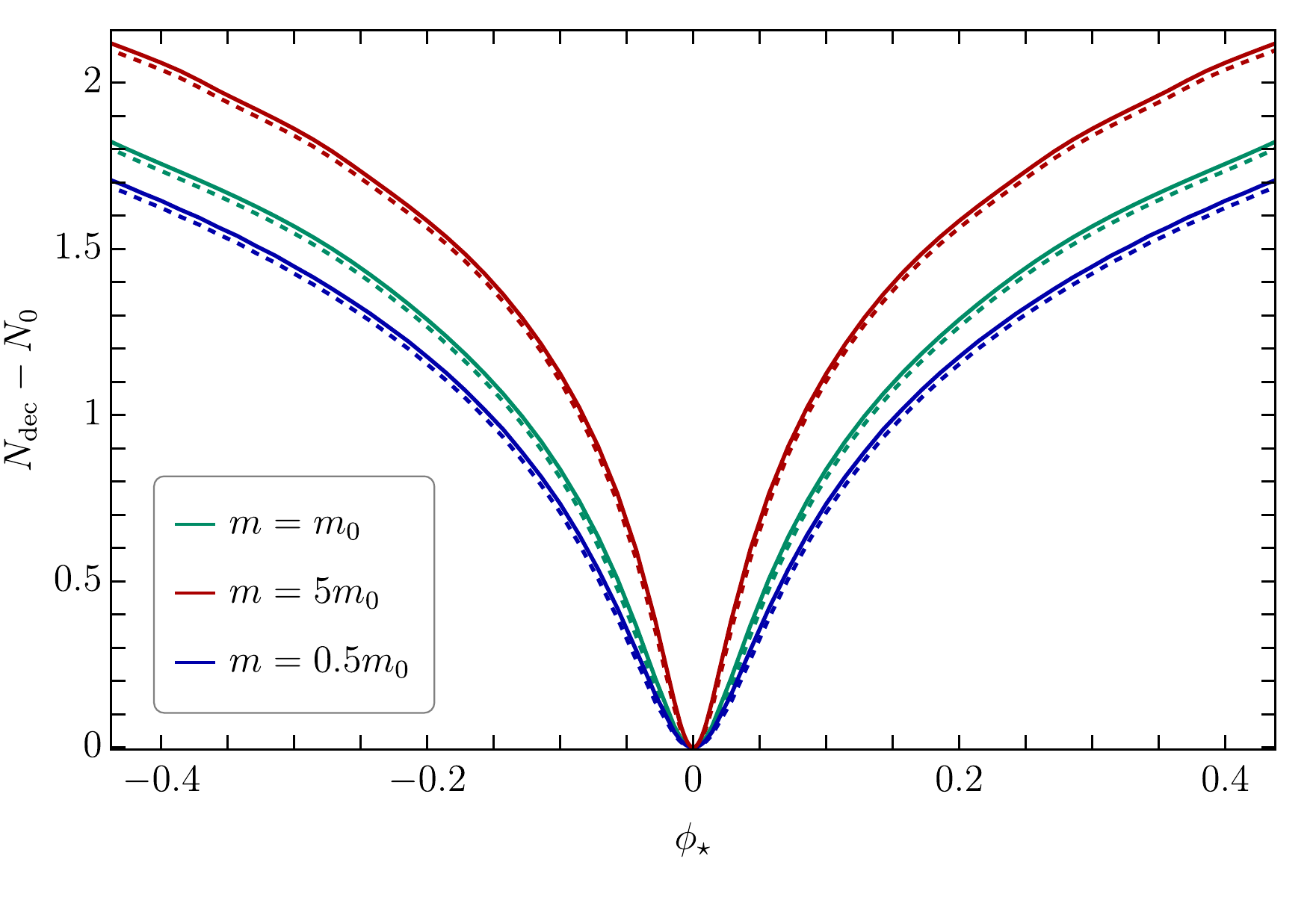}
\end{minipage}\hfill
\begin{minipage}{.5\textwidth} 
  \centering
  \includegraphics[width=\linewidth]{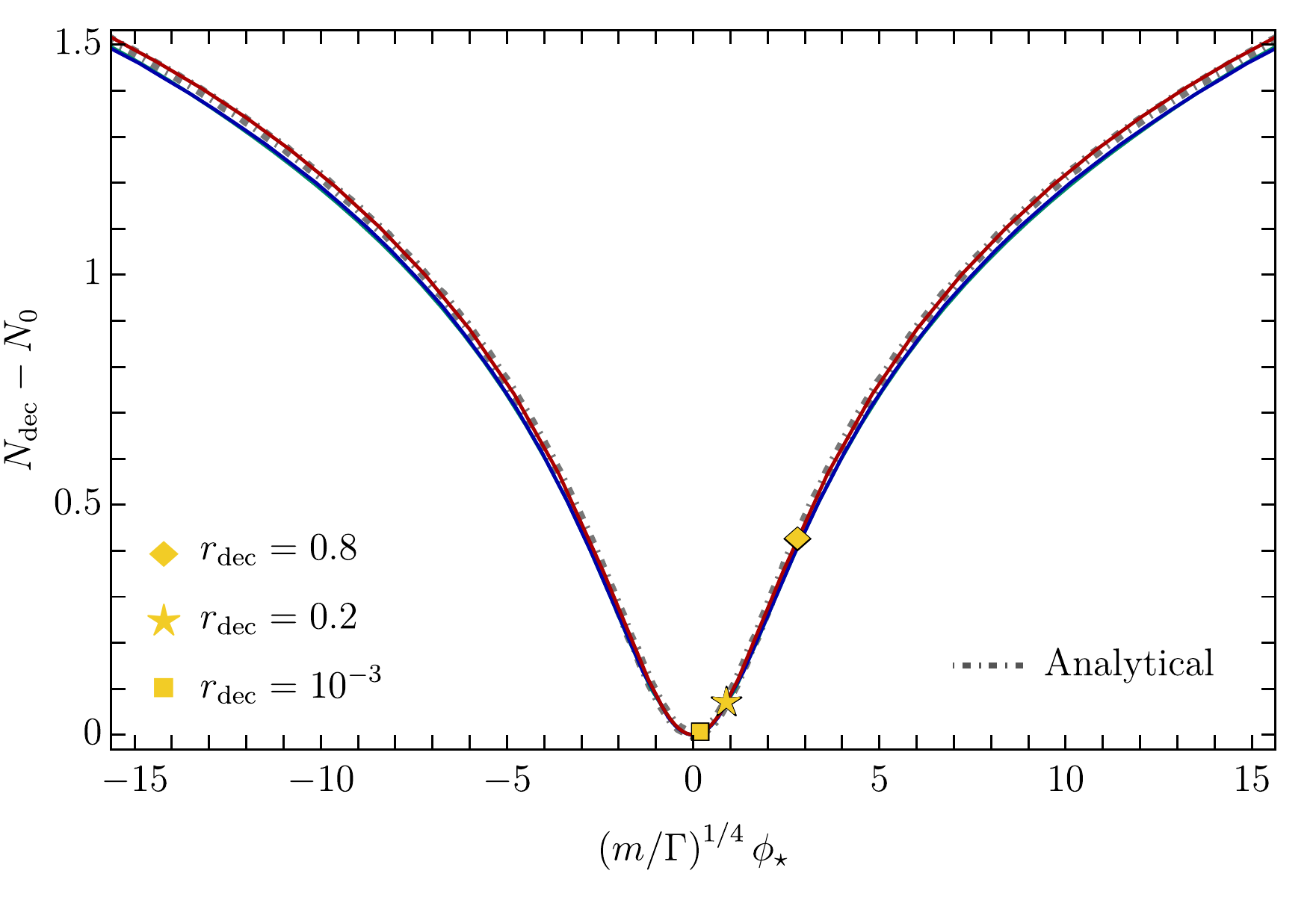} 
\end{minipage}
\caption{{\it Left panel}: Comparison between the numerical results for instantaneous {\it (dashed lines)} and non-instantaneous {\it (solid lines)} decay in the quadratic curvaton model~\eqref{eq:quadraticV}, for three different masses $m$. The numerical solutions are obtained for $m_0 = 10^{-8}$, $H_\star = 10^{-7}$, and $\Gamma_\phi = 10^{-15}$. {\it Right panel}: Comparison between the numerical curves in the left panel and the analytical universal scaling from Eq.~\eqref{eq:UniversalQuad}. The highlighted points mark the choices of the background field $\bar{\phi}_\star$ corresponding to the different $r_{\rm dec}$ shown in Fig.~\ref{fig:quad2}.}
    \label{fig:quad}
\end{figure*}

In the right panel of Fig.~\ref{fig:quad}, we compare the analytical solution of Eq.~\eqref{eq:UniversalQuad} with the corresponding numerically computed curves. The numerical $\delta N$ estimates are obtained by solving Eqs.~\eqref{eq:EOM} and~\eqref{eq:fluidEOM} for three different curvaton masses. In the instantaneous decay approximation, we stop the numerical integration at fixed $H=\Gamma_\phi$. The left panel of Fig.~\ref{fig:quad} demonstrates the accuracy of the instantaneous decay approximation.

To allow comparison with existing literature, it is useful to define the Gaussian variable $\zeta_{\rm G}$ by adopting the standard convention $\bar\phi_{\star} = \langle \phi_\star \rangle$ and writing
$\zeta = \zeta_{\rm G} + \ldots$ at the linear order~\eqref{eq:FirstExpansion}. Applying this to Eq.~\eqref{eq:W,Had_quad}, we recover the well-known result~\cite{Sasaki:2006kq, Pi:2021dft, Ferrante:2023bgz}
\be    \label{eq:zetaphiQuad}
    \zeta_{\phi} = \frac{2}{3}\ln \left|1+\frac{3}{2 r_{\rm dec}}\zeta_{\rm G}\right|\,,
    \qquad \mbox{where} \qquad
    \zeta_{\rm G} 
    = \frac{2 r_{\rm dec}}{3}\frac{\delta \phi}{\langle \phi_\star \rangle}\,.
\ee
The extra factor of $1/r_{\rm dec}$ arises from the expansion~\eqref{eq:zetaR}, and the absolute value appears because the right-hand side of the last equation in Eq.~\eqref{eq:W,Had_quad} is always positive.
As discussed in Sec.~\ref{sec:background}, the efficiency parameter $r_{\rm dec}$ corresponds to the curvaton energy density on the decay surface when $\phi_{\star} = \langle \phi_\star \rangle$, rather than to the average energy density, which would read instead
\be\label{eq:omega_phi_quad}
    \Omega_{\phi} 
    = \bar\Omega_{\phi} \left(1 + \sigma^2_{\phi}/\langle \phi_\star\rangle^2 \right)\,.
\ee
In particular, if $\sigma_{\phi} \gg \langle \phi_\star\rangle$, then $\bar\Omega_{\phi}$ is only weakly related to the cosmological average $\Omega_{\phi}$. The commonly used convention can be modified by defining $r_{\rm dec}$ via~\eqref{eq:omega_phi_quad}. This comes at the cost of introducing a dependence on $\sigma_{\phi}$ in $F$.

Inserting Eq.~\eqref{eq:zetaphiQuad} into the closed-form solution~\eqref{eq:MasterX} allows us to find an analytical expression for $F(\zeta_{\rm G})$ in the free-curvaton scenario.
The closed-form solution in Eq.~\eqref{eq:MasterX} exhibits a characteristic logarithmic behaviour at large $\zeta_{\rm G}$, with a non-trivial dependence on both $r_{\rm dec}$ and $\zeta_{\rm G}$.
Remarkably, in the limit $r_{\rm dec}\rightarrow 1$, Eq.~\eqref{eq:sasaki} yields $\zeta=\zeta_\phi$, so that $F(\zeta_{\rm G})$ reduces to a single logarithm of $\zeta_{\rm G}$ and develops the exponential-tail behaviour typical of ultra-slow roll (USR) models of inflation~\cite{Atal:2018neu,Tomberg:2023kli}. This is not a coincidence: as pointed out in Ref.~\cite{Pi:2022ysn}, the logarithmic structure of the $\delta N$ map in single-field inflation with a piecewise-quadratic potential and that of the curvaton scenario in the sudden-decay approximation share the same algebraic origin.
This exact solution develops a singularity at $\zeta_G = -2r_{\rm dec}/3$ in Eq.~\eqref{eq:zetaphiQuad}. For $r_{\rm dec}<1$, the exact solution~\eqref{eq:MasterX} regularises this divergence. As a consequence, when the curvaton oscillates near the minimum of a quadratic potential, the function $F(\zeta_{\rm G})$ can be accurately approximated by the regularised logarithmic ansatz
\be\label{eq:template}
    \zeta = F(\zeta_{\rm G}) \approx \frac{1+\epsilon^2}{2 \beta} \ln\left[ \left(1+\beta\zeta_{\rm G}\right)^2 + \epsilon^2\right] + F_0\,,
\ee
where $F_0$ is fixed by imposing $\langle \zeta \rangle = 0$, and the parameters $\beta$ and $\epsilon$ are obtained by fitting the exact results. In particular, to reproduce the concave logarithmic behaviour and the presence of a minimum, we have $\beta>0$. The coefficient in front of the logarithm is fixed by requiring that $\zeta = \zeta_{\rm G} + \ldots$ at linear order.

\begin{figure*}[t!]
\centering
\begin{minipage}{.5\textwidth}
  \centering
  \includegraphics[width=\linewidth]{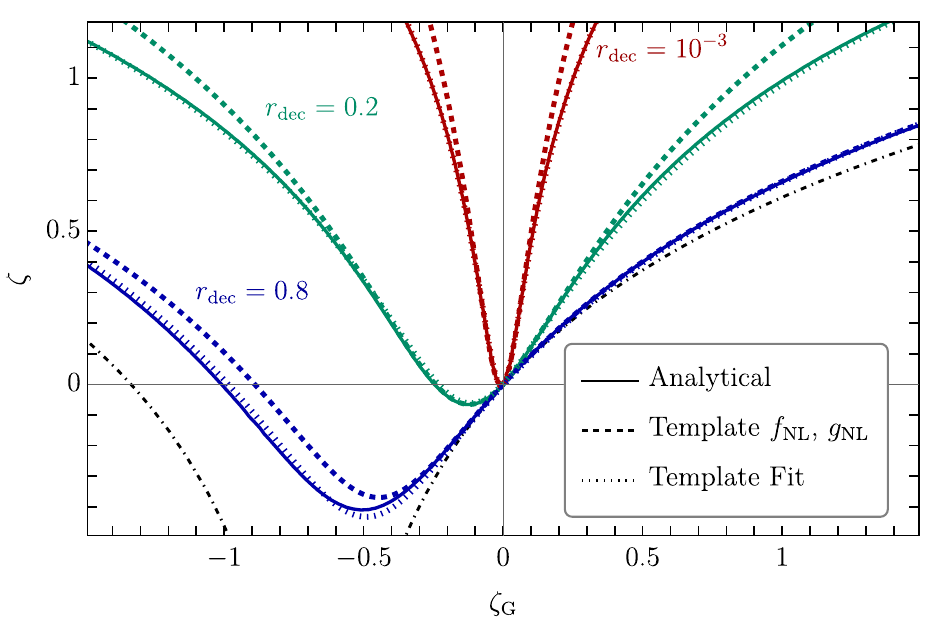}
\end{minipage}\hfill
\begin{minipage}{.5\textwidth} 
  \centering
  \includegraphics[width=\linewidth]{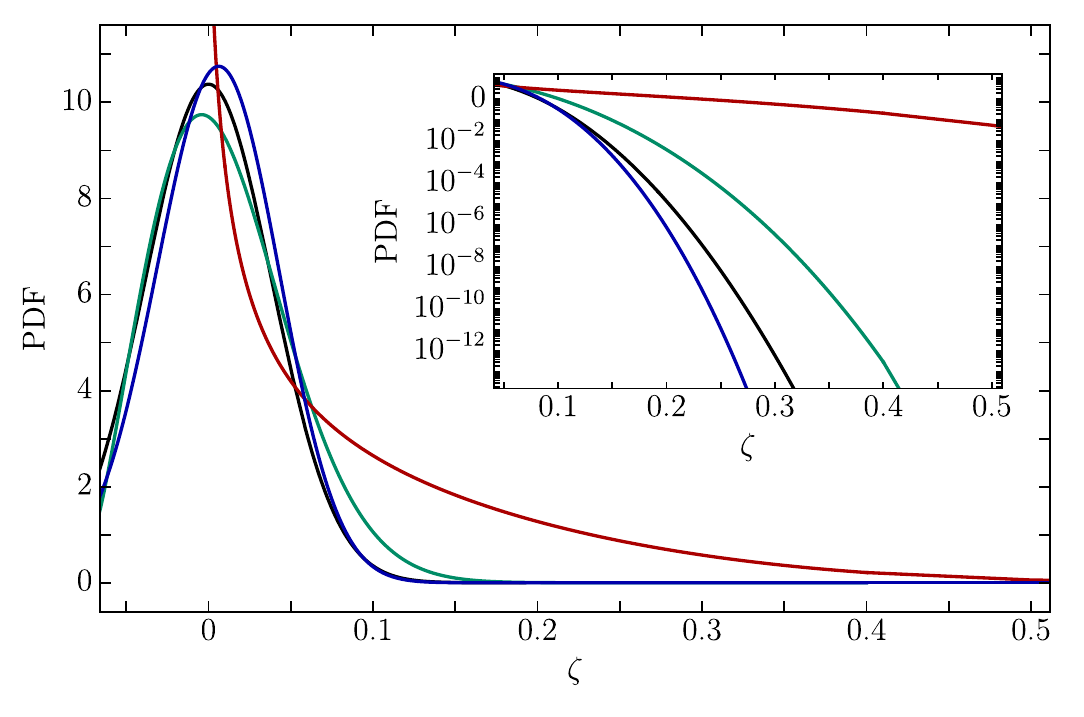} 
\end{minipage}
\caption{{\it Left panel:} Comparison between $F(\zeta_{\rm G})$ obtained from the analytical result~\eqref{eq:MasterX} for different $r_{\rm dec}$ in the quadratic scenario and the template~\eqref{eq:template}, with parameters obtained from a numerical fit (dotted) and from matching the Taylor series at $\zeta_{\rm G}=0$ (dashed). The dot-dashed line highlights the limit $r_{\rm dec} \rightarrow 1$. {\it Right panel:} Comparison between the non-Gaussian PDFs of $\zeta$ and the Gaussian PDF of $\zeta_{\rm G}$ (black line) for the same models shown in the left panel. We fix the initial variance of the curvaton field to be $\sigma_\phi=0.005$.}
    \label{fig:quad2}
\end{figure*}

As shown in Fig.~\ref{fig:quad2}, the template~\eqref{eq:template} fits the analytical solution~\eqref{eq:MasterX} well. The parameters of \eqref{eq:template} can be determined analytically in terms of $r_{\rm dec}$ by matching the Taylor expansion of $F(\zeta_{\rm G})$ around $\zeta_{\rm G}=0$. This procedure reproduces the local behaviour of the exact solution at the cost of reducing the accuracy of the fit away from the expansion point (see the dashed line in the left panel of Fig.~\ref{fig:quad2}). In particular, it establishes a direct relation between the template parameters $\beta$ and $\epsilon$, and the non-linearity parameters $f_{\rm NL}$ and $g_{\rm NL}$ of a free curvaton scenario~\cite{Lyth:2005fi, Sasaki:2006kq}
\bea \label{eq:fNLgNL2}
    \frac{3}{5}f_{\rm NL}
&   = \frac{\beta(\epsilon^2-1)}{2(1+\epsilon^2)}
    = \left(
    -1
    +\frac{3}{4r_{\rm dec}}
    -\frac{r_{\rm dec}}{2}
    \right)\,,
\\[6pt]
     \frac{9}{25}g_{\rm NL}
&   = \frac{\beta^2(1-3\epsilon^2)}{3(1+\epsilon^2)^2}
    = \left(
    \frac{1}{12}
    -\frac{3}{2r_{\rm dec}}
    +\frac{5r_{\rm dec}}{3}
    +\frac{r_{\rm dec}^2}{2}
    \right)\,.
\eea

The right panel of Fig.~\ref{fig:quad2} compares the Gaussian PDF of $\zeta_{\rm G}$ with the PDF of the NG field $\zeta = F(\zeta_{\rm G})$, given by
\be\label{eq:PDFNG}
    P(\zeta) 
    = \sum_{p \in \{\pm\}} P_{\rm G}\!\left[F^{-1}_{p}(\zeta)\right] 
    \left| \frac{\td F^{-1}_{p}(\zeta)}{\td \zeta} \right|\,,
\ee
where the sum runs over the two branches of the inverse function of $F$.
By construction, $\zeta_{\rm G}$ is centred around zero, $\langle \zeta_{\rm G} \rangle=0$, and has a variance $\sigma_{\rm 0} \propto r_{\rm dec} \sigma_\phi / \bar{\phi}_\star$. 
The figure illustrates how NGs reshape the tail of the distribution, enhancing or suppressing the probability of large perturbations depending on $r_{\rm dec}$. 

In particular, for the ansatz~\eqref{eq:template}, Eq.~\eqref{eq:PDFNG} admits a relatively compact analytic form. The two branches of the inverse map lead to the PDF
\bea
    P_\pm(\zeta)
    =
    \frac{e^{2y} 
    }
    {
    \sqrt{2\pi}\,(1+\epsilon^2)\,\sigma_0\,
    \sqrt{e^{2y} - \epsilon^2
    }}\exp\!\left[
    -\dfrac{\left(1\pm\sqrt{e^{2y} - \epsilon^2}\right)^2}
    {2\beta^2\sigma_0^2}\right]\,, 
    \quad
    y \equiv \beta\dfrac{\zeta-F_0}{1+\epsilon^2}
\eea
In the case of $\beta>0$, the PDF is supported only for $\zeta \geq \zeta_{\rm min}=F_0 + (1+ \epsilon^2)\ln(\epsilon) / \beta$, which corresponds to the minimum of the function $F(\zeta_{\rm G})$. When $\epsilon \to 0$, the lower bound is pushed to $-\infty$ as the logarithm in~\eqref{eq:template} diverges when $\beta\zeta_G \to -1$. In that case, $P(\zeta)$ develops an exponential $P(\zeta) \propto \exp(y)$ tail when $\zeta \to -\infty$. In the opposite limit, $\zeta\rightarrow+\infty$, the tail is more strongly suppressed than in the Gaussian case, exhibiting double-exponential behaviour. This is a direct consequence of the logarithmic form of $F(\zeta_{\rm G})$. However, as shown in the right panel of Fig.~\ref{fig:quad2}, the onset of this suppression is controlled by the parameters $\beta$ and $\sigma_0$, and therefore it inherits their non-trivial dependence on $r_{\rm dec}$. As a result, despite the asymptotic suppression, the PDF can exhibit a significant enhancement relative to the Gaussian case over the range of $\zeta$ relevant for phenomenological applications, such as for the PBH abundance, as we will see in Sec.~\ref{sec:SMBH}.

The $\beta < 0$ case displays similar behaviour, but with the positive and negative asymptotes exchanged.

\subsection{Quartic self-interactions}
\label{sec:quartic_V}

In this subsection, we focus on the special case of the renormalizable potential
\be \label{eq:quarticV}
    V(\phi) = \frac{1}{2} m^2 \phi^2 + \lambda \phi^4 \,,
\ee
which, within our formalism, can be analysed using analytical estimates. The corresponding reduced action is 
\bea \label{eq:Wquartic}
    W(\rho_\phi) 
    & \simeq \sqrt{8\,V(\Phi)} \int_0^{\Phi} \td\phi \,
    \sqrt{1-\frac{V(\phi)}{V(\Phi)}} 
    \\
    & = \frac{m^3}{\lambda}s(\Phi) \sqrt{s(\Phi)}
    \int_0^1 \td x \,\sqrt{\left(x^2+\frac{1+s(\Phi)}{s(\Phi)}\right)\left(1-x^2\right)} \,,
    \\
    &= \frac{m^3 }{3\,\lambda} \sqrt{1+s(\Phi)} \left[ -E_s (\Phi)+ (1+2\, s(\Phi))\,K_s(\Phi) \right]\,.
\eea
where the energy density of the curvaton is related to the amplitude of the oscillation $\Phi$ by $\rho_\phi = V(\Phi)$, we introduced the elliptic functions
\be
    E_s (\Phi) \equiv E\!\left(-1+\frac{1}{1+s(\Phi)}\right)\,, 
    \qquad
    K_s (\Phi)  \equiv K\!\left(-1+\frac{1}{1+s(\Phi)}\right)\,.
\ee
and defined the dimensionless parameter~\cite{Enqvist:2009zf, Byrnes:2011gh, Enqvist:2010dt}
\be
    s(\Phi) = \frac{2\lambda}{m^2}\,\Phi^2 \,.
\ee
to quantify the strength of self-interactions at a given amplitude. This is the ratio of the quartic to quadratic terms in~\eqref{eq:quarticV}, and $s=1$ corresponds to the case where both terms are equal at the onset of oscillations. As the amplitude is damped by Hubble friction, the self-interactions are strongest initially, when $\Phi \approx \phi_{\star}$, but the quadratic term will eventually begin to dominate, and the role of self-interactions diminishes.
In particular, we have the asymptotics
\be
    W(\rho_\phi) \simeq
\begin{cases}
    \dfrac{\pi}{m}\,\rho_\phi & s \ll 1 \,, \\[10pt]
    \dfrac{\Gamma\!\left(\frac{1}{4}\right)^2}{3\sqrt{\pi}}\,
    \lambda^{-1/4}\,\rho_\phi^{3/4} 
    & s \gg 1 \,,
\end{cases}
\ee
so the behaviour of monomial quadratic and quartic  potentials~\eqref{eq:Wevol} is recovered for weakly and strongly interacting cases, respectively.

Given Eq.~\eqref{eq:Wquartic}, the evolution of the reduced action during the adiabatic regime, $W \approx W_\star \left( H/H_{\rm ad}\right)^{3/2}$, can be determined analytically. In particular, we will use $W_\star = W(s(\phi_\star))$ and $H_{\rm ad}=2 \pi /T(s(\phi_\star)) = \pi/\partial_{\rho_\phi} W(\rho_\phi) |_{\Phi=\phi_\star}$ that follows by analogy with the quadratic case. The analytical expression for $H_{\rm ad}$ in terms of the self-interaction strength $s(\phi_\star)$ is reported in Eq.~\eqref{eq:Hadquartic}.
Combining the analytical expression for $W$ with Eqs.~\eqref{eq:fluidEOM} and \eqref{eq:Wcontinuity} implies that the number of e-folds elapsed in a given patch, and hence the curvature perturbation $\zeta$, depends on the parameters of the potential through $s(\phi_\star)$ and on the decay rate $\Gamma_\phi$. Consequently, the shape of the NGs generated by a quartic potential is determined by the interplay between the decay rate $\Gamma_\phi$ and the self-interaction strength $s(\phi_\star)$.

\begin{figure*}[t!]
\centering 
\begin{minipage}{.5\textwidth}
  \centering
  \includegraphics[width=\linewidth]{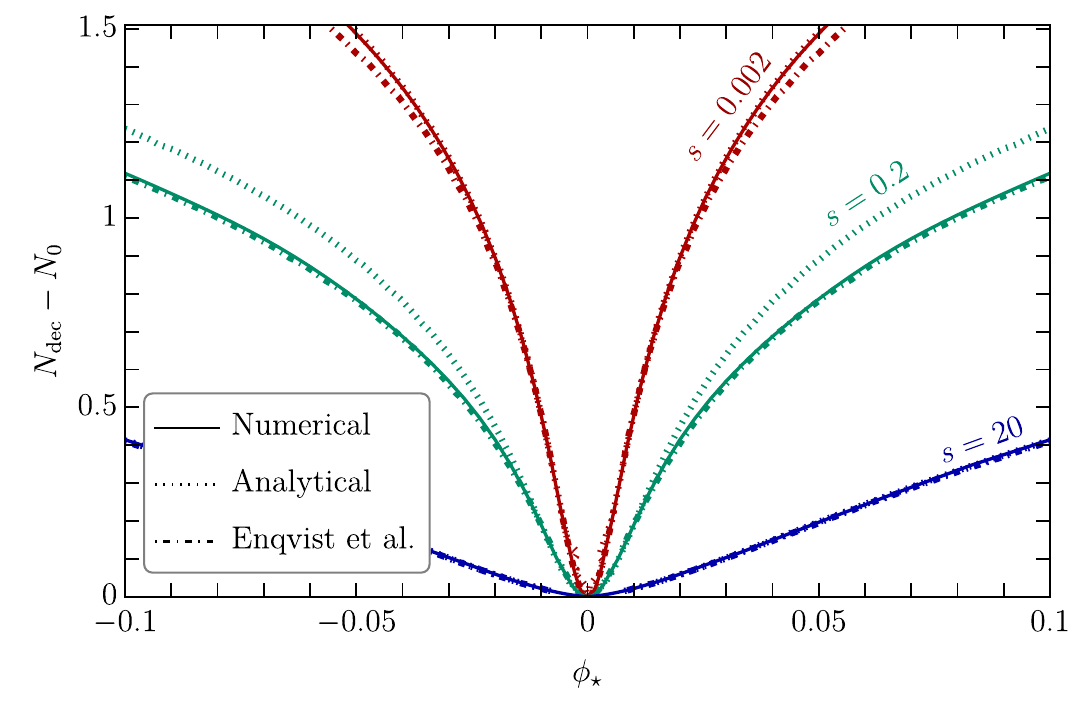}
\end{minipage}\hfill
\begin{minipage}{.5\textwidth} 
  \centering
  \includegraphics[width=\linewidth]{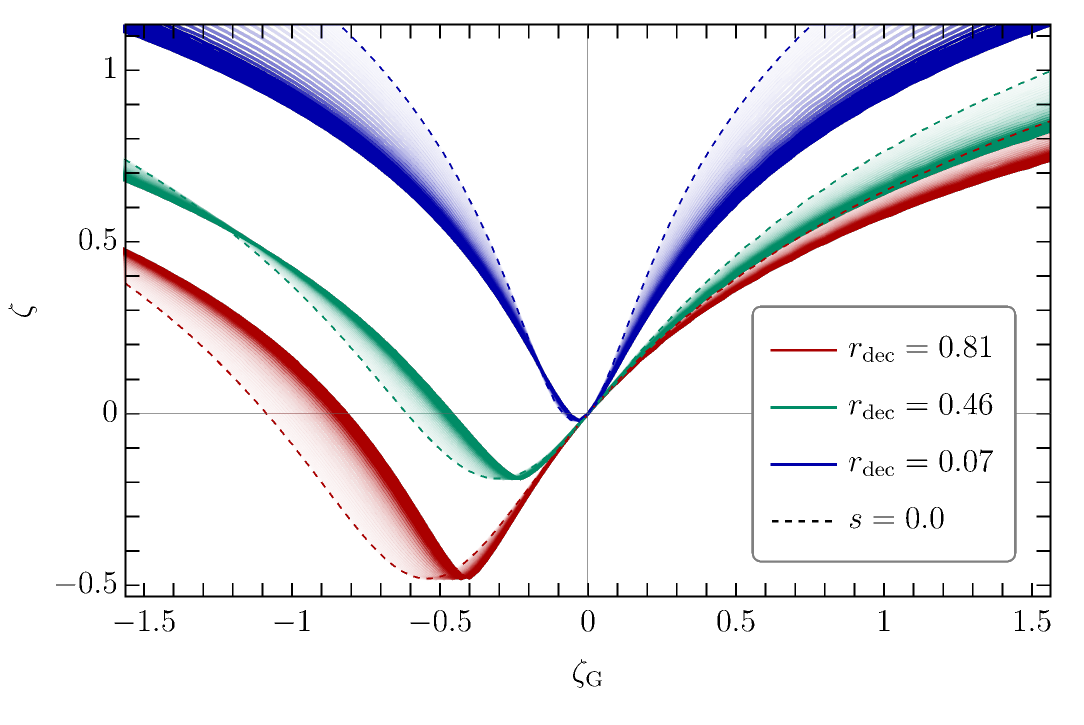} 
\end{minipage}
\caption{ {\it Left panel:} Comparison between numerical and analytical approaches for the computation of $\zeta = F(\phi_\star)$ in the presence of quartic self-interactions. We also include the analytical results from~\cite{Enqvist:2009ww, Enqvist:2009zf}, discussed in App.~\ref{app:quartic}. In all cases we fix $H_\star = 10^{-6.5}$, $\Gamma_{\phi} = 10^{-17}$, and $\lambda = 10^{-13}$, while varying the mass parameter to tune the self-interaction strength $s$. 
{\it Right panel:} Analytic scan of $F(\zeta_{\rm G})$ obtained from Eq.~\eqref{eq:logzeta} for three benchmark $r_{\rm dec}$ and varying the self-interaction strength across the full range between the purely quadratic and quartic limits, $s \in [0, +\infty)$.  }
    \label{fig:quartic}
\end{figure*}

The left panel of Fig.~\ref{fig:quartic} shows the comparison between the analytical and numerical results for a few benchmark configurations, varying the strength of the self-interactions. 
The analytical solution is obtained by plugging Eq.~\eqref{eq:zetaW1} into Eq.~\eqref{eq:sasaki} using the analytical expression of $W$ in terms of $s(\phi_\star)$ introduced above. 
To solve the equation, we fix the background at the reference value $\bar \phi_\star=0.01$. After that, the curvature $\zeta$ is rescaled in order to match $F(\phi_\star=0)=0$.
In Fig.~\ref{fig:quartic} and from now on, we denote by $s \equiv s(\bar\phi_\star)$ the self-interaction strength for the initial background field.

Our analytical approximation fails to accurately reproduce $N_{\rm dec}-N_0$ at large $\phi_\star$ for $s = 0.2$ (green line). This discrepancy arises when $s \sim 1$ and can be traced back to the analytical treatment of $H_{\rm ad}$. The resulting shift can be reabsorbed by fixing $H_{\rm ad}$ numerically. This issue, together with a detailed comparison with previous analytical results by Enqvist et al.~\cite{Enqvist:2009ww, Enqvist:2009zf}, is extensively discussed in App.~\ref{app:quartic}.

Adopting the standard convention for the background, $\bar \phi_\star = \langle \phi_\star \rangle$, Eq.~\eqref{eq:zetaW1} can be linearised for small perturbations around the central value
\be \label{eq:zetaW}
    e^{3\zeta_{\phi}} 
    \simeq
    \left( 1+ \frac{\td \ln W}{\td \Phi}  \delta \phi_\star \right)
    \left( 1+ \frac{\td \ln H_{\rm ad}}{\td \Phi} \delta \phi_\star \right)^{-3/2} 
    \Bigg|_{\Phi = \bar \phi_\star}
    \simeq 
    \left( 1 + \frac{\delta \phi_\star}{\bar \phi_\star}\right)^{G(\bar {\phi}_\star)}\,,
\ee
where we introduced the auxiliary function
\be \label{eq:Gauxiliary}
    G(\bar{\phi}_\star)
    \equiv \bar \phi_\star
    \left[
    \frac{\td \ln W}{\td \Phi}
    -
    \frac{3}{2}\frac{\td \ln H_{\rm ad}}{\td \Phi}
    \right]_{\Phi=\bar{\phi}_\star}\,.
\ee
This auxiliary function, $G(s)\equiv G(\bar \phi_\star)$, can be written as a unique function of the self-interaction strength (see Eq.~\eqref{eq:G(s)}).
Given the relation above, it is convenient to introduce the Gaussian curvature perturbation by again imposing the first-order relation $\zeta_{\phi} = \zeta_{\rm G}/r_{\rm dec}+ \ldots$.
Within this choice, Eq.~\eqref{eq:zetaW1} becomes
\be \label{eq:logzeta}
    \zeta_\phi 
    = \frac{G\left(s\right)}{3} \ln \left|  1 + \frac{3}{G \left(s\right) r_{\rm dec}} \zeta_{\rm G} \right|\,,\qquad \mbox{where} \qquad
    \zeta_{\rm G} 
    = \frac{G(s) \,r_{\rm dec}}{3}\frac{\delta \phi}{\langle \phi_\star \rangle}\,,
\ee
where the absolute value is justified by the fact that the right-hand side of Eq.~\eqref{eq:zetaW1} is strictly positive for a quartic potential. A closed-form analytical expression for $\zeta = F(\zeta_{\rm G})$ can be obtained by substituting Eq.~\eqref{eq:logzeta} into Eq.~\eqref{eq:MasterX}, yielding the full non-Gaussian mapping $F(\zeta_{\rm G},\, r_{\rm dec},\, s)$, which can be expanded perturbatively in $\zeta_{\rm G}$ to extract the non-linearity parameters (see Fig.~\ref{fig:fNLgNL}).
In the right panel of Fig.~\ref{fig:quartic}, we show a scan of $F(\zeta_{\rm G})$ for three benchmark $r_{\rm dec}$. As anticipated from the left panel, increasing $s$ suppresses curvature perturbations at large $\zeta_{\rm G}$. This behaviour indicates that stronger self-interactions tend to suppress positive NGs of quartic potentials.

\subsection{Axion-like curvaton: cosine potentials}
\label{sec:cosine_V}

A theoretically well-motivated example for a self-interacting curvaton is the axion-like curvaton model, where the curvaton is a pseudo-Nambu-Goldstone boson of a global U(1) broken symmetry~\cite{Kawasaki:2011pd, Kawasaki:2012wr, Kawasaki:2013xsa, Ando:2017veq, Ando:2018nge, Inomata:2023drn}. If the symmetry is explicitly broken by non-perturbative effects, the curvaton obtains the following potential
\be \label{eq:PotCos}
    V(\phi) = V_0^2 \left[ 1- \cos \left(\frac{\phi}{f}\right) \right]\,.
\ee

The integral in Eq.~\eqref{eq:EoS_W} can be written in a simpler form by performing the change of variable $y = (2V_0^2/\rho)\sin\phi'$, with $\phi' = \phi/(2f)$, and obtaining
\be \label{eq:WevolCos}
    W(\rho) \simeq 4 V_0 f \, g\!\left(\frac{\rho}{2 V_0^2}\right)\,, 
    \qquad 
    g(x) \equiv x \int_0^{1/2} \td y \, \frac{\sqrt{1-y}}{y(1 - x y)}\,.
\ee
Expanding around the minimum, we find $g(x) \sim \left(\pi/4 + 1/2\right)x + \mathcal{O}(x^2)$ for $\rho \ll 2V_0^2$, which reproduces the expected quadratic behaviour with effective mass $m = V_0/f$.
In contrast to the quadratic case, the dynamics here depend on two independent parameters, $V_0$ and $f$.

\begin{figure}
	\begin{center}
		\includegraphics[width=\linewidth]{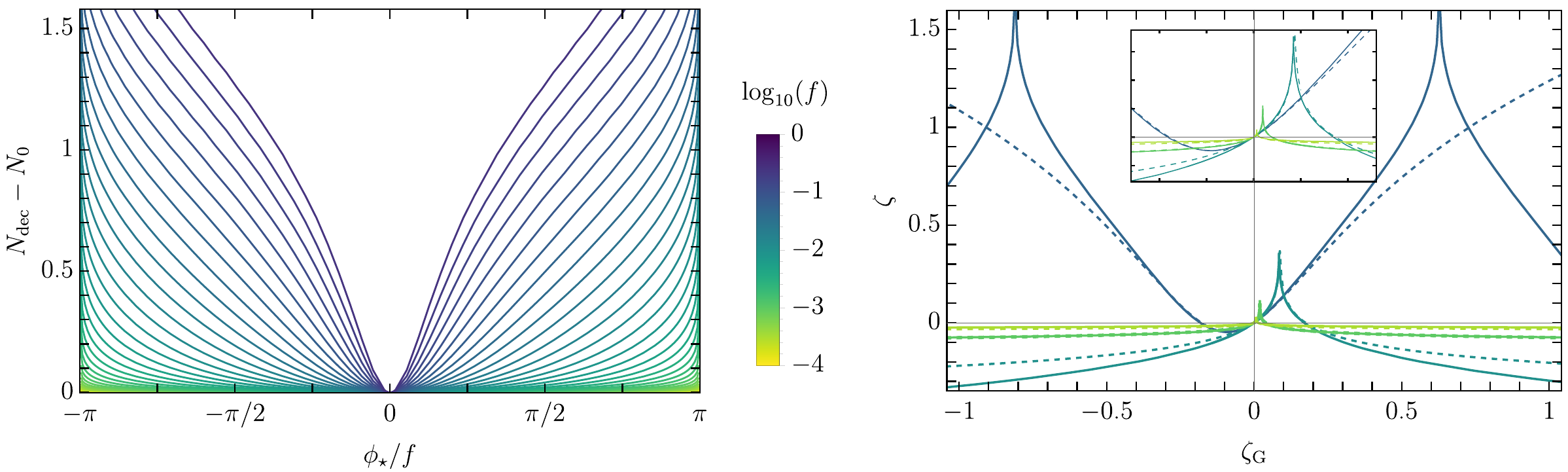}
		\caption{ {\it Left panel:} Numerical scan showing the behaviour of $N_{\rm dec}(\phi_\star)-N_0$ for the cosine potential with different $f$. For all scenarios we fix $H_\star = 10^{-7}$, $\Gamma_{\phi} = 10^{-15}$, and the height of the potential $V_0 = 10^{-10}$. {\it Right panel:} $F(\zeta_\textrm{G})$ for four benchmark cases with different potentials and choices of the background $\bar{\phi}_\star=\langle \phi_\star\rangle$. In particular, we show $f=\left[\,10^{-1.3},\,10^{-2},\, 10^{-3},\, 10^{-3.5} \,\right]$ with $\bar{\phi}_\star$ chosen to respectively have $r_{\rm dec} = \left[ \, 0.1, \, 0.8,\,0.3,\, 0.1 \, \right]$. The dashed lines display the comparison against the  ansatz~\eqref{eq:template} for the different cases. In particular, the plotted curves have $\beta \simeq \left[11, \,- 11, \,- 52,\, -154\,\right]$ in order of decreasing $f$.}
	\label{fig:cosine}  
	\end{center}
\end{figure}

This additional freedom in the choice of model parameters, compared to the quadratic case, allows for a wider variety of shapes for the function $F(\zeta_{\rm G})$. As a representative example, in the left panel of Fig.~\ref{fig:cosine}, we show a numerical scan of $N_{\rm dec}(\phi_\star)$ obtained by varying the shape of the potential through the parameter $f$. For a fixed height of the potential $V_0$, this parameter controls the period and, therefore, the flatness of the potential. A smaller $f$ corresponds to a steeper and less quadratic potential away from the minimum, which suppresses the magnitude of the curvature perturbation generated in the patch. For higher $f$, the potential approaches the quadratic limit and can lead to configurations in which the curvaton easily dominates the energy density before the start of oscillations if the initial field is placed close to the hilltop of the potential.

In the right panel, we show $F(\zeta_G)$ for four benchmark scenarios, adopting the standard convention for the background field $\bar{\phi}_\star = \langle \phi_\star \rangle$. Unfortunately, for the potential treated in this section, we lack a closed form expression for $W$ in the adiabatic regime. Consequently, the normalisation of the variable $\zeta_G$ is determined numerically via Eq.~\eqref{eq:EoS_W}.

Fig.~\ref{fig:cosine} shows that for larger $f$ (dark blue line) and a background field located close to the minimum of the potential, the shape of the mapping $F(\zeta_G)$ resembles the standard logarithmic behaviour found in polynomial potentials. In particular, the mapping $F(\zeta_{\rm G})$ is well described by the quadratic ansatz~\eqref{eq:template} with $\beta>0$ (dark blue dashed line). In this case, the curve follows the same shape as in a free curvaton scenario with mass $m=V_0/f$, and the parameters of the template are fixed via Eqs.~\eqref{eq:fNLgNL2}.

However, for smaller $f$, the quadratic limit fails to capture the true shape of $F(\zeta_{\rm G})$, and the mapping develops an exponential behaviour, similar to that found in USR inflation~\cite{Atal:2018neu,Tomberg:2023kli,Iovino:2024sgs}.
Interestingly, we found that, for small $f$, the mapping around the cusp can be very well approximated by the same ansatz used in the quadratic case~\eqref{eq:template} (represented by the inset in the right panel of Fig.~\ref{fig:cosine}).
In the plot, $F_0$ is fixed by imposing $F(\zeta_G=0)=0$, while the parameters $\beta$ and $\epsilon$ are numerically fitted around $\zeta_{\rm G}=0$. In contrast to the quadratic case, we find $\beta<0$, which is required to reproduce the presence of a convex cusp.

Within the classical picture, given that the hilltop at $\phi_\star \rightarrow \pm f\pi$ is a stationary point, we have $\epsilon = 0$ when modelling the cusps with the $\beta<0$ ansatz. Thus, when located exactly at the hilltop, a classical curvaton can never roll down the potential, and the corresponding patch eventually enters a second phase of inflation, causing $\delta N$ to diverge. In Fig.~\ref{fig:cosine}, the seemingly non-vanishing value of $\epsilon$ is a numerical artefact arising from the finite resolution of the grid in $\phi_\star$.

However, when the full quantum evolution is taken into account, quantum fluctuations of the scalar field can experience exponential growth caused by tachyonic instabilities around the hilltop $V'' \simeq - V_0^2/f^2$~\cite{Felder:2000hj, Felder:2001kt, Koivunen:2022mem,Tomberg:2021bll}. This mechanism will cause the field to roll down the potential in a finite amount of time, resulting in a finite $\zeta$ even when $\phi_\star$ is placed at the hilltop. In this sense, $\epsilon$ may be thought of as an effective regulator that can be computed once quantum effects are properly included.\footnote{Quantum effects for a curvaton field lying around the hilltop of the potential, like fragmentation rates~\cite{Kusenko:1997si,Dine:2003ax,Kallosh:2013hoa,Cotner:2017tir,Kim:2017duj,Cotner:2018vug,Cotner:2019ykd} and  exponential growth of quantum modes~\cite{Felder:2000hj, Felder:2001kt,Koivunen:2022mem,Tomberg:2021bll} could also  potentially modify the results obtained within our classical picture. The study of these non-perturbative effects lies beyond the scope of this work.}

The results shown in Fig.~\ref{fig:cosine} imply that, unlike standard polynomial potentials, this axion-like curvaton scenario can generate large NGs also for large $r_{\rm dec}$. Small $r_{\rm dec}$, corresponding to $\bar \phi_\star$ being located near a minimum of $N_{\rm dec}(\phi_\star)$, reproduce large NGs in agreement with the standard polynomial potential. However, as illustrated by the three cases with smaller $f$ in the right panel, large $r_{\rm dec}$ can also lead to a strongly non-Gaussian mapping when the chosen $\bar \phi_\star$ lies in the vicinity of the hilltop of the cosine. In Sec.~\ref{sec:SMBH} we will study the phenomenological implications of this non-trivial effect.

\subsection{Axion monodromy: linear and monomial potentials} 
\label{sec:linear_V}

As a last trivial example, we consider the case of monomial potentials
\be
    V \propto |\phi|^n\,.
\ee
In this simple case, Eqs.~\eqref{eq:EoS_W} and \eqref{eq:Wevol} imply that (see also App.~\ref{app:mono})
\be \label{eq:monoWRel}
    w = \frac{n-2}{n+2}\,, 
    \qquad\qquad
    W \propto \rho_{\phi}^{\frac{1}{1+w}}\,.
\ee
Therefore, whenever $n\geq4$ or, equivalently, $w\geq1/3$, the curvaton fluid is diluted at the same rate as, or faster than, radiation. Thus, an initially subdominant curvaton would never reach a significant abundance in a radiation-dominated universe, and its contribution to curvature perturbations would also be negligible. On the other hand, potentials characterised by $n < 2$ imply an effective equation-of-state parameter $w < 0$, resulting in an energy density that redshifts more slowly than non-relativistic matter. This slower dilution leads to a more efficient amplification of the curvaton field fluctuations. Given this, the general treatment we present here will only be needed to study the linear toy model $n=1$, presented in the second part of the section. Notably, Eq.~\eqref{eq:monoWRel} also holds whenever the curvaton can be described as a perfect fluid with an equation-of-state parameter $w$.

For a generic monomial, Eq.~\eqref{eq:decayW} simplifies to
\be\label{eq:masterMono}
    e^{4\zeta} - \bar \Omega_{\phi,\rm dec} \, e^{3(1+w)\zeta_{\phi} + (1-3w)\zeta} + \bar \Omega_{\phi,\rm dec} - 1 = 0 \,,
\ee
where $w$ denotes the barotropic equation-of-state parameter. At the leading order, this generalizes Eq.~\eqref{eq:zetaR} so that
\be
    \zeta = r_{\rm dec} \zeta_{\phi} + \mathcal{O}(\zeta_\phi^2)\,, 
    \qquad\quad
    r_{\rm dec} = \frac{3 (1+w) \bar \Omega_{\phi,\rm dec} }{4 - (1-3 w) \bar \Omega_{\phi,\rm dec}}\,.
\ee
As a consistency check, the efficiency parameter $r_{\rm dec}$ agrees with the quadratic case~\eqref{eq:rdec} when $w=0$, as expected.

To relate the initial curvaton fluctuation to the curvature fluctuations, we must also consider the relation between $\zeta_{\phi}$ and $\phi_\star$. By applying \eqref{eq:zetaW1} to monomial potentials \eqref{eq:Wevol}, \eqref{eq:Hosc}, we obtain
\be \label{eq:zetaphiMono1}
    e^{3\zeta_{\phi}} 
    \propto W(V(\phi_\star)) H_{\rm ad}(\phi_\star)^{-3/2} 
    \propto |\phi_{\star}|^{\frac{10 - n}{4}} = |\phi_{\star}|^{\frac{2-3w}{1-w}}\,,
\ee
so that
\be
    e^{4\zeta} - \bar \Omega_{\phi,\rm dec} \, \left|\phi_\star/\bar \phi_\star \right|^{\frac{(1+w)(2-3w)}{1-w}} e^{(1-3w)\zeta} +  \bar \Omega_{\phi,\rm dec} - 1 = 0\,,
\ee
where $\bar \phi_\star$  is the reference background. Adopting the choice $\bar \phi_\star=\langle \phi_\star \rangle$, from~\eqref{eq:zetaphiMono1} we can define the Gaussian variable $\zeta_{\rm G}$ as 
\be    \label{eq:zetaphiMono}
    \zeta_{\phi} = \frac{2-3 w }{3(1-w)}\ln \left|1+\frac{3(1-w)}{(2-3 w)\, r_{\rm dec}}\zeta_{\rm G}\right|\,,
    \qquad \mbox{where} \qquad
    \zeta_{\rm G} 
    = \frac{(2-3 w) \,r_{\rm dec}}{3(1-w)}\frac{\delta \phi}{\langle \phi_\star \rangle}\,.
\ee

As an explicitly worked out case, we consider the linear potential, for which $w = -1/3$. In particular, we consider a potential of the form
\be \label{eq:monomial}
    V(\phi) = \lambda \sqrt{\phi^2 + \mu^2} \;\overset{\mu \ll 1}{\simeq}\; \lambda |\phi| \,,
\ee
where $\mu$ acts as a regulator, smoothing the derivative at $\phi = 0$. Although this potential can arise in certain axion monodromy scenarios~\cite{Ashoorioon:2008pj,Brandenberger:2008kn,Silverstein:2008sg,Hannestad:2009yx,Berg:2009tg,McAllister:2008hb,Marchesano:2014mla}, we will treat it as a toy model without connecting it to any specific model. 
We are interested in the case where the curvaton decays before reaching the quadratic regime $m_{\phi} = \sqrt{\lambda/\mu}\ll H_{\rm dec}$, so we can neglect the contribution of $\mu$. This potential provides the simplest extension beyond the standard quadratic potential in which the curvaton can dominate the radiation component at decay due to the associated scaling of the energy density (see Fig.~\ref{fig:density}).

Remarkably, the linear model admits an analytic piecewise solution for the early oscillatory phase, allowing for a precise determination of the onset of oscillations. Defining the auxiliary function $f(a) = a\, \phi / \phi_\star$, the first of Eq.~\eqref{eq:EOM} can be rewritten in terms of the scale factor as
\be
    \ddot{f}(a) \pm c\, a^3 
    = 0 \,, 
    \quad \text{with} \quad 
    c = \frac{\lambda}{\phi_\star H_\star^2} \,,
    \quad {\rm sign}(\phi_\star) = \pm1\,,
\ee
which, for the first oscillation, admits the initial conditions $f(1)=1$ and $\dot{f}(1)=1$ giving the solution
\be
    \phi(N) = \phi_\star \left( 1 \pm \frac{c}{4} \mp \frac{c}{5} e^{-N} \mp \frac{c}{20} e^{4N} \right)\,,
\ee
where we fixed $N=0$ at the reference scale $H=H_\star$.

Given this solution, the transition from the frozen regime to adiabatic oscillations is determined by the first turning point, defined by $\phi(N_{\rm ad}) = 0$, where $N_{\rm ad}$ fixes the instantaneous thawing scale through $H_{\rm ad} = H_\star e^{-2N_{\rm ad}}$. Since this equation does not admit a closed-form solution, we determine $N_{\rm ad}$ numerically for a given $c$.
During the subsequent adiabatic regime, Eq.~\eqref{eq:Wad} implies that the reduced action scales as $W = W_{\star}\,(H/H_{\rm ad})^{3/2}$, with $W_\star$ obtained from Eq.~\eqref{eq:Wevol}, and the $\phi_\star$ dependence of $N_{\rm ad}$ approximately given by Eq.~\eqref{eq:Hosc}.

\begin{figure*}[t!]
\centering 
\begin{minipage}{.5\textwidth}
  \centering
  \includegraphics[width=\linewidth]{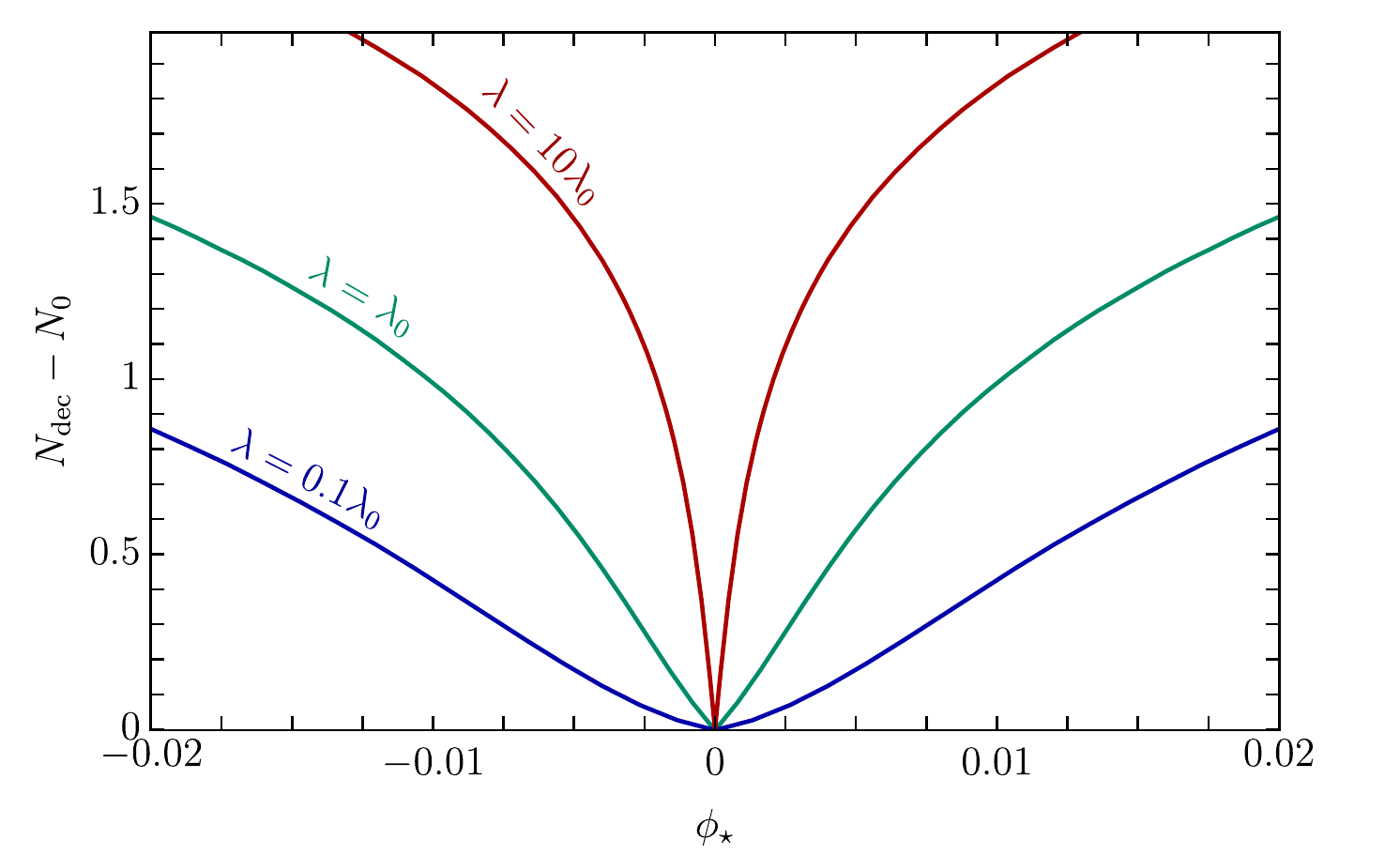}
\end{minipage}\hfill
\begin{minipage}{.5\textwidth} 
  \centering
  \includegraphics[width=\linewidth]{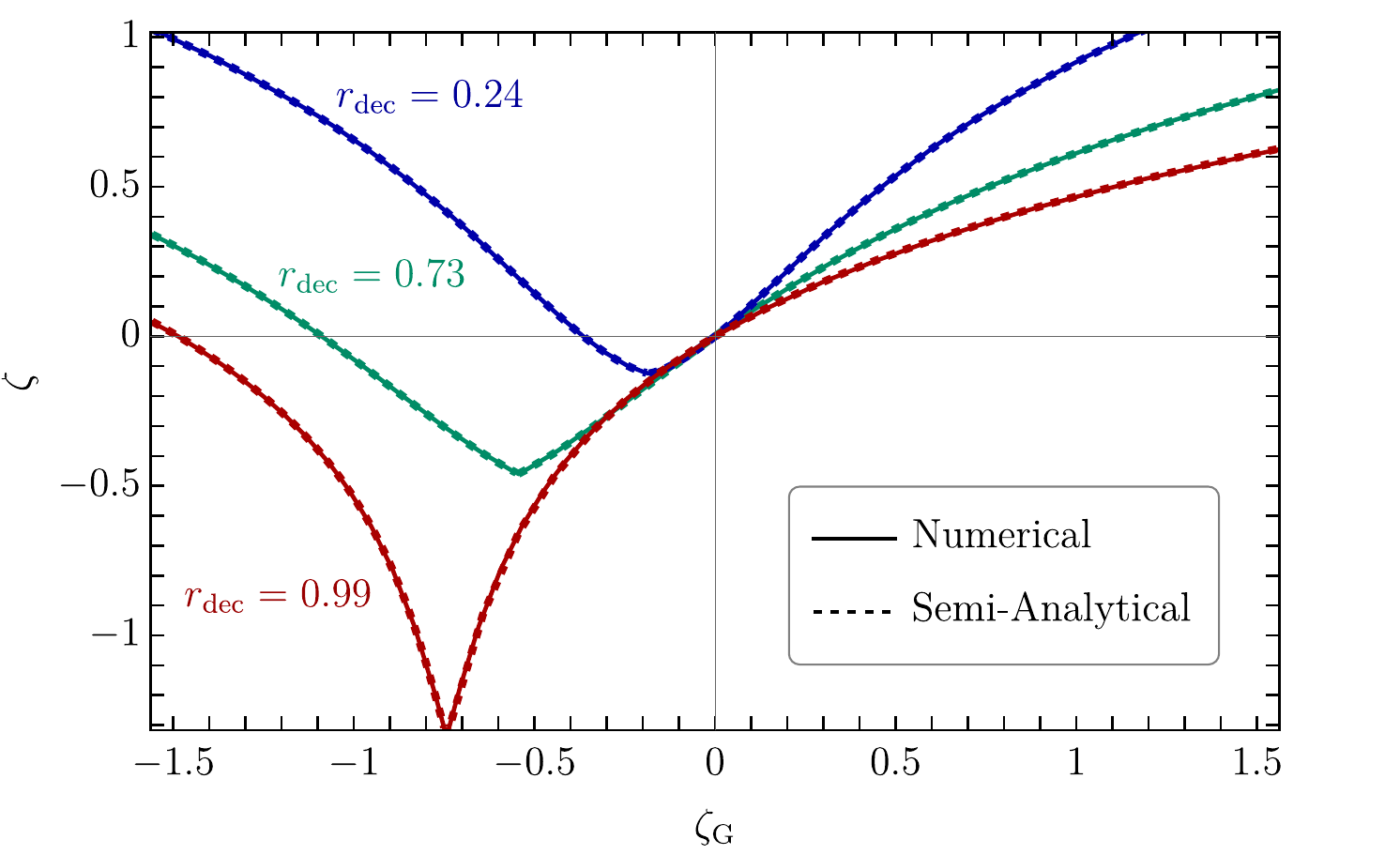} 
\end{minipage}
\caption{ {\it Left panel:} Numerical results for $N_{\rm dec}(\phi_\star)-N_0$ for different couplings $\lambda$. For all curves, we fix $H_\star = 10^{-7}$, $\Gamma = 10^{-12}$, and $\lambda_0 = 10^{-16}$. {\it Right panel:} Comparison between the numerical and semi-analytical approaches for computing $F(\zeta_{\rm G})$ in the case of a monomial potential. The Gaussian variable $\zeta_{\rm G}$ is defined via Eq.~\eqref{eq:zetaphiMono}, with the background fixed to $\bar{\phi}_\star = \langle \phi_\star \rangle = 0.003$.}
    \label{fig:linear}
\end{figure*}

In Fig.~\ref{fig:linear}, we demonstrate the robustness of our semi-analytical approach in the case of the linear potential. The left panel shows the mapping $F(\phi_\star)$ determined numerically, while the right panel shows the corresponding mapping $F(\zeta_{\rm G})$.
The latter can be determined analytically by solving Eq.~\eqref{eq:masterMono}. In the linear case, for $w=-1/3$, this equation reduces to a quadratic equation in $\exp(\zeta)$, so that
\be
    \zeta 
    = \zeta_{\phi} + \frac{1}{2}\ln \left(1 + \sqrt{4\frac{1-\bar \Omega_\phi}{\bar \Omega_\phi^2}e^{-4\zeta_{\phi}}  + 1}\right)
\ee
Given this result, after fixing a background $\bar \phi_\star$, we can substitute $\zeta_\phi$ with our semi-analytical expressions of $W_\star$ and $H_{\rm ad}$ through Eq.~\eqref{eq:zetaW1}, thereby obtaining the relation $\zeta = F(\phi_\star)$.
In the right panel of Fig.~\ref{fig:linear}, we show the corresponding function $F(\zeta_{\rm G})$ where the Gaussian curvature perturbation is defined through Eq.~\eqref{eq:zetaphiMono} for $w=-1/3$. Fig.~\ref{fig:linear} shows that the semi-analytical procedure allows us to determine the mapping $F(\zeta_{\rm G})$ with excellent accuracy, which would not be achievable when fixing $H_{\rm ad}$ analytically through Eq.~\eqref{eq:Hosc}\footnote{According to Eq.~\eqref{eq:Hosc}, $H_{\rm ad}$ diverges for small values of $\phi_\star$, implying that the field would start oscillating well before the end of inflation. In this regime, the semi-analytical treatment fails. However, this issue does not affect the overall behaviour of the curves shown in Fig.~\ref{fig:linear} around their minima.}.

\subsubsection{Decay of the curvaton}

The curvaton decay depends on the shape of the potential. In the simplest case of a quadratic potential, the late-time evolution is dominated by the mass term, and the curvaton naturally settles into a quadratic regime well before $H=\Gamma_{\phi}$. The oscillation frequency is then field-independent, and the curvaton behaves as a non-relativistic particle of well-defined mass $m_\phi$. A perturbative decay into Standard Model degrees of freedom can be implemented through a dimension-five operator $(g_{a\gamma\gamma}/4f)\,\phi\, F_{\mu\nu}\tilde F^{\mu\nu}$, yielding $\Gamma_{\phi}=g_{a\gamma\gamma}^2 m_\phi^3/(64\pi f^2)$~\cite{Ferrante:2023bgz}. In this simple scenario, $\Gamma_{\phi}$ is fixed by the mass and a single coupling constant, so that it can be treated as an effective free phenomenological parameter, bounded from below only by the requirement that the curvaton decays well before BBN.

For potentials in which a mass term is not generated by the bare Lagrangian, as in the linear case $V(\phi)\propto |\phi|$ discussed here, this picture must be revisited. Once the field enters the adiabatic oscillatory regime, the effective frequency of the oscillations becomes amplitude-dependent and the conventional notion of a single fixed mass loses its meaning. In particular, the perturbative decay rate cannot be obtained from a simple two-body amplitude with a well-defined external mass, and a different decay channel must be invoked.

A natural possibility is to assume a quadratic coupling to a second scalar field $\chi$ of the form $\mathcal{L}\supset \tfrac{1}{2}g^2\phi^2\chi^2$. The coherent oscillations of $\phi$ then induce a periodic time-dependent effective mass for $\chi$, which can lead to a copious production of $\chi$ quanta through parametric resonance, in close analogy with the preheating mechanism that follows the inflaton oscillations~\cite{Traschen:1990sw,Kofman:1994rk,Kofman:1997yn}. Within these resonant decay channels, a sizeable fraction of the curvaton energy can be transferred into the $\chi$ sector on timescales much shorter than the standard perturbative decay time\,\cite{Enqvist:2008be,Chambers:2009ki}.

A crucial question, in the context of our analysis, is whether the additional couplings that trigger the resonance can spoil the non-linear map $\zeta = F(\zeta_{\rm G})$ that we derive from the bare curvaton dynamics. To address this, it is useful to recall how $F$ enters our framework. In the $\delta N$ approach, the curvature perturbation is given by the difference in the number of $e$-folds elapsed between a perturbed patch and the background up to a surface of uniform total energy density. The functional form of $F(\zeta_{\rm G})$ is therefore fixed by two ingredients: \emph{(i)} the dynamics of the curvaton between the end of inflation and the onset of the adiabatic oscillatory regime, which determines $\zeta_\phi$ as a function of the initial fluctuation $\delta\phi_\star$ through Eq.~\eqref{eq:zetaphiMono}, and \emph{(ii)} the matching condition at the decay hypersurface, Eq.~\eqref{eq:masterMono}, which relates $\zeta$ to $\zeta_\phi$ and depends on the curvaton-to-radiation transition only through the single parameter $r_{\rm dec}$. During inflation and throughout the frozen-spectator phase, the curvaton is effectively decoupled from $\chi$: the coupling $g^2\phi^2\chi^2$ becomes operative only when $\phi$ starts to oscillate coherently at $H\sim H_{\rm ad}$, so that the relation $\zeta_\phi(\zeta_{\rm G})$ obtained from the post-inflationary evolution is unaffected.

There is, however, one important caveat. The reasoning above relies on the assumption that, throughout the local separate-universe evolution, $\zeta$ remains a function of the single Gaussian variable $\zeta_{\rm G}\propto\delta\phi_\star$. If the secondary scalar $\chi$ were also a light field during inflation, it would develop its own long-wavelength Gaussian fluctuations $\delta\chi_\star$, statistically independent of $\delta\phi_\star$. These would enter the post-inflationary evolution as a second source of curvature perturbations on super-horizon scales, and $\zeta$ would become a function of two independent Gaussian variables, breaking the single-variable mapping $\zeta=F(\zeta_{\rm G})$. This is prevented as long as $\chi$ is sufficiently heavy during inflation, so that its long-wavelength modes are exponentially suppressed by Hubble friction and $\chi$ effectively sits in its vacuum on super-horizon scales. Only short-wavelength vacuum fluctuations are then present at the onset of oscillations, and assuming this hierarchy on the spectator field $\chi$ ensures that the non-linear relation $\zeta=F(\zeta_{\rm G})$ derived is preserved.

A complete analysis of the resonant decay regime, including the strongly non-linear backreaction dynamics at the end of the resonance, lies beyond the scope of the present work. In the following, we will therefore restrict our discussion to the instantaneous perturbative decay approximation, treating $\Gamma_{\phi}$ as a free phenomenological parameter.

\section{Non-Gaussian power spectra}
\label{sec:NGPS}

Having derived the relation $\zeta = F(\zeta_{\rm G})$ between the physical curvature perturbation and its Gaussian counterpart for different curvaton potentials in Sec.~\ref{sec:NonGaussianities}, we can now discuss the impact of the associated NGs on the curvature power spectrum. As is common in curvaton models, we use a log-normal function as an ansatz for the Gaussian power spectrum 
\be\label{eq:PS_LN}
    \calP_{\zeta_{\rm G}}(k) 
    = \frac{\mathcal{A}}{\sqrt{2\pi}\,\Delta}\exp\left(-\frac{1}{2\Delta^2}\ln^2\left(\frac{k}{k_*}\right)\right)\,,
\ee
where $\mathcal{A}$ sets the overall prefactor, $k_*$ denotes the peak scale, and $\Delta$ controls the width of the spectrum.

Once the Gaussian power spectrum~\eqref{eq:PS_LN} and the non-linear map $\zeta = F(\zeta_{\rm G})$ are specified, the physical two-point function $\Pz(k)$ can be computed non-perturbatively following  Ref.~\cite{Veermae:2026yzz}.
The key object is the non-linear two-point function in real space,
\be\label{eq:xiNG}
    \xi(x)\equiv\langle\zeta(\vx)\zeta({\bf 0})\rangle = \mathcal{G}_2\bigl(\xi_{\rm G}(x)/\sigma_0^2\bigr)\,,
    \qquad
    \xi_{\rm G}(x) = \int_0^{\infty} \frac{\td k}{k} \frac{\sin(kx)}{kx} \calP_{\rm G}(k)
\ee
where $\xi_{\rm G}(x)$ denotes the Gaussian two-point correlator corresponding to the power spectrum~\eqref{eq:PS_LN} and $\sigma_0^2\equiv\xi_{\rm G}(0)$ the variance of the Gaussian field. In the case of a log-normal spectrum defined as in Eq.~\eqref{eq:PS_LN}, $\xi_{\rm G}(0)=\mathcal{A}$, i.e., the variance of $\zeta_{\rm G}$ is fixed by the overall amplitude of the spectrum. The function $\mathcal{G}_2$ is given by the bivariate integral\footnote{ In case $\left\langle F\left(\zeta_{\rm G}\right)\right\rangle \neq 0$, we make the shift $F\left(\zeta_{\rm G}\right) \rightarrow F\left(\zeta_{\rm G}\right)-\left\langle F\left(\zeta_{\rm G}\right)\right\rangle$.}
\be\label{eq:G2}
    \frac{\mathcal{G}_2(\psi) }{\sigma_0^2}
    = \int \frac{\td \zeta_1\,\td \zeta_2\, F(\zeta_1 \sigma_0)F(\zeta_2 \sigma_0)}{2\pi\sqrt{1-\psi^2}}\,\exp\left(-\frac{\zeta_1^2+\zeta_2^2-2\,\psi\,\zeta_1\zeta_2}{2(1-\psi^2)}\right)\,.
\ee
representing a Gaussian average over standard Gaussians with correlation $\psi \equiv \xi_{\rm G}/\sigma_0^2$. Here we reabsorbed the variance through the definitions $\zeta_i\rightarrow\sigma _0\zeta_i$ and $F(\zeta_i)\rightarrow\sigma _0F(\zeta_i)$ . The non-Gaussian power spectrum is then obtained through
\be\label{eq:PkNG}
    \Pz(k) = \frac{2k^2}{\pi}\int_0^\infty \td x\, x\,\sin(kx)\, \mathcal{G}_2\bigl(\xi_{\rm G}(x)/\sigma_0^2\bigr)\,.
\ee
Eqs.~\eqref{eq:xiNG}--\eqref{eq:PkNG} provide a fully non-perturbative prescription: given the Gaussian power spectrum~\eqref{eq:PS_LN} and the curvaton-specific function $F(\zeta_{\rm G})$ derived in Sec.~\ref{sec:NonGaussianities}, the physical power spectrum can be computed directly, without the need for a perturbative expansion~\eqref{eq:FirstExpansion}.

It is instructive to compare this non-perturbative result with its perturbative counterpart. A natural semi-perturbative formulation of the correlation functions can be obtained by expanding $F$ in a basis of probabilistic Hermite polynomials ${\rm He}_n$,~\cite{Veermae:2026yzz}
\be\label{eq:Hermite_expansion}
    F(\zeta_{\rm G}) - \langle F\rangle 
    = \sum_{n=1}^{\infty}C_n\,\sigma_0^{n}\,{ \rm He}_n\bigl(\zeta_{\rm G}/\sigma_0\bigr)\,,
    \qquad
    C_n = \frac{\sigma_0^{-n}}{ n!}\bigl\langle F(\zeta_{\rm G})\,{ \rm He}_n(\zeta_{\rm G}/\sigma_0)\bigr\rangle\,,
\ee
where the average is taken over a Gaussian distribution with variance $\sigma_0^2$. In particular, the orthogonality of Hermite polynomials under a Gaussian measure reduces the non-linear two-point function~\eqref{eq:G2} to the series
\be\label{eq:G2_Mehler}
    \mathcal{G}_2(\psi) = \sum_{n=1}^{\infty}n! C_n^2\,\sigma_0^{2 n}\,\psi^n\,.
\ee
In momentum space, each power $x^n$ translates into the $n$-fold convolution of $\calP_{\zeta_{\rm G}}$ with itself, so that
\be\label{eq:PkNG_pert}
    \Pz(k) = \sum_{n=1}^{\infty}n! C_n^2\,\calP_{\zeta_{\rm G}}^{*n}(k)\,,
    \qquad
    \calP_{\zeta_{\rm G}}^{*n}(k)\equiv \frac{2k^2}{\pi}\int_0^\infty \td x\, x\,\sin(kx)\,\bigl[\xi_{\rm G}(x)\bigr]^n\,.
\ee

\begin{figure*}[t!]
\centering 
\begin{minipage}{.5\textwidth}
  \centering
  \includegraphics[width=\linewidth]{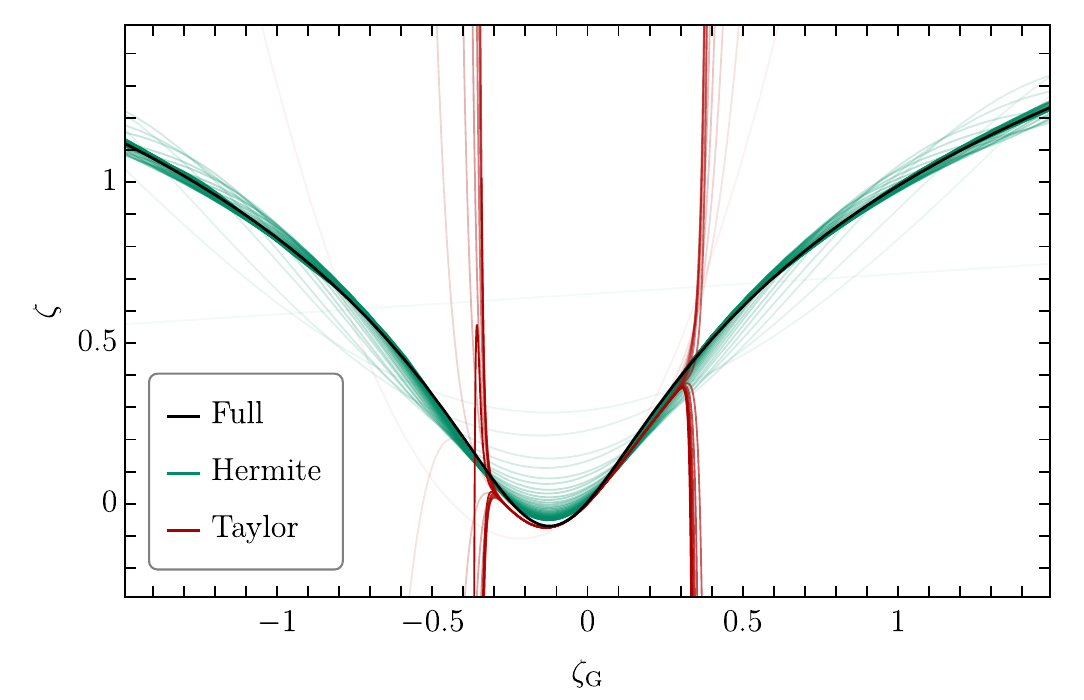}
\end{minipage}\hfill
\begin{minipage}{.5\textwidth} 
  \centering
  \includegraphics[width=\linewidth]{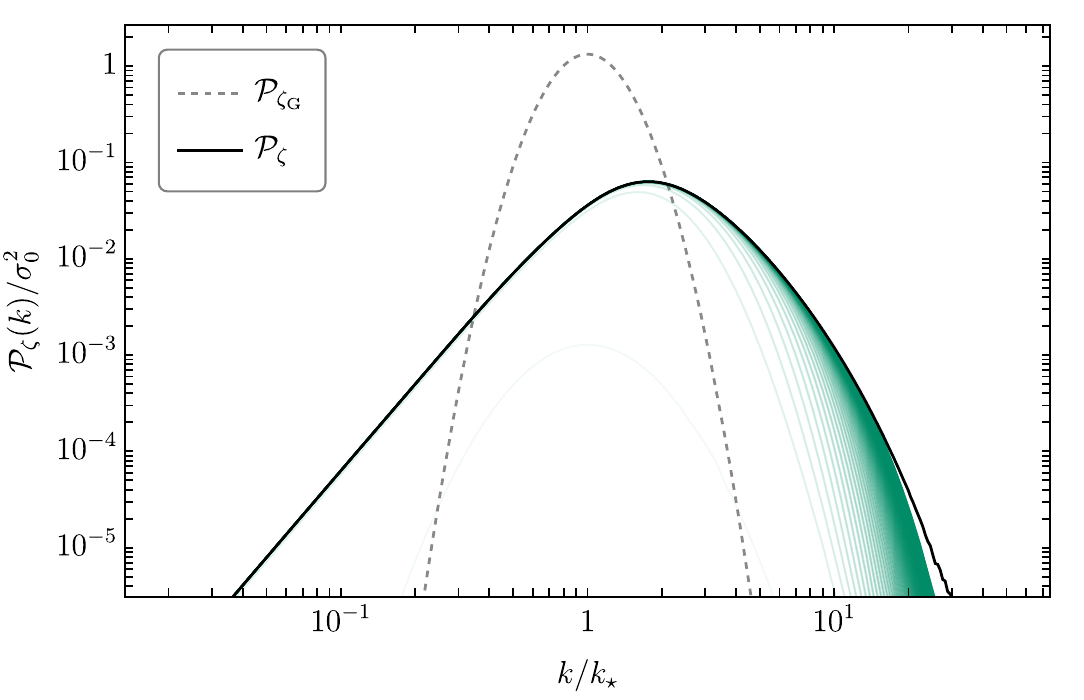} 
\end{minipage}
\caption{{\it Left panel:} $F(\zeta_G)$ for the quadratic curvaton template~\eqref{eq:template}, compared with the power series expansion~\eqref{eq:FirstExpansion} and the Hermite expansion~\eqref{eq:Hermite_expansion}. The opacity of the lines increases with the truncation order of the series, $N$. We set the template parameters to $F_0=-0.63$, $\beta=8.5$, and $\epsilon =2.2$. {\it Right panel:} Non-Gaussian curvature power spectrum $\Pz(k)$ obtained from a log-normal Gaussian power spectrum~\eqref{eq:PS_LN} with $\Delta=0.3$ and the template shown in the left panel. $\zeta_{G}$ has unit variance, fixed by $\sigma_0=1$. The black solid line shows the full non-perturbative result from Eq.~\eqref{eq:PkNG}, while the green lines correspond to the truncated series~\eqref{eq:G2_Mehler} at different orders, $N \in \left[1,100\right]$.}
    \label{fig:series}
\end{figure*}

In Fig.~\ref{fig:series} we compare the full non-perturbative results with the truncated perturbative series expressed in terms of the $C_n$ coefficients. The left panel displays a direct comparison between the Hermite series expansion~\eqref{eq:Hermite_expansion} of $F(\zeta_G)$ and the more commonly used Taylor series~\eqref{eq:FirstExpansion}.  The coefficients of the two series are directly related through~\cite{Veermae:2026yzz}
\begin{equation} \label{eq:seriesCn}
    F_{{\rm NL}, n} =
    \sum_{k \geq 0}
    \frac{(n+2k)!}{2^k\,n!\,k!}
    \, (-\sigma_0^2)^k\, C_{n+2k}\ 
    \quad \Leftrightarrow \quad
    C_n =
    \sum_{k \geq 0}
    \frac{(n+2k)!}{2^k\,n!\,k!}
    \, \sigma_0^{2k}\,F_{{\rm NL},\,n+2k}\,.
\end{equation}
This provides a direct correspondence with the common perturbative treatment in Eq.~\eqref{eq:FirstExpansion} which is based on the Taylor series. In particular, up to the $g_{\rm NL}$ order, the first Hermite coefficients are related to the non-linearity parameters by
\be\label{eq:Cn_fNL}
    C_1 = 1 + \tfrac{27}{25} \sigma_0^2 g_{\rm NL} + \ldots
    \,,\qquad
    C_2 = \tfrac{3}{5}\,f_{\rm NL} + \ldots
    \,,\qquad
    C_3 = \tfrac{9}{25}\,g_{\rm NL} + \ldots
    \,,
\ee
where $f_{\rm NL} \equiv (5/3) F_{\rm NL,2}$, $g_{\rm NL} \equiv (5/3)^2 F_{\rm NL,3}$. The standard polynomial approximation can be obtained by truncating the sum~\eqref{eq:PkNG_pert} and the expressions~\eqref{eq:seriesCn}. 
However, at higher orders, since the typical logarithmic behaviour of $F(\zeta_{\rm G})$ causes the non-linearity parameters $F_{{\rm NL},n}$ to diverge (see the left panel of Fig.~\ref{fig:series}), the coefficients $C_n$ cannot be obtained perturbatively from Eq.~\eqref{eq:seriesCn}. Conversely, by deriving the coefficients $F_{{\rm NL},n}$ perturbatively from the first relation in Eq.~\eqref{eq:seriesCn}, we obtain a power series that approximates the Hermite expansion of $F(\zeta_{\rm G})$ at a given order, rather than the standard Taylor expansion around $\zeta_{\rm G}=0$. As a result, no divergence appears in the series of $F_{{\rm NL},n}$ coefficients for large $\zeta_{\rm G}$, once they are derived from the coefficients $C_n$.

The right panel of Fig.~\ref{fig:series} displays a comparison between the non-Gaussian power spectrum obtained non-perturbatively from Eq.~\eqref{eq:PkNG} and the perturbative series~\eqref{eq:PkNG_pert} truncated at different orders.
In the infrared region (IR) ($k\ll k_*$), the truncated series rapidly converges to the full result already at the lowest orders considered. This is not surprising, since Eq.~\eqref{eq:PkNG_pert} shows that the $n=1$ term coincides with the Gaussian spectrum $\calP_{\zeta_{\rm G}}(k)$, while the higher-order convolutions $\calP_{\zeta_{\rm G}}^{*n}$ contribute a universal $k^3$ IR tail~\cite{Veermae:2026yzz}, which differs from that of the log-normal Gaussian power spectrum.
A qualitatively different picture emerges in the ultraviolet (UV) region ($k\gtrsim k_*$), where the truncated series fails to reproduce the full result for small $N$, while increasing the order $N$ generates a progressively larger tail that eventually matches the full non-perturbative spectrum. An analytical understanding of these tails is discussed in Sec.~\ref{sec:tails}.
\begin{figure*}[t!]
\centering 
\begin{minipage}{.5\textwidth}
  \centering
  \includegraphics[width=\linewidth]{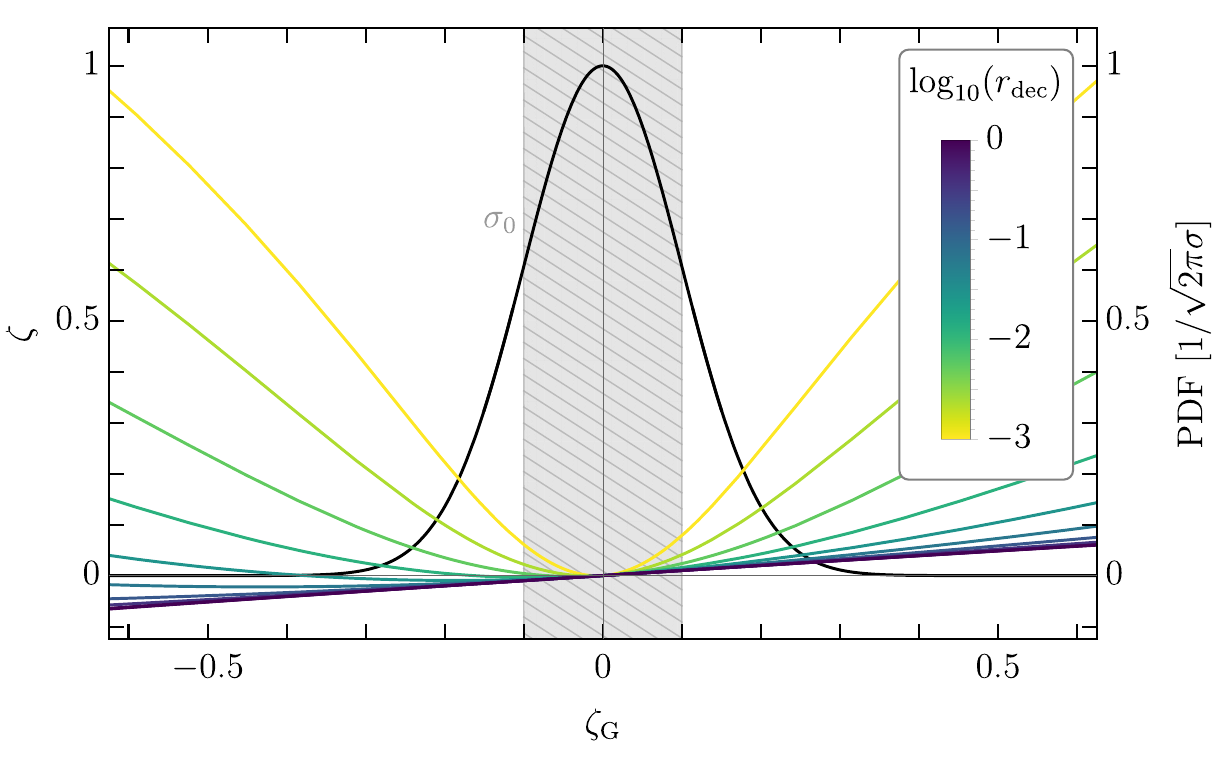}
\end{minipage}\hfill
\begin{minipage}{.5\textwidth} 
  \centering
  \includegraphics[width=\linewidth]{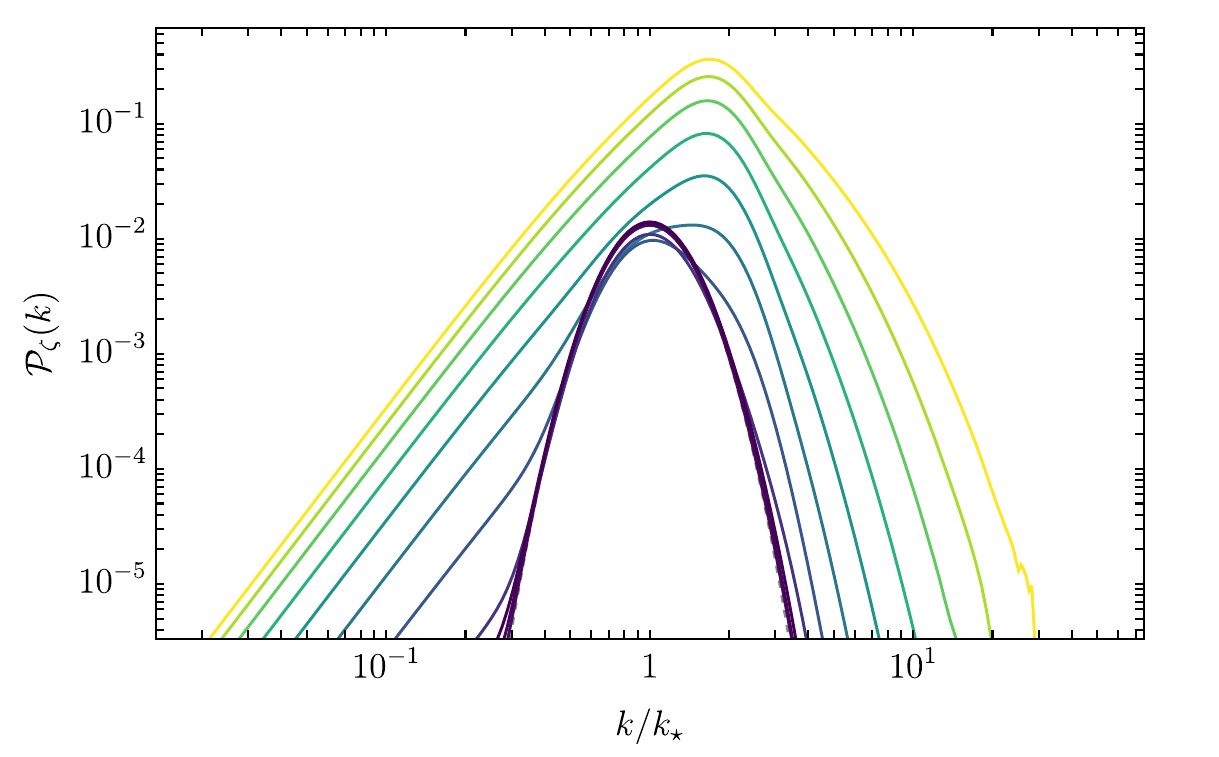} 
\end{minipage}
\caption{ {\it Left panel:} $F(\zeta_G)$ for the quadratic curvaton template~\eqref{eq:template} for different $r_{\rm dec}$, together with the Gaussian PDF of $\zeta_{\rm G}$. The variance of $\zeta_{\rm G}$ is fixed by $\sigma_0=\sqrt{\xi_{\rm G}(0)}$, and the template parameters are computed using Eqs.~\eqref{eq:fNLgNL2}. {\it Right panel:} Full non-Gaussian power spectrum~\eqref{eq:PkNG} obtained for the $F(\zeta_G)$ templates shown in the left panel, starting from a log-normal Gaussian power spectrum~\eqref{eq:PS_LN} with $\Delta=0.3$ and $\mathcal{A}=10^{-2}$ (gray dashed line). }
    \label{fig:spectra}
\end{figure*}

In Fig.~\ref{fig:spectra} we show how the full non-Gaussian power spectrum is sensitive to different amounts of NG in the quadratic curvaton scenario. 
In the left panel, we tune the template parameters~\eqref{eq:template} by varying $r_{\rm dec}$~\eqref{eq:fNLgNL2}.
As expected, a smaller $r_{\rm dec}$ corresponds to increasingly positive NGs, leading to a higher peak shifted to the right with respect to the Gaussian power spectrum, along with the appearance of a UV tail of increasing amplitude.
In contrast, a larger $r_{\rm dec}$ suppresses the amount of NGs, leading to a spectrum that approaches the Gaussian result in the limit $r_{\rm dec} \rightarrow1$.

\subsection{Tails of spectra for curvaton-like NGs}
\label{sec:tails}

Besides rescaling the peak amplitude, primordial NGs also induce non-trivial modifications in both the UV and IR regimes of the power spectrum. 

\paragraph{IR regime.} In the IR, the non-Gaussian power spectrum develops a universal $k^3$ behaviour~\cite{Veermae:2026yzz}. From~\eqref{eq:PkNG}, in the limit $k \rightarrow 0$ we obtain
\be\label{eq:k3_tail}
    \mathcal{P}(k) 
    \stackrel{k \ll k_{\rm IR}}{\sim} \mathcal{G}_2(1) \left(\frac{k}{k_{\rm IR}}\right)^3\,,
    \qquad
    k_{\rm IR} \equiv \left(\frac{2}{\pi}\int^{\infty}_0 \td x\, x^2 \frac{\mathcal{G}_2(\xi_{\rm G}(x)/\sigma_0^2)}{\mathcal{G}_2(1)}\right)^{-1/3}\,,
\ee
where $k_{\rm IR}$ characterises the correlation volume of the field. For sufficiently strong NGs, $k_{\rm IR} \sim k_\star$ causes the power spectrum to be completely dominated by the $k^3$ in the IR. For milder NGs, instead, the spectrum departs from the $k^3$ behaviour before the peak and approaches the Gaussian spectrum for $k \gtrsim k_{\rm IR}$. This IR feature is clearly visible in Fig.~\ref{fig:spectra} for almost all lines whose spectra significantly deviate from the Gaussian one. In particular, we note that the $k^{3}$ scaling defines the entire IR region before the peak for the spectra with $r_{\rm dec} \ll 1$.
From Eq.~\eqref{eq:PkNG}, this property is a general feature of locally non-Gaussian fields that only depends on the shape of $F(\zeta_{\rm G})$ through the definition of the scale $k_{\rm IR}$. In the right panel of Fig.~\ref{fig:series}, we see that the $k^3$ tail already appears at second order in the series expansion. Indeed, convolutions naturally generate such causal IR tails. Thus, the IR behaviour is expected to be captured perturbatively.

\paragraph{UV regime.} The large-$k$ tails are more difficult to pin down in general. However, if the Gaussian two-point correlation function $\xi_{\rm G}({\bf r})$ is irregular at the origin and regular away from the origin, the UV tail can be captured by the ${\bf r} \to 0$ limit. In detail, isotropy dictates that when $\xi_{\rm G}({\bf r})$ is regular around the origin, it admits an expansion in terms of even powers of $r$, i.e., $\xi_{\rm G}({\bf r}) \sim 1 - B r^{2}$. Let us consider deviations from this behaviour. When the two-point function asymptotes to
\be
    \xi_{\rm G}({\bf r}) \stackrel{r \rightarrow 0}{\sim} \sigma_0^2(1 - B\, r^{\alpha})\,,
\ee
where $B$ and $0 < \alpha < 2$ are positive constants, the corresponding Gaussian power spectrum behaves as 
\be \label{eq:UVPS}
    \mathcal{P}_{\rm G} (k)  
    \stackrel{k \rightarrow \infty}{\sim} \sigma_0^2 \, B \gamma_\alpha \,k^{-\alpha}\,,
\ee
where we defined $\gamma_\alpha = (2 / \pi ) \Gamma(2 + \alpha)\sin(\pi \alpha / 2)$. Note that $\gamma_\alpha = 0$ when $\alpha$ is an even integer, recovering the regular case, where the UV behaviour cannot be inferred from the ${\bf r} \to 0$ limit. In the following, our aim is to estimate how the exponent $\alpha$ is modified when the auxiliary Gaussian field gets mapped into the non-Gaussian one by $\mathcal{G}_2$.

In the ${\bf r} \to 0$ limit, the field becomes maximally correlated $\psi \to 1$, so it is sufficient to study the behaviour of $\mathcal{G}_2$ at that point.
Although the integral Eq.~\eqref{eq:G2} converges in this limit, the integrand diverges, complicating numerical evaluation.
To ensure numerical stability, it is more convenient to work with the quantity
\be\label{def:Q}
    Q(\psi) 
    = \frac{1}{2} \left \langle  (\zeta({\bf x})-\zeta({\bf y}))^2\right \rangle 
    = \mathcal{G}_2 (1) - \mathcal{G}_2 (\psi)\,,
\ee
which naturally vanishes at $\psi=1$. In particular, if 
\be
    Q(\psi) \stackrel{\psi \rightarrow 1}{\sim} \mathcal{G}_2 (1) Q_1 (1-\psi)^{\lambda}
\ee
when $\psi \to 1$, then the correlation function transforms as $\psi({\bf r}) = 1 - B\, r^{\alpha}$, $\xi({\bf r}) = \mathcal{G}_2 (1) (1 - Q_1 B^{\lambda}\, r^{\alpha \lambda})$ and the power spectrum acquires a $k^{-\alpha \lambda}$ tail
\be
    \mathcal{P}_{\zeta} (k)  
    \stackrel{k \rightarrow \infty}{\sim} \mathcal{G}_2 (1) Q_1 \gamma_{\alpha \lambda} \, B^{\lambda} \,k^{-\alpha \lambda}\,.
\ee
Importantly, for regular Gaussian spectra $\alpha = 2$, the non-Gaussian power spectrum acquires a power-law UV tail when $0 < \lambda < 1$.

Let us consider some examples for the NGs of the curvaton. For the ansatz \eqref{eq:template} approximating the quadratic curvaton, we find
\bea
    \frac{Q(\psi)}{\sigma^2_0} 
    = \frac{1}{4 \pi}\left( \frac{1+\epsilon^2}{2 \beta}\right)^2  
    &\int \td \zeta_+ \td \zeta_- \,e^{-\frac{1}{2}(\zeta_+^2+\zeta_-^2)}
\\
    \times&\ln^2 \left[
    \frac{
    \epsilon^2+\frac{1}{2}
    \left(
    \sqrt{2}\!-\!\sigma_0\beta\zeta_{-} \sqrt{1-\psi}\!-\!\sigma_0\beta\zeta_+ \sqrt{1+\psi}
    \right)^2
    }{
    \epsilon^2+\frac{1}{2}
    \left(
    \sqrt{2}\!+\!\sigma_0\beta\zeta_{-} \sqrt{1-\psi}\!-\!\sigma_0\beta\zeta_+ \sqrt{1+\psi}
    \right)^2
    }
    \right]
    \,,
\eea
where we introduced the independent unit Gaussian variables
$\zeta_\pm = (\zeta_1 \pm \zeta_2)/\sqrt{2(1\pm \psi)}$, that is, $\langle \zeta_{\pm}^2\rangle=1$, $\langle \zeta_{+}\zeta_{-}\rangle=0$. With this choice, all dependence on $\psi$ is explicit in the integrand. We note that this integral provides a numerically stable way to compute $\mathcal{G}_2$ via \eqref{def:Q}. 
The $\psi \rightarrow1$ asymptotic can be obtained analytically,
\be \label{eq:curvatonUV}
    \frac{Q(\psi)}{\sigma^2_0}\stackrel{\psi \rightarrow 1}{\sim}
    \sigma_0^2\left( 1+\epsilon^2\right)^2
    \left\langle
    \frac{(1-\sigma_0 \beta\zeta_+)^2}{\bigl(\epsilon^2 + (1-\sigma_0\beta\zeta_+)^2\bigr)^2}
    \right\rangle
    (1-\psi)
    \equiv \mathcal{G}_2 (1) Q_1(1-\psi)\,,
\ee
where $\zeta_-$ has been averaged out. The coefficient $Q_1$ depends on $\beta$, $\epsilon$ and $\sigma_0$ and can be determined numerically through a Gaussian average.

For the log-normal spectrum used in Figs.~\ref{fig:series}--\ref{fig:spectra}, we have
$
\psi_{\rm LN}
\sim
1-e^{2 \Delta^2}(k_\star r)^2/6,
$
when ${\bf r} \rightarrow 0$ so that the 2-point function is regular at the origin and thus  $\alpha=2$. As a result, the coefficient $\gamma_\alpha$ in Eq.~\eqref{eq:UVPS} vanishes identically. This indicates that, for a log-normal spectrum, the small-$x$ behaviour does not contain information about the UV tail of the power spectrum.
This can also be understood from the right panel of Fig.~\ref{fig:spectra}, where the non-Gaussian spectra appear to follow the same UV scaling as the Gaussian spectrum. The UV behaviour is therefore not captured by a simple power-law scaling, and Eq.~\eqref{eq:UVPS} does not provide useful information in this case. 

We note that for $r_{\rm dec}=1$, the template~\eqref{eq:template} reduces, up to a constant shift, to the USR-like NGs~\cite{Atal:2018neu,Tomberg:2023kli,Pi:2022ysn,Veermae:2026yzz}
\be \label{eq:USRTemplate}
    \zeta
    = F(\zeta_{\rm G}) = \frac{1}{\beta} \ln \left | 1+\beta \zeta_{\rm G}\right|\,.
\ee
In particular, from~\eqref{eq:fNLgNL2}, we find that in the limit $r_{\rm dec}\to1$ one has $\beta=3/2$. However, the power spectra shown in Figs.~\ref{fig:series}--\ref{fig:spectra} exhibit qualitatively different UV behaviour from that characteristic of~\eqref{eq:USRTemplate}. In the latter case, the spectrum develops a power-law UV tail, which can be understood as a direct consequence of the divergence of the mapping $F(\zeta_{\rm G})$. The strongly correlated regime asymptotes to~\cite{Veermae:2026yzz}
\be \label{eq:UVUSR}
Q(\psi)
\propto \sqrt{1-\psi}
\qquad \implies \qquad
\mathcal{P}_\zeta (k)  \propto k^{-\alpha/2} \,.
\ee
where $\mathcal{G}_2 (1)\, Q_1 =  \pi / (\sqrt{2} \beta^2) \exp(-1/(2 \sigma_0^2\beta^2))$.
Therefore, for USR-like NGs and a log-normal spectrum with $\alpha=2$, the non-Gaussian power spectrum should naturally develop a $k^{-1}$ UV tail. However, in the limiting case $r_{\rm dec}=1$, shown in the right panel of Fig.~\ref{fig:spectra} (dark purple line), this tail is absent. This reflects the fact that, over the range of $\zeta_{\rm G}$ relevant to our analysis, the mapping $F(\zeta_{\rm G})$ is effectively Gaussian, so that the divergence does not play a significant role; see the left panel of Fig.~\ref{fig:spectra}.

In conclusion, when the divergence in Eq.~\eqref{eq:USRTemplate} is cured by introducing a regulator $\epsilon$, as, for example, in the case of the quadratic curvaton, we expect the $k^{-1}$ UV tail present in the USR-like ansatz to disappear, and the non-Gaussian spectra to follow the same UV behaviour as the Gaussian one.
This can be readily demonstrated by considering simple Gaussian spectra with a UV tail $\mathcal{P}_{\rm G}(k) \stackrel{k \gg k_{\rm UV}}{\propto} k^{-1}$ ($\alpha =1$).
We see that, for this simple case, in the curvaton scenario~\eqref{eq:curvatonUV}, the non-Gaussian spectrum recovers the same UV scaling, $\mathcal{P}(k) \propto k^{-1}$ ($\lambda=1$).
If the mapping is instead of the USR-like~\eqref{eq:UVUSR}, the divergence leads to an enhanced UV tail in the non-Gaussian spectrum,
$\mathcal{P}(k)\propto  k^{-1/2}$ ($\lambda=1/2$). 

\section{Can Supermassive PBHs evade $\mu$-distortion constraints?}
\label{sec:SMBH}

PBHs can serve as candidates for dark matter (DM)~\cite{Green:2020jor,Carr:2020gox,Carr:2026hot}~(for a recent review of the DM problem, see Ref.~\cite{Cirelli:2024ssz}), probe phase transitions during the evolution of the Universe~\cite{Escriva:2023nzn,Escriva:2024ivo,Pritchard:2025pcn,Blas:2026xws, Boccia:2026azi}, be progenitors of LIGO–Virgo–KAGRA GW events~\cite{Bird:2016dcv,Raidal:2017mfl, Raidal:2018bbj,Vaskonen:2019jpv,Hutsi:2020sol,Wong:2020yig,DeLuca:2021wjr,Franciolini:2022tfm,Miller:2024fpo,Andres-Carcasona:2024wqk,Andres-Carcasona:2026avd}, or seeds of the supermassive black holes (SMBHs) that populate most galactic nuclei~\cite{Carr:2018rid}. In this section, we focus on the last scenario.

\subsection{PBH abundance from non-Gaussian threshold statistics}
\label{sec:thr}

Let us begin by reviewing the estimate of the PBH abundance. The relation between the mass of the resulting PBH $M_\text{\tiny PBH}$ and the horizon mass $M_\text{\tiny H}$ is dictated by the following critical scaling law~\cite{Choptuik:1993,Evans:1994}
\be\label{Eq:MassPBH}
    M_\text{\tiny PBH}(\mathcal{C}) = \mathcal{K} M_\text{\tiny H} (\mathcal{C} - \mathcal{C}_{\text{\tiny c}})^{\gamma}\,,
\ee
where $\gamma=0.38$~\cite{Musco:2008hv,Ianniccari:2024ltb} and $\mathcal{C}$ is the compaction defined as twice the local mass $M$ excess relative to the background mass $M_b$, divided by the areal radius $R(r,t)=a(t)\, e^\zeta\, r$ (in terms of the scale factor $a$ and the comoving curvature perturbation $\zeta$)
\begin{align}\label{eq:DefinitionCompaction}
    \mathcal{C}(r,t) 
    = \frac{2\left[M(r,t) - M_b(r,t)\right]}{R(r,t)} 
    = \frac{2}{R(r,t)}\int_{V_{R}} \td^{3}\vec{x}\,
    \rho_b(t)\delta(\vec{x},t)\,,
\end{align}
where $\delta$ is the density contrast. The horizon mass corresponds to the $k$ with the relation
\be \label{eq:MHtok}
    M_{\text{\tiny H}} \simeq 17 \Msun\left(\frac{g_{\star}}{10.75}\right)^{-1 / 6}\left(\frac{k / \kappa}{10^6 \mathrm{Mpc}^{-1}}\right)^{-2}\,,
\ee
where the factor $\kappa$ depends on the shape of the power spectrum (see Ref.~\cite{Musco:2020jjb}) and in this work, it is fixed to $\kappa=4.5$. 

The mass fraction $\beta$ of the total radiation density that collapses into PBHs is obtained from the probability distribution of the compaction ${\rm P}(\mathcal{C})$ as 
\be\label{eq:beta}
    \beta(M_\text{\tiny PBH},M_{H}) = 
    \int_{\mathcal{C} > \mathcal{C}_{\text{\tiny c}}} \frac{M_\text{\tiny PBH}}{M_{\rm H}}
    \delta\left[ \ln\frac{M_\text{\tiny PBH}}{M_\text{\tiny PBH}(\mathcal{C})} \right] 
    {\rm P}\left(\mathcal{C}\right)
    \td\mathcal{C}\,.
\ee
We follow Ref.~\cite{Musco:2020jjb} to compute the threshold $\mathcal{C}_{\text{\tiny c}}$ and the position of the maximum of the compaction function $r_m$, which depend on the shape of the power spectrum. 
The present-day mass function of PBHs is then given by
\bea \label{eq:fpbh1}
    f_{\text{\tiny PBH}}\left(\MPBH\right)
    \simeq
    \frac{1}{\Omega_{\rm  DM}}
    \int \td \ln M_{H} \left (\frac{M_{H}}{M_\odot}\right )^{-1/2}
    \left(\frac{g_{*s}^4/g_*^3}{106.75} \right)^{-\frac14}\!
    \left (\frac{\beta(M_\text{\tiny PBH},M_\text{\tiny H})}{10^{-8}} \right)\,,
\eea
where $\Omega_{\rm  DM} = 0.264$ is the present-day cold dark matter density parameter, and we integrate the mass fraction $\beta$ over the horizon masses contributing to a given PBH mass.
The prefactor inside the integral arises from computing the present-day abundance by noting that the ratio between the PBH energy density and the entropy density is conserved in a comoving volume. Evaluating this ratio at the horizon crossing of the scale that collapses, one has
\be
  \left.\frac{\rho_{\rm PBH}}{s}\right|_{\rm form}
  =\frac{3}{4}\,\frac{g_\star}{g_{\star s}}\,T_{\rm f}\;\beta(M_{\rm PBH},M_H)\,,
\ee
where the formation temperature is determined by the horizon mass, $M_H=(4\pi/3)\rho\,H^{-3}\propto g_\star^{-1/2}T_{\rm f}^{-2}$, so that
\be
  T_{\rm f}\simeq 0.22\ {\rm GeV}\,
  \left(\frac{g_\star}{106.75}\right)^{-1/4}\left(\frac{M_H}{M_\odot}\right)^{-1/2}.
\ee
Dividing by the present-day dark-matter-to-entropy ratio, $\rho_{{\rm DM},0}/s_0=\Omega_{\rm DM}\rho_{c,0}/s_0$, one obtains Eq.~\eqref{eq:fpbh1}.
Finally, the total abundance of PBHs relative to dark matter is
\be
    f_{{\text{\tiny PBH}}} = \int f_{{\text{\tiny PBH}}}\left(\MPBH\right) \td \ln \MPBH\,.
\ee

The critical component of the estimate is the distribution ${\rm P}(\mathcal{C})$ of the compaction. To obtain the non-Gaussian \emph{threshold statistics} of the compaction function, we will follow the formalism developed by~\cite{Ferrante:2022mui}. On super-horizon scales, adopting the gradient expansion approximation and assuming spherical symmetry, the density contrast is~\cite{Harada:2015yda}
\begin{align}\label{eq:SphericalDelta}
    \delta(r,t) = 
    -\frac{4}{9}
    \left(\frac{1}{aH}\right)^2 
    e^{-2\zeta(r)}\left[
    \zeta^{\prime\prime}(r) + \frac{2}{r}\zeta^{\prime}(r) + \frac{1}{2}\zeta^{\prime 2}(r)
    \right]\,,
\end{align}
where $' \equiv d/dr$, the factor 4/9 is for a radiation-dominated universe, and $\zeta(r)$ is assumed to be constant on super-horizon scales\footnote{We can use the super-horizon expansion since we are interested in the correlation at scales larger than the Hubble radius within which PBHs form.}. In substituting Eq.~\eqref{eq:SphericalDelta} into Eq.~\eqref{eq:DefinitionCompaction} and performing the volume integral, the compaction function takes the form~\cite{Harada:2015yda}
\begin{align}\label{eq:CompactionFull}
    \mathcal{C}(r) 
    = 
    -\frac{4}{3}\,r\,\zeta^{\prime}(r)\left[
    1 + \frac{r}{2}\zeta^{\prime}(r)
    \right] 
    = 
    \mathcal{C}_1(r) - \frac{3}{8}\mathcal{C}^2_1(r),
    \qquad
    \mathcal{C}_1(r) 
    = -\frac{4}{3} r\zeta^{\prime}(r).
\end{align} 
Notice that $\mathcal{C}$ becomes time-independent and Eq.~\eqref{eq:CompactionFull} includes the full non-linear relation between $\delta$ and $\zeta$. 
We define $r_m$ as the scale at which the compaction function is maximised. Therefore, it verifies the condition 
\be
    \mathcal{C}^{\prime}(r_m) = 0
    \qquad\text{that is}\qquad
    \zeta^{\prime}(r_m) + r_m\zeta^{\prime\prime}(r_m) = 0    
\ee
in terms of  the comoving curvature perturbation. 
If we define $\mathcal{C}_{\text{\tiny max}} = \mathcal{C}(r_m)$ as the compaction at the position of the maximum,  PBHs form only if the maximul of the compaction function exceeds some threshold, $\mathcal{C}_\text{\tiny max} > \mathcal{C}_{\text{\tiny c}}$. At the horizon crossing of the relevant scale $r_m = (aH)^{-1}$, the compaction function at its peak becomes equal to the fully non-linear density contrast smoothed over the horizon volume, which is the quantity we will compute the correlation of.

In a radiation-dominated universe, the compaction function can be expressed as
\begin{align}\label{eq:CCgau}
    \mathcal{C}(r) = 
    \mathcal{C}_{\rm G}(r)\,
    F' - \frac{3}{8}
     \mathcal{C}^2_{\rm G}(r)
    \left(F'
    \right)^2\,, 
    \qquad
    \mathcal{C}_{\rm G}(r) \equiv
    -\frac{4}{3}r\zeta_{\rm G}^{\prime}(r)\,
\end{align}
where $F' \equiv \pd F/\pd \zeta_{\rm G}$. The compaction function thus depends on both the Gaussian linear component $\mathcal{C}_{\rm G}$ and the Gaussian curvature perturbation $\zeta_{\rm G}$. 
Both are Gaussian random variables since $\zeta_{\rm G}$ is Gaussian by definition, while $\mathcal{C}_{\rm G}$ is proportional to the derivative $\zeta_{\rm G}'$. 
We write~\cite{Young:2022phe}
\be
    \mathcal{C}_{\rm G}(\vec x, r)
    = -\frac{4}{9}r^2\int \td^3y\,\nabla^2\zeta_{\rm G}(\vec y)\,W(\vec x-\vec y,r)
\ee
and 
\be
    \zeta_{\rm G}(\vec x, r)
    = \int \td^3y \,\zeta_{\rm G}(\vec y)\,W_{\text{\tiny s}}(\vec x-\vec y,r)\,,
\ee
where $W_{\text{\tiny s}}$ is the spherical-shell window and $W$ is the Heaviside-step window with Fourier transforms
\be
    W_{\text{\tiny s}}(k,r) 
    = \frac{\sin(kr)}{kr}\,,
    \qquad
    W(k,r)   
    =  3\frac{\sin(kr) - kr\cos(kr)}{(kr)^3}\,.
\ee
The two-dimensional joint PDF of $\vec{Y} = (\zeta_{\rm G},\mathcal{C}_{\rm G})$ can be written as
\bea\label{eq:PDFCompa}
    {\rm P}_{\rm G}(\mathcal{C}_{\rm G},\zeta_{\rm G}) 
    &= \frac{1}{2\pi\sqrt{\det\Sigma}}
    \exp\left(
    -\frac{1}{2}\vec{Y}^{\rm T}\Sigma^{-1}\vec{Y}
    \right)\\
    &=
    \frac{1}{2\pi\sigma_{\text{\tiny c}}\sigma_{\text{\tiny r}}\sqrt{1-\gamma_{\text{\tiny cr}}^2}}
    \exp\left(
    -\frac{\zeta_{\rm G}^2}{2\sigma_{\text{\tiny r}}^2}
    \right)
    \exp\left[
    -\frac{1}{2(1-\gamma_{\text{\tiny cr}}^2)}\left(
    \frac{\mathcal{C}_{\rm G}}{\sigma_{\text{\tiny c}}} - \frac{\gamma_{\text{\tiny cr}}\zeta_{\rm G}}{\sigma_{\text{\tiny r}}}
    \right)^2
    \right],  
\eea
where $\gamma_{\text{\tiny cr}} \equiv \sigma_{\text{\tiny cr}}^2/(\sigma_{\text{\tiny c}} \sigma_{\text{\tiny r}})$ and the covariance matrix
\bea
    \Sigma
    \equiv 
    \langle \vec{Y}\vec{Y}^{\rm T}\rangle
    =
    \left(
    \begin{array}{cc}
    \sigma_{\text{\tiny c}}^2 &
    \sigma_{\text{\tiny cr}}^2 \\ \sigma_{\text{\tiny cr}}^2 &\sigma_{\text{\tiny r}}^2    
    \end{array}
    \right)\,,
\eea
is given by
\bea\label{eq:Var}
    \sigma_{\text{\tiny c}}^2  
    &=\langle\mathcal{C}_{\rm G}\mathcal{C}_{\rm G}\rangle  
    = \frac{16}{81}\int_0^{\infty}\frac{\td k}{k}(kr_m)^4 W^2(k,r_m)  \calP^{T}_{\zeta_{\rm G}}(k)\,,
\\
    \sigma_{\text{\tiny cr}}^2
    &= \langle\mathcal{C}_{\rm G}\zeta_{\rm G}\rangle  =  
    \frac{4}{9}\int_0^{\infty}\frac{\td k}{k}(kr_m)^2 W(k,r_m) W_{\text{\tiny s}}(k,r_m)  \calP^{T}_{\zeta_{\rm G}}(k)\, ,
\\
    \sigma_{\text{\tiny r}}^2  
    &= \langle\zeta_{\rm G}\zeta_{\rm G}\rangle  
    = \int_0^{\infty}\frac{\td k}{k}
    W_{\text{\tiny s}}^2(k,r_m) \calP^{T}_{\zeta_{\rm G}}(k)\,,
\eea
with $\calP^{T}_{\zeta_{\rm G}}(k) = T^2\left(k, r_m\right) \calP_{\zeta_{\rm G}}(k)$, where $T\left(k, r_m\right)$ is the radiation transfer function, and all the entries are evaluated at $r_m=1/aH$. 

Up to this point, a few simplifications have been made on top of spherical symmetry and the gradient expansion. First, applying the window and transfer functions to the auxiliary Gaussian field instead of $\zeta$ is a simplification that enables computability. An analytic prescription for windowed non-Gaussian fields is currently missing from the literature, and the accuracy of this simplification should be tested using more accurate but computationally expensive numerical methods. Second, the constraint~\eqref{eq:CompactionFull}, as well as the requirement that there is a peak at $r=0$, can affect the distribution of $\mathcal{C}_{\rm G}$ and $\zeta_{\rm G}$ when either of these quantities is correlated with $\mathcal{C}'(r_m)$ or other constrained quantities~\cite{Germani:2019zez,Germani:2023ojx}. These constraints are not accounted for in the Gaussian PDF~\eqref{eq:PDFCompa}.

\begin{figure}
	\begin{center}
\includegraphics[width=0.74\textwidth]{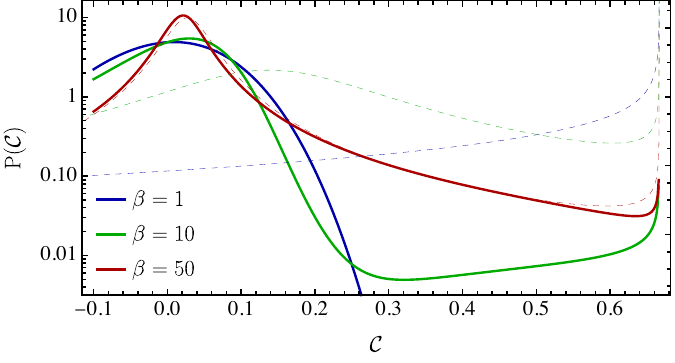}
		\caption{The distribution of the compaction~\eqref{eq:Domain} for the ansatz~\eqref{eq:template}. The solid lines show ${\rm P}\left(\mathcal{C}\right)$ computed numerically from~\eqref{eq:Domain} for $\beta \in \{1,10,50\}$, $\epsilon = 0.05$, and a log-normal $\zeta_{\rm G}$ spectrum~\eqref{eq:PS_LN} with $\mathcal{A} = 10^{-2}$, $\Delta = 0.5$, and $r_m k_{\star} = 2.5$. The dashed lines show the corresponding approximate PDF in the large-$\beta$ limit~\eqref{eq:appPC}.}
	\label{fig:P(C)}  
	\end{center}
\end{figure}

The distribution of the compaction function at any given point can now be expressed as
\bea
\label{eq:Domain} 
     {\rm P}\left(\mathcal{C}\right) 
&    =
    \int_{\mathcal{C}_{\rm G}F'(\zeta_{\rm G}) < 4/3} \delta(\mathcal{C} - \mathcal{C}(\mathcal{C}_{\rm G},\zeta_{\rm G})) {\rm P}_{\rm G}(\mathcal{C}_{\rm G},\zeta_{\rm G})\td\mathcal{C}_{\rm G} \td\zeta_{\rm G}\,
\\
&   =
    \frac{1}{\sqrt{1- 3\mathcal{C}/2}} \int \frac{\td\zeta_{\rm G}}{|F'(\zeta_{\rm G})|} {\rm P}_{\rm G}\left[\frac{4}{3 F'(\zeta_{\rm G})}\left(1 - \sqrt{1- 3\mathcal{C}/2}\right),\zeta_{\rm G} \right]  \,,
\eea
where we integrated out $\mathcal{C}_{\rm G}$ in the second row. Further integration depends on the shape of $F$ and must be evaluated numerically in most cases. In numerical practice, it can be more convenient to plug the first expression of \eqref{eq:Domain} into \eqref{eq:beta} and take the integral over $\mathcal{C}$ analytically, leaving a numerical Gaussian double integral over $\mathcal{C}_{\rm G}$ and $\zeta_{\rm G}$. This approach was used in Ref.~\cite{Ferrante:2022mui}, and because of that, the distribution of the compaction was never considered there.

The ansatz~\eqref{eq:template} in the limit of $\beta \to \infty$ allows us to take the remaining integral analytically since $F'(\zeta_{\rm G}) \propto 1/\zeta_{\rm G}$. This yields
\bea\label{eq:appPC}
    {\rm P}\left(\mathcal{C}\right) 
    &\sim 
    \frac{\sqrt{1-\gamma_{\text{\tiny cr}}^2} }{\pi\sqrt{1 - 3\mathcal{C}/2}} \frac{\tilde \beta}{1 - \frac{8}{3}\gamma_{\text{\tiny cr}} \tilde \beta (1 - \sqrt{1- 3\mathcal{C}/2}) + \frac{16}{9}\tilde \beta^2 (1 - \sqrt{1- 3\mathcal{C}/2})^2}
\eea
where $\tilde\beta = \sigma_{\text{\tiny r}}\beta/(\sigma_{\text{\tiny c}}(1 + \epsilon^2))$. This distribution is defined for $\mathcal{C} < 2/3$ by construction. It is peaked around $\mathcal{C} \approx \gamma_{\text{\tiny cr}}/\tilde\beta \sim 0$ and decreases polynomially away from that peak.

When accounting for the statistics of the curvaton, not just the shape of $F$, we see that the strength of NGs is controlled by $\sigma_r \beta$ instead of just $\beta$.  Numerical estimates (e.g., in Fig.~\ref{fig:P(C)}) indicate that NGs can be considered strong already when $\sigma_r \beta \gtrsim 0.1$.

Some examples of the distribution are shown in Fig.~\ref{fig:P(C)} assuming the logarithmic ansatz \eqref{eq:template} and a log-normal spectrum~\eqref{eq:PS_LN} with $\mathcal{A} = 10^{-2}$ and $\Delta = 0.5$. We observe that when the NGs are mild ($\beta = 1$, $\sigma_r\beta = 0.04$), the distribution is almost Gaussian. However, for stronger NGs, a long tail develops. Although the large $\beta$ approximation~\eqref{eq:appPC} (indicated by the dashed lines) fails at the onset of the strongly non-Gaussian regime ($\beta = 10$, $\sigma_r\beta = 0.4$), it agrees with the asymptotic estimate quite well already when ($\beta = 50$, $\sigma_r\beta = 1.9$). At this stage, the probability for PBH formation ceases to be exponentially suppressed, and the overproduction of PBHs becomes more likely. Another artefact of this behaviour can be observed in Fig.~\ref{fig:QuadAbu}, where the flattened tail causes $f_{\rm PBH}$ to depend weakly on $\beta$.  

\subsection{$\mu$-distortion constraints on PBHs in light of strong non-Gaussianities}

In the range of scales $1\,{\rm Mpc}^{-1}\lesssim k\lesssim 10^{5}\,{\rm Mpc}^{-1}$, the primordial power spectrum is strongly constrained by the CMB spectral distortions~\cite{Chluba:2012we,Chluba:2013dna}. Indeed, at redshifts $z\lesssim 10^6$, energy injections into the primordial plasma give rise to persistent spectral distortions in the CMB, which are divided into chemical potential $\mu$-type distortions created at early times and Compton $y$-type distortions created at $z\lesssim 5\times 10^4$. For a given curvature power spectrum $\Pz(k)$, the spectral distortions read~\cite{Chluba:2012we,Chluba:2013dna}
\be\label{eq:Xdist}
    X 
    = \int_{k_{\rm min}}^{\infty}\frac{\td k}{k}\,\Pz(k)\,W_X(k)\,,
\ee
with $X=\mu,y$, $k_{\rm min}=1\,{\rm Mpc}^{-1}$ and window functions approximated by
\bea\label{eq:Wmu,Wy}
    W_\mu(k) 
    &= 2.2\left[e^{-\frac{(\hat k/1360)^2}{1+(\hat k/260)^{0.6}+\hat k/340}}-e^{-(\hat k/32)^2}\right]\,,\\
    W_y(k) 
    &= 0.4\,e^{-(\hat k/32)^2}\,,
\eea
with $\hat k = k/(1\,{\rm Mpc}^{-1})$. The COBE/FIRAS observations constrain $\mu \leq 4.7\times 10^{-5}$~\cite{Bianchini:2022dqh} and $y\leq 1.5\times 10^{-5}$~\cite{Fixsen:1996nj,Fabbian:2025txv} at $95\%$ confidence level.\footnote{Since the non-Gaussian power spectrum is derived through the non-Gaussian two point function in coordinate space, it is computationally more efficient to express \eqref{eq:Xdist} in coordinate space 
\be
    X = \int^{\infty}_{0} \td r \,\bar{W}_{X}(r) \mathcal{G}_2\left(\xi_{\rm G}(r)/\sigma_0^2\right)
\ee
and circumvent the computation of the non-Gaussian power spectrum. In this case, $\bar{W}_{X}(r) = (2 r/\pi)\int^{\infty}_{k_{\rm min}} \td k \,k \sin(k r) W_X(k)$.}

The origin of SMBHs represents an outstanding puzzle in astrophysics. While it is well established that they reside at the centres of most galaxies~\cite{Kormendy:1995er,Magorrian:1997hw,Richstone:1998ky}, the processes driving their formation remain unclear. Even a small population of heavy PBHs within the mass range $10^3$--$10^6\,\Msun$ could provide viable seeds for SMBHs~\cite{Duechting:2004dk,Kohri:2014lza,Bernal:2017nec,Iovino:2024tyg}. Following Refs.~\cite{Vaskonen:2020lbd,Serpico:2020ehh}, the required primordial black holes abundance $f_{\rm PBH}$ in order to be the seeds for SMBHs can be estimated as
\be\label{eq:SMBH}
    \fpbh \sim \mathcal{O}(10^{-6}) \times \langle M_{\rm PBH}\rangle/M_{\rm SMBH}\,,
\ee
with $\langle M_{\rm PBH}\rangle \gtrsim 10^3\,\Msun$. In the Gaussian approximation, the FIRAS bound on the power spectrum amplitude translates into a PBH abundance that is too small to provide a primordial origin for SMBH seeds.

In the absence of primordial NGs, generating a sizeable PBH abundance requires enhancing the power spectrum up to $\mathcal{O}(10^{-2})$, an enhancement that, on scales $k\lesssim 8\cdot10^{4}$ Mpc$^{-1}$, is in clear tension with the COBE/FIRAS constraints. Remarkably, using the ansatz~\eqref{eq:template}, the curvaton scenario generates sizeable positive NGs that substantially enhance the high-$\zeta$ tail of the PDF, so that a non-negligible population of heavy PBHs can form even at peak amplitudes $\mathcal{A}_{\rm pk}\sim 10^{-3}$, compatible with the FIRAS bound.

\begin{figure}
	\begin{center}
		\includegraphics[width=0.99\textwidth]{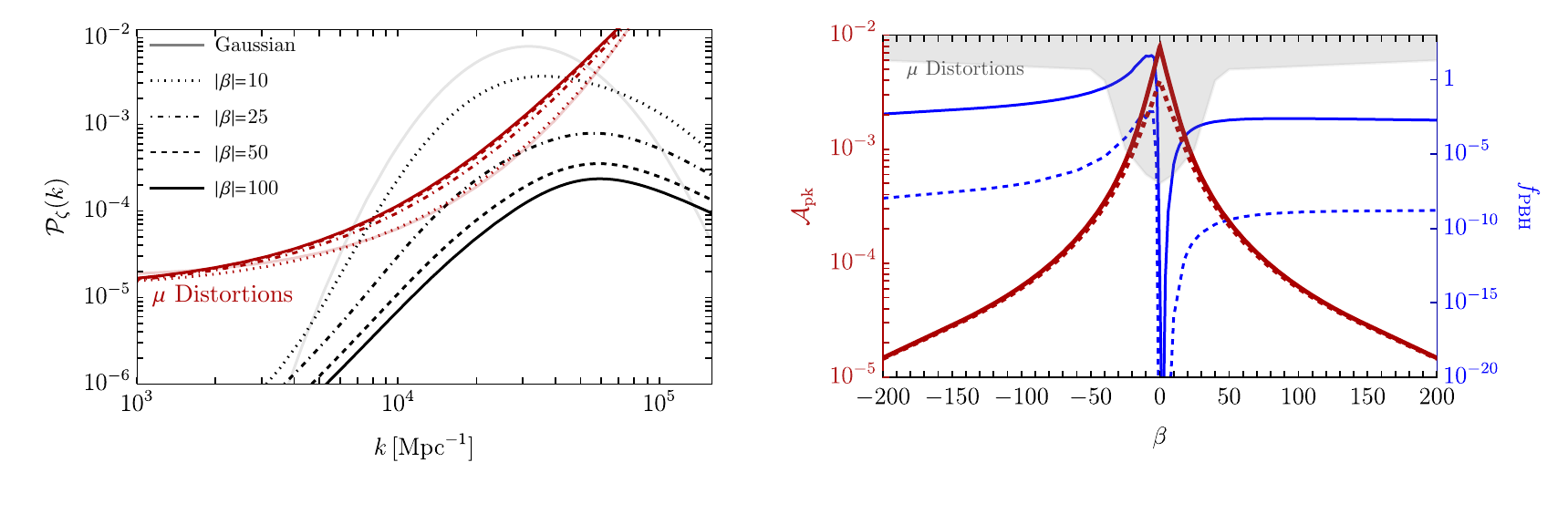}
		\caption{{\it Left panel:} Full non-Gaussian power spectra~\eqref{eq:PkNG} (black lines) obtained for the $F(\zeta_G)$ templates defined in Eq.~\eqref{eq:template}, with varying $\beta$ and $\epsilon=\sqrt{0.05}$, starting from a Gaussian power spectrum~\eqref{eq:PS_LN} with $\Delta=0.5$, $\mathcal{A}=10^{-2}$, and $k_{\star}=3\cdot10^{4}$ Mpc$^{-1}$ (gray line). The red lines show the $\mu$-distortion constraints on the amplitude of the peak of the power spectrum, $\mathcal{A}_{\rm pk}$, for the Gaussian (light red line) and non-Gaussian (red lines) power spectra. The different line styles correspond to different $|\beta|$. {\it Right panel:} The red line shows the amplitude of the peak of the power spectrum, $\mathcal{A}_{\rm pk}$, as a function of $\beta$, while the gray regions represent the $\mu$-distortion constraints at the relevant scale. The blue line shows the PBH abundance, $f_{\rm PBH}$, as a function of $\beta$. The solid (dashed) lines are obtained using a log-normal power spectrum~\eqref{eq:PS_LN} with $\Delta=0.5$, $\mathcal{A}=10^{-2}$ ($\mathcal{A}=0.005$), and $k_{\star}=3\cdot10^{4}$ Mpc$^{-1}$. }
	\label{fig:QuadAbu}  
	\end{center}
\end{figure}

As a benchmark\footnote{Through Eqs.~\eqref{Eq:MassPBH} and~\eqref{eq:beta}, this benchmark corresponds to a mass function peaked around $\mathcal{O}(10^{5})\,\Msun$, which is therefore the PBH mass discussed throughout this section.}, in Fig.~\ref{fig:QuadAbu} we assume a log-normal Gaussian power spectrum~\eqref{eq:PS_LN} with $\mathcal{A}=10^{-2}$, $\Delta=0.5$ and $k_*=3\cdot 10^{4}\,{\rm Mpc}^{-1}$. Fixing the amplitude of the Gaussian power spectrum, i.e. the gray line in the left panel of Fig.~\ref{fig:QuadAbu}, we find that for $\epsilon= \sqrt{0.05}$ and $|\beta|\gtrsim 30$ the power spectrum is suppressed below the FIRAS bound (red lines in the same panel). Since the resulting shape of the power spectrum is different, given~\eqref{eq:Xdist}, the constraint on $\mathcal{A}_{\rm pk}$ differs slightly between the Gaussian (light red line) and the non-Gaussian (red line) case.

As shown in the right panel of Fig.~\ref{fig:QuadAbu}, the red line represents the suppression of the power spectrum peak as a function of $\beta$, while the blue line gives the corresponding abundance $f_{\rm PBH}$. For $\beta\gtrsim 50$ ($\beta\lesssim -50$), the abundance saturates around $f_{\rm PBH}\simeq 10^{-3}$ ($f_{\rm PBH}\simeq 10^{-1}$). As discussed in Sec.~\ref{sec:thr}, this happens because the quantity relevant for the PBH abundance is not the PDF of $\zeta$, which was used in Ref.~\cite{Hooper:2023nnl}, but the density contrast field $\delta$: suppressing the PDF of $\zeta$ does not necessarily imply a suppression of the PDF of $\delta$.

We stress that this statement is scale-dependent, since $W_\mu$ in Eq.~\eqref{eq:Wmu,Wy} falls exponentially for $k\gtrsim10^{3}\,{\rm Mpc}^{-1}$. For a log-normal spectrum~\eqref{eq:PS_LN} with $\Delta=0.5$, the FIRAS limit constrains the peak amplitude to $\mathcal{A}_{\rm pk}\lesssim2\cdot10^{-5}$ at $k_\star=10^{3}\,{\rm Mpc}^{-1}$, $\mathcal{A}_{\rm pk}\lesssim6\cdot10^{-5}$ at $k_\star=10^{4}\,{\rm Mpc}^{-1}$ and $\mathcal{A}_{\rm pk}\lesssim5\cdot10^{-4}$ at $k_\star=3\cdot10^{4}\,{\rm Mpc}^{-1}$, which is the benchmark scale adopted below; this is the bound displayed by the red lines in the left panel of Fig.~\ref{fig:QuadAbu}. For $|\beta|\lesssim50$ amplitude $\mathcal{A}_{\rm pk}\sim10^{-3}$ is therefore allowed only for $k_\star\gtrsim3\cdot10^{4}\,{\rm Mpc}^{-1}$, while for $k\simeq10^3$ Mpc$^{-1}$, only $|\beta|\gtrsim180$ can evade the bounds.
It is worth clarifying which portion of the SMBH-seed mass range $10^3$--$10^6\,\Msun$ is actually affected by spectral distortions. Since $M_{\text{\tiny H}}\propto k^{-2}$ and $W_\mu$ falls exponentially, the FIRAS bound on the peak amplitude degrades very rapidly with decreasing mass. For $M_{\text{\tiny PBH}}\lesssim10^{4}\,\Msun$, i.e. $k_\star\gtrsim10^{5}\,{\rm Mpc}^{-1}$, the bound is weaker than $\mathcal{A}_{\rm pk}\lesssim5\cdot10^{-2}$ and is therefore irrelevant: Gaussian perturbations alone already produce a sizeable abundance and no non-Gaussianity is needed. Repeating the estimate of Sec.~\ref{sec:thr} for a Gaussian field saturating the FIRAS bound, we find that the abundance required by Eq.~\eqref{eq:SMBH} can be reached only up to $M_{\text{\tiny PBH}}\simeq{\rm few}\times10^{3}\,\Msun$; above that mass the required amplitude falls so deep in the Gaussian exponential tail that $f_{\rm PBH}$ drops by tens of orders of magnitude within a factor of a few in mass. This is consistent with the findings of Ref.~\cite{Nakama:2017xvq}. At the opposite end, for $M_{\text{\tiny PBH}}\gtrsim{\rm few}\times10^{6}\,\Msun$ ($k_\star\lesssim6\cdot10^{3}\,{\rm Mpc}^{-1}$) the bound tightens to $\mathcal{A}_{\rm pk}\lesssim4\cdot10^{-5}$, and the required suppression of the spectrum can only be achieved with values of $|\beta|$ substantially larger than those shown in Fig.~\ref{fig:QuadAbu}. The window in which the mechanism discussed here is both necessary and effective is therefore $10^{4}\lesssim M_{\text{\tiny PBH}}/\Msun\lesssim10^{6}$.

These configurations with large negative $\beta$ and relatively small $\epsilon$ can be realized in the case of an axion-like curvaton (Sec.~\ref{sec:cosine_V}). In particular, for the benchmark configurations corresponding to a steep potential (small $f$) shown in Fig.~\ref{fig:cosine}, we obtain $\beta\simeq -52,\,-154$, with $\epsilon$ acting as a regulator of the field instability around the hilltop.
By contrast, in the standard free curvaton scenario, the parameter $\epsilon$ is constrained by Eq.~\eqref{eq:fNLgNL2}. Therefore, a large positive $\beta$ is always accompanied by a relatively large $\epsilon$, e.g., for $r_{\rm dec}=10^{-2}$ we have $\beta \simeq 150$ and $\epsilon\simeq 12$. This results in an enhanced peak in the power spectrum, as shown in Fig.~\ref{fig:spectra}. Consequently, improving upon the results of~\cite{Hooper:2023nnl,Byrnes:2024vjt, Pritchard:2025yda}, we find that, beyond the computation of $f_{\rm PBH}$ in the presence of NGs, the possibility of evading the $\mu$-distortion constraints should also take into account the modifications of the physical power spectrum induced by NGs.
The enhancement of the peak amplitude shown in Fig.~\ref{fig:spectra} for large NGs makes it extremely challenging to evade the $\mu$-distortion constraints within the standard quadratic curvaton scenario.

Moreover, the PBH mass is set by the peak scale $k_\star$ of the spectrum of $\delta\phi$ through Eqs.~\eqref{Eq:MassPBH} and~\eqref{eq:MHtok}, and is therefore independent of $\bar\phi_\star$: heavier seeds require a smaller $k_\star$, not a smaller VEV. Note here that the Gaussian power spectrum $\calP_{\zeta_{\rm G}}$ (e.g. the ansatz~\eqref{eq:PS_LN}) is linearly related to the power spectrum of perturbations $\calP_{\delta\phi}$ via Eq.~\eqref{def:zeta_G}. The background value enters instead through the normalisation of the Gaussian variable, $\sigma_0\propto r_{\rm dec}\,\sigma_\phi/\bar\phi_\star$, and, via $r_{\rm dec}$, through the shape parameters $\beta$ and $\epsilon$ of the map~\eqref{eq:template}. The amplitude $\mathcal{A}=10^{-2}$ adopted in Fig.~\ref{fig:QuadAbu} requires $\sigma_\phi/\bar\phi_\star=\mathcal{O}(1)$, i.e.\ a background comparable to the size of the curvaton fluctuations at the peak scale rather than exponentially close to zero; this is precisely the regime in which the distinction between $\bar\phi_\star$ and $\langle\phi_\star\rangle$ discussed in Sec.~\ref{sec:background} becomes relevant. In the free curvaton, a large $f_{\rm NL}$ requires $r_{\rm dec}\ll1$ and hence a small background field, but Eq.~\eqref{eq:fNLgNL2} then forces a large $\epsilon$, which enhances rather than suppresses the peak of the physical spectrum. In the axion-like curvaton, instead, the parameter driving $\beta$ to large negative values is the steepness of the potential, i.e.\ the decay constant $f$ (see Fig.~\ref{fig:cosine}), and the only requirement on $\bar\phi_\star$ is that it lies within the cusp region of $N_{\rm dec}(\phi_\star)$, whose width is regulated by $\epsilon$. Since the abundance required to seed SMBHs, Eq.~\eqref{eq:SMBH}, is many orders of magnitude below the saturation values reached in Fig.~\ref{fig:QuadAbu}, the scenario does not sit at a fine-tuned point in parameter space\footnote{The level of fine-tuning required to achieve the abundance in PBH models has been recently discussed in Refs.~\cite{Cole:2023wyx,Iovino:2025tcv,Profumo:2026dek,Aljazaeri:2026tji}}.

In full agreement with Ref.~\cite{Veermae:2026yzz}, we find that the inclusion of large NGs shifts the position of the main peak by an $\mathcal{O}(2)$ factor with respect to the Gaussian case. Since the main peak at $k_*=3\cdot10^{4}$ Mpc$^{-1}$, through Eq.~\eqref{Eq:MassPBH}, gives rise to a population of PBHs peaked around $\mathcal{O}(10^{5})\msun$, the scenario presented in Fig.~\ref{fig:QuadAbu}, is also relevant for the interpretation of the \emph{Little Red Dots} (LRDs) recently detected by JWST~\cite{Matthee:2023utn,2023ApJ...959...39H,Greene:2024phl, 2023ApJ...954L...4K, Akins:2025iqo,Umeda:2025bha}, which are believed to host over-massive black holes at high redshift. In Ref.~\cite{DeLuca:2025nao}, it was argued that the direct formation of the BHs hosted in LRDs is excluded by $\mu$-distortion constraints, and that LRDs can only arise from lighter PBHs grown through hierarchical mergers or gas accretion. In our setup, instead, the strong positive NGs generated by the axion-like curvaton dynamics allow heavy PBHs to form directly via gravitational collapse of curvature perturbations compatible with FIRAS constraints, providing a potential primordial channel for the LRDs~\footnote{Several astrophysical interpretations of the LRDs have been suggested, ranging from (mildly) super-Eddington accretion onto the central BH~\cite{Pacucci:2024tws,Madau:2024fdv,Lambrides:2024ugh,Liu:2025lrd,Inayoshi:2025lsf,Maiolino:2024uon} to hierarchical growth within nuclear star clusters~\cite{Kritos:2025aqo}. See Ref.~\cite{Inayoshi:2025crit} for a recent critical assessment of these scenarios. Additional primordial black holes interpretations of the LRDs include strongly clustered scenarios~\cite{Zhang:2025tgm,Zhang:2026vjk} and collapse from catalysed dark phase transition~\cite{Guo:2026cuv}.}.

As a conclusive remarks, it is useful to clarify how these results compare with the robust $\mu$-distortion bounds on supermassive PBHs derived in Refs.~\cite{Nakama:2017xvq,Byrnes:2024vjt}. At fixed $\epsilon$, the large-$|\beta|$ regime does correspond to large $|f_{\rm NL}|$, as dictated by Eq.~\eqref{eq:fNLgNL2}, but it is \emph{not} the same as the large-$f_{\rm NL}$ limit of an expansion of $F(\zeta_{\rm G})$ truncated at the quadratic order. Refs.~\cite{Nakama:2017xvq,Byrnes:2024vjt} consider convex, tail-enhancing maps, either the polynomial family $\zeta\propto\zeta_{\rm G}^n$ or, equivalently, a stretched-exponential PDF $P(\zeta)\propto\exp[-(|\zeta|/\sqrt{2}\tilde\sigma)^p]$ with $p<2$, for which the tail of $P(\zeta)$ is heavier than Gaussian. The map \eqref{eq:template} is instead concave and saturating, and, as discussed below Eq.~\eqref{eq:PDFNG}, its large-$\zeta$ tail is more suppressed than in the Gaussian case. What makes PBH formation efficient here is not the tail of $P(\zeta)$ but that of the compaction: from Eq.~\eqref{eq:CCgau} the compaction is controlled by $\mathcal{C}_{\rm G}F'(\zeta_{\rm G})$, and $F'$ is enhanced up to $(1+\epsilon^2)/(2\epsilon)$ around the cusp at $\beta\zeta_{\rm G}=-1$, so that ${\rm P}(\mathcal{C})$ develops the heavy tail~\eqref{eq:appPC}. This is why we evaluate the abundance from threshold statistics of the compaction rather than from a Press--Schechter integral over $P(\zeta)$, as done in Refs.~\cite{Hooper:2023nnl,Nakama:2017xvq,Byrnes:2024vjt}. Where the two analyses overlap we find agreement: in the Gaussian case the FIRAS bound excludes PBH seeds above a few $\times10^{3}\,\Msun$, and the standard quadratic curvaton is unable to evade the constraint.

\section{Conclusions}
\label{sec:conc}

In this work, we have studied the curvaton scenario beyond the standard quadratic potential, deriving the full local non-Gaussian map $\zeta({\bf x}) = F(\zeta_{\rm G}({\bf x}))$ for a class of self-interacting potentials in which the onset of the oscillatory phase is itself a field-dependent quantity. We provided a simple, systematic way to numerically compute the non-Gaussian mapping $ F(\zeta_{\rm G})$ within the $\delta N$ formalism for any curvaton potential.

We showed how self-interactions affect the onset of oscillations and modify the effective equation of state once the curvaton enters the adiabatic regime. In particular, self-interactions can cause causally disconnected patches of the Universe to begin oscillating at slightly different times. This introduces a new source of non-linearity absent in the quadratic case and leaves a characteristic imprint on the non-Gaussian statistics of $\zeta$. To capture this effect systematically, we introduced the abbreviated action $W(\rho)$ as an adiabatic invariant, which allows one to bypass the non-adiabatic oscillatory phase by introducing the instantaneous thawing approximation and matching the frozen regime directly to the subsequent adiabatic fluid description.

We compared the numerical results for $\zeta = F(\zeta_{\rm G})$ with analytical and semi-analytical estimates obtained in the instantaneous thawing and instantaneous decay approximations for several benchmark potentials — quadratic, monomial, quartic, and cosine. Our results showed that self-interactions can either suppress or enhance non-Gaussianity depending on the shape and parameters of the potential and $r_{\rm dec}$.
As a companion result, we provided a new, simple logarithmic ansatz that accurately approximates the mapping $F(\zeta_{\rm G})$ and reproduces the non-Gaussian mapping in the quadratic and axion-like curvaton scenarios. In particular, this ansatz can be interpreted as a regularised version of the mapping commonly encountered for non-Gaussianity in ultra-slow roll models of inflation.

In addition, we studied the non-perturbative aspects of primordial non-Gaussianity. We computed the non-Gaussian curvature power spectrum non-perturbatively and demonstrated that the truncated perturbative series fails to reproduce the correct UV tail of the spectrum even at a relatively high order of the perturbative expansion, i.e., $N\sim 50$. Moreover, we showed how, in the free curvaton scenario, already mild non-Gaussianity can shift and enhance the peak of the power spectrum of $\zeta$ when compared to the power spectrum of auxiliary $\zeta_{\rm G}$. These findings indicate that it is often necessary to go beyond the standard perturbative techniques in curvaton like scenarios.

As an application, we considered heavy PBH seeds for supermassive black holes with masses around $\mathcal{O}(10^5)\,M_\odot$, demonstrating that the strong positive non-Gaussianity arising from the curvaton dynamics makes it possible to obtain a non-negligible PBH abundance even for power spectrum amplitudes as small as $\mathcal{A}_{\rm pk} \sim 10^{-3}$, which remain consistent with the COBE/FIRAS $\mu$-distortion constraint. We showed that, in addition to determining the abundance of primordial supermassive black hole seeds in the presence of primordial non-Gaussianities, it is essential to incorporate the modifications to the physical power spectrum induced by the non-Gaussian mapping. These modifications can substantially affect the peak amplitude of the spectrum and, consequently, either tighten or relax its compatibility with the constraints from CMB $\mu$-distortions. Because of this, the standard free curvaton scenario is not a viable framework for evading the spectral distortion constraints. Instead, the axion-like curvaton scenario can provide a natural arena for the production of supermassive PBHs. These scenarios offer a potential primordial explanation for the over-massive black holes observed at high redshift by JWST as \emph{Little Red Dots}, completely bypassing the need for extreme accretion or hierarchical merger histories.

Although the formation of PBHs constituted our sole explicit phenomenological application, the implications of our results are considerably more general. Any observable that is sensitive to the high-$\zeta$ tail of the probability distribution is affected in an analogous manner and can be treated consistently within the same non-perturbative framework. A compelling example, left as future work, is the scalar-induced gravitational wave (SIGW) background sourced at second order in perturbation theory~\cite{Acquaviva:2002ud, Mollerach:2003nq, Ananda:2006af, Baumann:2007zm,Iovino:2025xkq}: the non-Gaussian corrections to the power spectrum computed here will directly feed into the SIGW spectrum, potentially leaving a distinctive imprint detectable by current and future GW observatories.

\section*{Acknowledgements}
The authors thank G. Franciolini for participating in the early discussions of this work. 
The work of S.A., G.P. and H.V. is supported by the Estonian Research Council grants PSG869, KOHTO34, TARISTU24-TK3, TARISTU24-TK10, and the Center of Excellence program TK202. A.J.I. is funded by Tamkeen under the research grant to NYUAD ADHPG-AD457.

\appendix
\section*{Appendix}

\section{Curvaton evolution with the abbreviated action}
\label{app:potentials}

\subsection{Monomial potentials}
\label{app:mono}

We can consider the evolution of the scalar field when a single term of a generic polynomial potential dominates, i.e., $V = \lambda |\phi|^n$, with $\lambda$ being the self-coupling constant.
In the oscillatory regime, for each cycle, the energy density can be approximated as $\rho_\phi \simeq \lambda \Phi^n$, where $\Phi$ is the amplitude of the scalar field oscillation. Since the potentials considered here are symmetric, the positive and negative amplitudes are the same, and therefore we can restrict ourselves to $\Phi > 0$ and drop the absolute value.
Given this and the definition in Eq.~\eqref{eq:EoS_W}, the abbreviated action reads
\be
    W(\rho_\phi) \simeq \sqrt{8 \lambda} \int_0^{\Phi} \td \phi \sqrt{\Phi^n-\phi^n} 
    = \frac{4 \sqrt{2}}{2+n} \sqrt{\lambda} \Phi^{\frac{2+n}{2}}\,,
\ee
or in terms of the energy density~\eqref{eq:Wevol}, which we repeat here for convenience
\be\label{eq:Wevol_app}
    W(\rho_\phi) \simeq \frac{4 \sqrt{2}}{2+n} \lambda^{-1/n} \rho_\phi^{\frac{n+2}{2n}} \,.
\ee
For monomial potentials, combining this result with Eq.~\eqref{eq:Wconserved} allows us to recover the known scaling $\rho_\phi\propto a^{-6n/(n+2)}$~\cite{Turner:1983he}.

The half-period of an oscillation can be computed by solving the integral
\be
    \tau 
    = \partial_\rho W(\rho) 
    = 2 \int_0^\Phi \frac{\td \phi}{\sqrt{2(\rho - \lambda \phi^n)}} 
    = \sqrt{\frac{2}{\rho}} \left(\frac{\rho}{\lambda}\right)^{1/n}\int_0^{x_{\rm max}} \frac{\td x}{\sqrt{1-x^n}}\,,
\ee 
where we performed the change of variable $x^n=(\lambda/\rho) \phi^n$. Since $x_{\rm max}=1$, the integral simply reduces to the Beta function, and the period reads
\be\label{eq:perioOsc}
    T 
    =  2 \tau = \sqrt{8 \pi} \frac{\lambda^{-1/n}}{n} \frac{\Gamma(\frac{1}{n})}{\Gamma(\frac{1}{2}+\frac{1}{n})} \rho_\phi^{\frac{2-n}{2n}}\,.
\ee
For $n=2$ and $\lambda=m^2/2$, this gives the well-known result of the harmonic oscillator $T = (2 \pi)/m$.

In the approximation of a subdominant curvaton up to decay, we obtained $H\simeq \sqrt{\rho_R/3}$ and, therefore, 
\be
    HT
    = \sqrt{\frac{8}{3}\pi} \frac{\lambda^{-1/n}}{n} \frac{\Gamma(\frac{1}{n})}{\Gamma(\frac{1}{2}+\frac{1}{n})} \rho_\phi^{1/n} \sqrt{\frac{\rho_R}{\rho_\phi}}\,.
\ee
In the limit of a rapidly oscillating scalar field, the last expression should satisfy the condition $HT \ll 1$. For each oscillation, the damping term can be neglected in the equations of motion and $\rho, \Phi \simeq {\rm const.}$.

This condition becomes very simple in the case of a generic polynomial potential. Given that a single term of the potential dominates, $V(\phi)\simeq \lambda \left |\phi\right|^n$, at the beginning of the oscillating regime, we have that the initial period is determined by Eq.~\eqref{eq:perioOsc}.
Given that we can determine the Hubble rate at this time as
\be \label{eq:Hosc_app}
    H_{\rm ad} 
    = A(n)\, n\, \sqrt{\frac{\pi}{2}}\, \lambda^{\frac{1}{n}}\,
    \frac{\Gamma\!\left(\tfrac{1}{2} + \tfrac{1}{n}\right)}{\Gamma\!\left(\tfrac{1}{n}\right)}\,
    \rho_{\phi\star}^{\frac{n-2}{2n}}\,,
\ee
where $A(n)$ is an $\mathcal{O}(1)$ constant that gives a correction dependent on the order of the specific self-interactions. For monomial potentials with even $n$, $A(n) = \sqrt{(n-1)/3}$ is an excellent approximation. In Fig.\,\ref{fig:density}, we show the comparison between our prescription and the numerical solution for various orders of self-interactions.

\subsection{The quartic potential \\and comparison with existing literature}
\label{app:quartic}

Here we report some analytic expressions used in Sec.\,\ref{sec:quartic_V} and compare the results with existing literature.

As described in Sec.~\ref{sec:insta_thawing}, during adiabatic oscillations, the scaling of the reduced action can be analytically determined by $W = W_\star \left( H/H_{\rm ad}\right)^{3/2}$. In analogy with the quadratic case, we can fix $H_{\rm ad}$ with the condition $H_{\rm ad}T=2 \pi$, such that
\be \label{eq:Hadquartic}
    H_{\rm ad}
    =
    \left.\frac{\pi}{\partial_{\rho_\phi} W(\rho_\phi)}\right|_{\Phi=\phi_\star}
    =
    \left.\frac{\pi m}{2}\frac{\sqrt{1+s(\Phi)}}{K_s(\Phi)}\right|_{\Phi=\phi_\star} .
\ee

\begin{figure*}[t!]
\centering
\begin{minipage}{.5\textwidth}
  \centering
  \includegraphics[width=\linewidth]{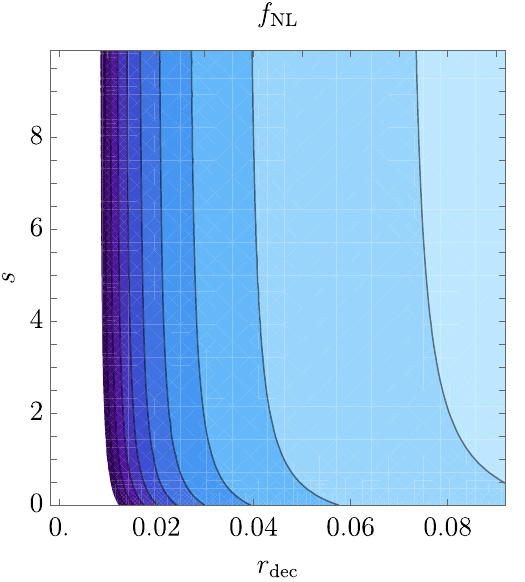}
\end{minipage}\hfill
\begin{minipage}{.5\textwidth} 
  \centering
  \includegraphics[width=\linewidth]{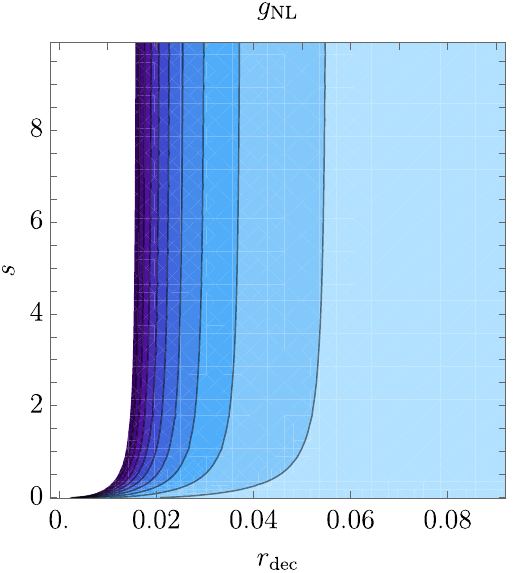} 
\end{minipage}
\caption{Left panel: $f_{\rm NL}$ plotted against the strength of self-interactions $s$ and $r_{\rm dec}$. Contours run from 10 (lightest) to 100 (darker) with a spacing of 10. Right panel: Same but for $g_{\rm NL}$. Contours run from $-200$ (lightest) to $-2000$ (darker) with a spacing of 200.}
     \label{fig:fNLgNL}
\end{figure*}

Consequently, from Eq.~\eqref{eq:Gauxiliary}, we find that the function $G(s)$ takes the following analytical form
\be \label{eq:G(s)}
G(s) =
\frac{
-3(1+s)\,E_s^{2}
+6(1+s)(1+2s),E_s K_s
+3(-1+s)(1+2s)^2\,K_s^{2}
}{
2(1+s)(1+2s)\,K_s\bigl[-E_s+(1+2s)K_s\bigr]
}\,,
\ee
where all quantities are evaluated at $\Phi=\bar{\phi}_\star$. From this point onwards, and in Fig.~\ref{fig:fNLgNL}, we suppress the explicit dependence of $s$ on $\Phi$, since $s$ will always denote the quantity evaluated at the initial background field.

For a direct comparison with the existing literature, it is useful to derive the non-linearity parameters. For this, we expand $F(\zeta_{\rm G})$ to a given order. 
In particular, defining $\Phi = g(\phi_\star) = \phi_\star^{G(s)}$ as the initial value of the envelope of curvaton oscillations during the quadratic regime, we can check that our formalism is consistent with Refs.~\cite{Lyth:2005fi,Sasaki:2006kq} for $f_{\rm NL}$ and $g_{\rm NL}$.

Given this definition, the first two non-linearity parameters can be expressed as
\bea \label{eq:fNLgNL}
    f_{\rm NL}
&   = \frac{5}{3}\left(
    -\frac{r_{\rm dec}}{2}
    -1
    +\frac{3}{4r_{\rm dec}}\,(1+F_2)
    \right)
\\[6pt]
    g_{\rm NL}
&   = \frac{25}{54}\left(
    3r_{\rm dec}^2 + 10r_{\rm dec}
    +\frac{1}{2}(1-9F_2)
    -\frac{9}{r_{\rm dec}}\,(1+F_2)
    +\frac{9}{4r_{\rm dec}^{2}}\,(G_{2}+3F_2)
    \right)
\\
&   \quad \ \textrm{with} \quad
    F_2(s) 
    = \frac{g\, g''}{g'^2} 
    \quad \textrm{and} \quad 
    G_2(s) =
    \frac{g^2 g'''}{g'^3}\,,
\eea
where the complete expressions for $F_2(s)$ and $G_2(s)$ in terms of the self-interaction strength are
\bea
    D 
&   = 3\Bigl[
    (1+s)E_s^{2}
    -2(1+s)(1+2s)\,E_s K_s
    +(1+3s-4s^{3})K_s^{2}
    \Bigr]\, 
\\
    F_2(s) 
&   = \frac{1}{D}\Bigl[3(1+s)\,E_s^{2}
    -10(1+s)(1+2s)\,E_s K_s
    +(7+s)(1+2s)^2\,K_s^{2}\Bigr]\,,
\\
    G_2(s) 
&   = \frac{1}{D^2}\Bigl[ 9(1+s)^2 E_s^{4}
    -72(1+s)^2(1+2s)\,E_s^{3}K_s +2(1+s)(1+2s)^2(97+79s)\,E_s^{2}K_s^{2}
\\
&   \qquad 
    -16(1+s)(1+2s)^3(13+4s)\,E_s K_s^{3}
    +(7+s)(1+2s)^4(11+5s)\,K_s^{4}\Bigr]\,.
\eea
In Fig.\,\ref{fig:fNLgNL}, we show the two non-linearity parameters from Eq.~\eqref{eq:fNLgNL} over a wide range of $s$ and $r_{\rm dec}$. 

In Fig.~\ref{fig:quartic}, we compare our results with the analytic expressions for the quartic case derived in Refs.~\cite{Enqvist:2009ww, Enqvist:2009zf}. In particular, following the procedure outlined in~\cite{Enqvist:2009zf}, one can show that in the $s \gtrsim 1$ limit, the envelope of the oscillations during the quadratic regime is given by
\be 
    g(\phi_\star) 
    = \phi_\star \frac{b \sqrt{\pi}\, e^{-\pi c /8}2^{-3/4}}{\left|\Gamma\left(\frac{3}{4}+i\frac{c}{4}\right) \right|} \,, \qquad 
    c(\phi_\star) 
    = b^2 \sqrt{s} 
    = \frac{b^2 \sqrt{2 \lambda}\, \phi_\star}{m}\,,
\ee
where $b = \pi (2K(-1))^{-1} \simeq 1.1981$.

Given this result, we can apply the same formalism used in our analysis by solving Eq.~\eqref{eq:MasterX} with $e^{3 \zeta_\phi} = (g/\bar{g})^2$~\cite{Sasaki:2006kq}. 
In the left panel of Fig.~\ref{fig:quartic}, we show the solution of $F(\phi_\star)$ rescaled to be zero at its minimum. The solution is in good agreement with the numerical ones and correctly reproduces the behaviour, aside from an expected small deviation for $s \ll 1$.

Comparing our results in Fig.~\ref{fig:fNLgNL} with those of Refs.~\cite{Enqvist:2009ww, Byrnes:2011gh}, we find that our analytic result does not reproduce the non-monotonic behaviour of $f_{\rm NL}$ and $g_{\rm NL}$ around $s \sim 1$ reported in those works, although the magnitudes in other regimes are in agreement. This is consistent with what is observed in Fig.~\ref{fig:quartic}, where for $s = 0.2$ (green line) our analytical prediction deviates from the numerical result.

We stress that this discrepancy originates from the inaccurate behaviour of $H_{\rm ad}$ around $s \sim 1$. Since $H_{\rm ad}$ is fixed through the empirical ansatz $H_{\rm ad} T = 2\pi$, this issue can be resolved by determining $H_{\rm ad}$ either numerically, by solving the non-adiabatic phase of the dynamics, or semi-analytically, as done for the linear case in Sec.~\ref{sec:linear_V}.

\bibliographystyle{JHEP}
\bibliography{main}

@article{Profumo:2026dek,
    author = "Profumo, Stefano",
    title = "{Primordial black hole dark matter: A quantitative parameter sensitivity comparison across formation mechanisms and particle candidates}",
    eprint = "2606.12775",
    archivePrefix = "arXiv",
    primaryClass = "hep-ph",
    doi = "10.1103/nk1q-5k51",
    journal = "Phys. Rev. D",
    volume = "114",
    number = "6",
    pages = "063025",
    year = "2026"
}

@article{Aljazaeri:2026tji,
    author = "Aljazaeri, Amirah and Byrnes, Christian T.",
    title = "{Fine-tuning in mixed Dark Matter models with Primordial Black Hole relics}",
    eprint = "2608.10977",
    archivePrefix = "arXiv",
    primaryClass = "astro-ph.CO",
    month = "8",
    year = "2026"
}

@article{DESI:2024mwx,
    author = "Adame, A. G. and others",
    collaboration = "DESI",
    title = "{DESI 2024 VI: cosmological constraints from the measurements of baryon acoustic oscillations}",
    eprint = "2404.03002",
    archivePrefix = "arXiv",
    primaryClass = "astro-ph.CO",
    reportNumber = "FERMILAB-PUB-24-0154-PPD",
    doi = "10.1088/1475-7516/2025/02/021",
    journal = "JCAP",
    volume = "02",
    pages = "021",
    year = "2025"
}

@article{Iovino:2025tcv,
    author = "Iovino, A. J. and Riotto, A.",
    title = "{Are primordial black holes truly fine-tuned?}",
    eprint = "2512.19668",
    archivePrefix = "arXiv",
    primaryClass = "astro-ph.CO",
    doi = "10.1088/1475-7516/2026/07/047",
    journal = "JCAP",
    volume = "07",
    pages = "047",
    year = "2026"
}

@article{Chabanier:2019eai,
    author = "Chabanier, Sol{\`e}ne and Millea, Marius and Palanque-Delabrouille, Nathalie",
    title = "{Matter power spectrum: from Ly$\alpha$ forest to CMB scales}",
    eprint = "1905.08103",
    archivePrefix = "arXiv",
    primaryClass = "astro-ph.CO",
    reportNumber = "Volume: 489, pages = 2247-2253",
    doi = "10.1093/mnras/stz2310",
    journal = "Mon. Not. Roy. Astron. Soc.",
    volume = "489",
    number = "2",
    pages = "2247--2253",
    year = "2019"
}

@article{Unal:2020mts,
    author = {{\"U}nal, Caner and Kovetz, Ely D. and Patil, Subodh P.},
    title = "{Multimessenger probes of inflationary fluctuations and primordial black holes}",
    eprint = "2008.11184",
    archivePrefix = "arXiv",
    primaryClass = "astro-ph.CO",
    doi = "10.1103/PhysRevD.103.063519",
    journal = "Phys. Rev. D",
    volume = "103",
    number = "6",
    pages = "063519",
    year = "2021"
}

@article{Hong:2026rcl,
    author = "Hong, Wencong and Pi, Shi and Wang, Ao and Zhang, Zhenyu",
    title = "{Constraining the Primordial Black Hole Abundance with Space-Based Detectors}",
    eprint = "2601.05069",
    archivePrefix = "arXiv",
    primaryClass = "astro-ph.CO",
    month = "1",
    year = "2026"
}

@article{Garcia-Bellido:2017aan,
    author = "Garcia-Bellido, Juan and Peloso, Marco and Unal, Caner",
    title = "{Gravitational Wave signatures of inflationary models from Primordial Black Hole Dark Matter}",
    eprint = "1707.02441",
    archivePrefix = "arXiv",
    primaryClass = "astro-ph.CO",
    reportNumber = "IFT-UAM-CSIC-17-056, UMN-TH-3630-17, IFT--UAM-CSIC-17-056",
    doi = "10.1088/1475-7516/2017/09/013",
    journal = "JCAP",
    volume = "09",
    pages = "013",
    year = "2017"
}

@article{Garcia-Bellido:2016dkw,
    author = "Garcia-Bellido, Juan and Peloso, Marco and Unal, Caner",
    title = "{Gravitational waves at interferometer scales and primordial black holes in axion inflation}",
    eprint = "1610.03763",
    archivePrefix = "arXiv",
    primaryClass = "astro-ph.CO",
    reportNumber = "IFT-UAM-CSIC-16-100, UMN-TH-3607-16",
    doi = "10.1088/1475-7516/2016/12/031",
    journal = "JCAP",
    volume = "12",
    pages = "031",
    year = "2016"
}

@article{Nakama:2017xvq,
    author = "Nakama, Tomohiro and Carr, Bernard and Silk, Joseph",
    title = "{Limits on primordial black holes from $\mu$ distortions in cosmic microwave background}",
    eprint = "1710.06945",
    archivePrefix = "arXiv",
    primaryClass = "astro-ph.CO",
    doi = "10.1103/PhysRevD.97.043525",
    journal = "Phys. Rev. D",
    volume = "97",
    number = "4",
    pages = "043525",
    year = "2018"
}

@article{Pritchard:2025pcn,
    author = "Pritchard, Xavier and Starbuck, Matthew and Leung, Wingfung",
    title = "{Beyond Standard Model equation of state and primordial black holes}",
    eprint = "2510.19629",
    archivePrefix = "arXiv",
    primaryClass = "astro-ph.CO",
    doi = "10.1088/1475-7516/2026/02/071",
    journal = "JCAP",
    volume = "02",
    pages = "071",
    year = "2026"
}

@article{Cembranos:2015oya,
    author = "Cembranos, J. A. R. and Maroto, A. L. and N{\'u}{\~n}ez Jare{\~n}o, S. J.",
    title = "{Cosmological perturbations in coherent oscillating scalar field models}",
    eprint = "1509.08819",
    archivePrefix = "arXiv",
    primaryClass = "astro-ph.CO",
    doi = "10.1007/JHEP03(2016)013",
    journal = "JHEP",
    volume = "03",
    pages = "013",
    year = "2016"
}

@article{Boccia:2026azi,
    author = "Boccia, Andrea and Chianese, Marco",
    title = "{Asteroid-mass Primordial Black Holes as Dark Matter from Supersymmetry}",
    eprint = "2604.26005",
    archivePrefix = "arXiv",
    primaryClass = "hep-ph",
    month = "4",
    year = "2026"
}

@article{Palma:2019lpt,
    author = "Palma, Gonzalo A. and Scheihing Hitschfeld, Bruno and Sypsas, Spyros",
    title = "{Non-Gaussian CMB and LSS statistics beyond polyspectra}",
    eprint = "1907.05332",
    archivePrefix = "arXiv",
    primaryClass = "astro-ph.CO",
    doi = "10.1088/1475-7516/2020/02/027",
    journal = "JCAP",
    volume = "02",
    pages = "027",
    year = "2020"
}

@article{Escriva:2023nzn,
    author = "Escriva, Albert and Tada, Yuichiro and Yoo, Chul-Moon",
    title = "{Primordial black holes and induced gravitational waves from a smooth crossover beyond standard model theories}",
    eprint = "2311.17760",
    archivePrefix = "arXiv",
    primaryClass = "astro-ph.CO",
    doi = "10.1103/PhysRevD.110.063521",
    journal = "Phys. Rev. D",
    volume = "110",
    number = "6",
    pages = "063521",
    year = "2024"
}

@article{Escriva:2024ivo,
    author = "Escriv{\`a}, Albert and Inui, Ryoto and Tada, Yuichiro and Yoo, Chul-Moon",
    title = "{LISA forecast on a smooth crossover beyond the standard model through the scalar-induced gravitational waves}",
    eprint = "2404.12591",
    archivePrefix = "arXiv",
    primaryClass = "astro-ph.CO",
    doi = "10.1103/PhysRevD.111.023528",
    journal = "Phys. Rev. D",
    volume = "111",
    number = "2",
    pages = "023528",
    year = "2025"
}

@article{Blas:2026xws,
    author = "Blas, D. and Foster, J. W. and Gouttenoire, Y. and Iovino, A. J. and Musco, I. and Trifinopoulos, S. and Vanvlasselaer, M.",
    title = "{The Dark Side of the Moon: Listening to Scalar-Induced Gravitational Waves}",
    eprint = "2602.12252",
    archivePrefix = "arXiv",
    primaryClass = "astro-ph.CO",
    reportNumber = "CERN-TH-2026-021",
    month = "2",
    year = "2026"
}

@article{Karam:2022nym,
    author = {Karam, Alexandros and Koivunen, Niko and Tomberg, Eemeli and Vaskonen, Ville and Veerm{\"a}e, Hardi},
    title = "{Anatomy of single-field inflationary models for primordial black holes}",
    eprint = "2205.13540",
    archivePrefix = "arXiv",
    primaryClass = "astro-ph.CO",
    doi = "10.1088/1475-7516/2023/03/013",
    journal = "JCAP",
    volume = "03",
    pages = "013",
    year = "2023"
}

@article{You:2023rmn,
    author = "You, Zhi-Qiang and Yi, Zhu and Wu, You",
    title = "{Constraints on primordial curvature power spectrum with pulsar timing arrays}",
    eprint = "2307.04419",
    archivePrefix = "arXiv",
    primaryClass = "gr-qc",
    doi = "10.1088/1475-7516/2023/11/065",
    journal = "JCAP",
    volume = "11",
    pages = "065",
    year = "2023"
}

@article{Iovino:2024tyg,
    author = {Iovino, A. J. and Perna, G. and Riotto, A. and Veerm{\"a}e, H.},
    title = "{Curbing PBHs with PTAs}",
    eprint = "2406.20089",
    archivePrefix = "arXiv",
    primaryClass = "astro-ph.CO",
    doi = "10.1088/1475-7516/2024/10/050",
    journal = "JCAP",
    volume = "10",
    pages = "050",
    year = "2024"
}

@article{Planck:2019kim,
    author = "Akrami, Y. and others",
    collaboration = "Planck",
    title = "{Planck 2018 results. IX. Constraints on primordial non-Gaussianity}",
    eprint = "1905.05697",
    archivePrefix = "arXiv",
    primaryClass = "astro-ph.CO",
    doi = "10.1051/0004-6361/201935891",
    journal = "Astron. Astrophys.",
    volume = "641",
    pages = "A9",
    year = "2020"
}

@article{Traschen:1990sw,
    author = "Traschen, Jennie H. and Brandenberger, Robert H.",
    title = "{Particle Production During Out-of-equilibrium Phase Transitions}",
    reportNumber = "BROWN-HET-731",
    doi = "10.1103/PhysRevD.42.2491",
    journal = "Phys. Rev. D",
    volume = "42",
    pages = "2491--2504",
    year = "1990"
}

@article{Kofman:1994rk,
    author = "Kofman, Lev and Linde, Andrei D. and Starobinsky, Alexei A.",
    title = "{Reheating after inflation}",
    eprint = "hep-th/9405187",
    archivePrefix = "arXiv",
    reportNumber = "UH-IFA-94-35, SU-ITP-94-13, YITP-U-94-15",
    doi = "10.1103/PhysRevLett.73.3195",
    journal = "Phys. Rev. Lett.",
    volume = "73",
    pages = "3195--3198",
    year = "1994"
}

@article{Chambers:2009ki,
    author = "Chambers, Alex and Nurmi, Sami and Rajantie, Arttu",
    title = "{Non-Gaussianity from resonant curvaton decay}",
    eprint = "0909.4535",
    archivePrefix = "arXiv",
    primaryClass = "astro-ph.CO",
    reportNumber = "IMPERIAL-TP-09-AC-01",
    doi = "10.1088/1475-7516/2010/01/012",
    journal = "JCAP",
    volume = "01",
    pages = "012",
    year = "2010"
}

@article{Kofman:1997yn,
    author = "Kofman, Lev and Linde, Andrei D. and Starobinsky, Alexei A.",
    title = "{Towards the theory of reheating after inflation}",
    eprint = "hep-ph/9704452",
    archivePrefix = "arXiv",
    reportNumber = "IFA-97-28, SU-ITP-97-18",
    doi = "10.1103/PhysRevD.56.3258",
    journal = "Phys. Rev. D",
    volume = "56",
    pages = "3258--3295",
    year = "1997"
}

@article{Mukaida:2014wma,
    author = "Mukaida, Kyohei and Nakayama, Kazunori and Takimoto, Masahiro",
    title = "{Suppressed Non-Gaussianity in the Curvaton Model}",
    eprint = "1402.1856",
    archivePrefix = "arXiv",
    primaryClass = "astro-ph.CO",
    reportNumber = "UT-14-03",
    doi = "10.1103/PhysRevD.89.123515",
    journal = "Phys. Rev. D",
    volume = "89",
    number = "12",
    pages = "123515",
    year = "2014"
}

@article{Liu:2023ymk,
    author = "Liu, Lang and Chen, Zu-Cheng and Huang, Qing-Guo",
    title = "{Implications for the non-Gaussianity of curvature perturbation from pulsar timing arrays}",
    eprint = "2307.01102",
    archivePrefix = "arXiv",
    primaryClass = "astro-ph.CO",
    doi = "10.1103/PhysRevD.109.L061301",
    journal = "Phys. Rev. D",
    volume = "109",
    number = "6",
    pages = "L061301",
    year = "2024"
}

@article{Vaskonen:2020lbd,
    author = {Vaskonen, Ville and Veerm{\"a}e, Hardi},
    title = "{Did NANOGrav see a signal from primordial black hole formation?}",
    eprint = "2009.07832",
    archivePrefix = "arXiv",
    primaryClass = "astro-ph.CO",
    doi = "10.1103/PhysRevLett.126.051303",
    journal = "Phys. Rev. Lett.",
    volume = "126",
    number = "5",
    pages = "051303",
    year = "2021"
}

@article{Gouttenoire:2025jxe,
    author = "Gouttenoire, Yann and Trifinopoulos, Sokratis and Vanvlasselaer, Miguel",
    title = "{Implications for Pulsar Timing Arrays of Sub-solar Black Hole Detections: From LVK to Einstein Telescope and Cosmic Explorer}",
    eprint = "2508.19328",
    archivePrefix = "arXiv",
    primaryClass = "astro-ph.CO",
    reportNumber = "CERN-TH-2025-169, MIT-CTP/5904, MITP-25-056",
    month = "8",
    year = "2025"
}

@article{Hajkarim:2019nbx,
    author = {Hajkarim, Fazlollah and Schaffner-Bielich, J{\"u}rgen},
    title = "{Thermal History of the Early Universe and Primordial Gravitational Waves from Induced Scalar Perturbations}",
    eprint = "1910.12357",
    archivePrefix = "arXiv",
    primaryClass = "hep-ph",
    doi = "10.1103/PhysRevD.101.043522",
    journal = "Phys. Rev. D",
    volume = "101",
    number = "4",
    pages = "043522",
    year = "2020"
}

@article{Planck:2018jri,
    author = "Akrami, Y. and others",
    collaboration = "Planck",
    title = "{Planck 2018 results. X. Constraints on inflation}",
    eprint = "1807.06211",
    archivePrefix = "arXiv",
    primaryClass = "astro-ph.CO",
    doi = "10.1051/0004-6361/201833887",
    journal = "Astron. Astrophys.",
    volume = "641",
    pages = "A10",
    year = "2020"
}

@article{Iovino:2024sgs,
    author = "Iovino, A. J. and Matarrese, S. and Perna, G. and Ricciardone, A. and Riotto, A.",
    title = "{How Well Do We Know the Scalar-Induced Gravitational Waves?}",
    eprint = "2412.06764",
    archivePrefix = "arXiv",
    primaryClass = "astro-ph.CO",
    month = "12",
    year = "2024"
}

@article{Zeng:2025cer,
    author = "Zeng, Xiang-Xi and Ning, Zhuan and Cai, Rong-Gen and Wang, Shao-Jiang",
    title = "{Scalar-induced gravitational waves with non-Gaussianity up to all orders}",
    eprint = "2508.10812",
    archivePrefix = "arXiv",
    primaryClass = "astro-ph.CO",
    month = "8",
    year = "2025"
}

@article{Andres-Carcasona:2024wqk,
    author = {Andr\'es-Carcasona, M. and Iovino, A. J. and Vaskonen, V. and Veerm\"ae, H. and Mart\'\i{}nez, M. and Pujol\`as, O. and Mir, Ll. M.},
    title = "{Constraints on primordial black holes from LIGO-Virgo-KAGRA O3 events}",
    eprint = "2405.05732",
    archivePrefix = "arXiv",
    primaryClass = "astro-ph.CO",
    doi = "10.1103/PhysRevD.110.023040",
    journal = "Phys. Rev. D",
    volume = "110",
    number = "2",
    pages = "023040",
    year = "2024"
}

@article{Taoso:2021uvl,
    author = "Taoso, Marco and Urbano, Alfredo",
    title = "{Non-gaussianities for primordial black hole formation}",
    eprint = "2102.03610",
    archivePrefix = "arXiv",
    primaryClass = "astro-ph.CO",
    doi = "10.1088/1475-7516/2021/08/016",
    journal = "JCAP",
    volume = "08",
    pages = "016",
    year = "2021"
}

@article{Lyth:2002my,
    author = "Lyth, David H. and Ungarelli, Carlo and Wands, David",
    title = "{The Primordial density perturbation in the curvaton scenario}",
    eprint = "astro-ph/0208055",
    archivePrefix = "arXiv",
    reportNumber = "PU-ICG-02-15",
    doi = "10.1103/PhysRevD.67.023503",
    journal = "Phys. Rev. D",
    volume = "67",
    pages = "023503",
    year = "2003"
}

@article{Riccardi:2021rlf,
    author = "Riccardi, Flavio and Taoso, Marco and Urbano, Alfredo",
    title = "{Solving peak theory in the presence of local non-gaussianities}",
    eprint = "2102.04084",
    archivePrefix = "arXiv",
    primaryClass = "astro-ph.CO",
    doi = "10.1088/1475-7516/2021/08/060",
    journal = "JCAP",
    volume = "08",
    pages = "060",
    year = "2021"
}

@article{Vaskonen:2019jpv,
    author = {Vaskonen, Ville and Veerm\"ae, Hardi},
    title = "{Lower bound on the primordial black hole merger rate}",
    eprint = "1908.09752",
    archivePrefix = "arXiv",
    primaryClass = "astro-ph.CO",
    reportNumber = "CERN-TH-2019-141, KCL-PH-TH/2019-69",
    doi = "10.1103/PhysRevD.101.043015",
    journal = "Phys. Rev. D",
    volume = "101",
    number = "4",
    pages = "043015",
    year = "2020"
}

@article{Kawasaki:2013xsa,
    author = "Kawasaki, Masahiro and Kitajima, Naoya and Yokoyama, Shuichiro",
    title = "{Gravitational waves from a curvaton model with blue spectrum}",
    eprint = "1305.4464",
    archivePrefix = "arXiv",
    primaryClass = "astro-ph.CO",
    reportNumber = "IPMU-13-0100, ICRR-REPORT-649-2012-38",
    doi = "10.1088/1475-7516/2013/08/042",
    journal = "JCAP",
    volume = "08",
    pages = "042",
    year = "2013"
}

@article{Escriva:2022pnz,
    author = "Escriv\`a, Albert and Tada, Yuichiro and Yokoyama, Shuichiro and Yoo, Chul-Moon",
    title = "{Simulation of primordial black holes with large negative non-Gaussianity}",
    eprint = "2202.01028",
    archivePrefix = "arXiv",
    primaryClass = "astro-ph.CO",
    doi = "10.1088/1475-7516/2022/05/012",
    journal = "JCAP",
    volume = "05",
    number = "05",
    pages = "012",
    year = "2022"
}

@article{Young:2014ana,
    author = "Young, Sam and Byrnes, Christian T. and Sasaki, Misao",
    title = "{Calculating the mass fraction of primordial black holes}",
    eprint = "1405.7023",
    archivePrefix = "arXiv",
    primaryClass = "gr-qc",
    doi = "10.1088/1475-7516/2014/07/045",
    journal = "JCAP",
    volume = "07",
    pages = "045",
    year = "2014"
}

@article{Harada:2015yda,
    author = "Harada, Tomohiro and Yoo, Chul-Moon and Nakama, Tomohiro and Koga, Yasutaka",
    title = "{Cosmological long-wavelength solutions and primordial black hole formation}",
    eprint = "1503.03934",
    archivePrefix = "arXiv",
    primaryClass = "gr-qc",
    reportNumber = "RUP-15-5, RESCEU-4-15",
    doi = "10.1103/PhysRevD.91.084057",
    journal = "Phys. Rev. D",
    volume = "91",
    number = "8",
    pages = "084057",
    year = "2015"
}

@article{Hutsi:2020sol,
    author = {H\"utsi, Gert and Raidal, Martti and Vaskonen, Ville and Veerm\"ae, Hardi},
    title = "{Two populations of LIGO-Virgo black holes}",
    eprint = "2012.02786",
    archivePrefix = "arXiv",
    primaryClass = "astro-ph.CO",
    doi = "10.1088/1475-7516/2021/03/068",
    journal = "JCAP",
    volume = "03",
    pages = "068",
    year = "2021"
}

@article{Raidal:2018bbj,
    author = {Raidal, Martti and Spethmann, Christian and Vaskonen, Ville and Veerm\"ae, Hardi},
    title = "{Formation and Evolution of Primordial Black Hole Binaries in the Early Universe}",
    eprint = "1812.01930",
    archivePrefix = "arXiv",
    primaryClass = "astro-ph.CO",
    reportNumber = "KCL-PH-TH/2018-70, CERN-TH-2018-266",
    doi = "10.1088/1475-7516/2019/02/018",
    journal = "JCAP",
    volume = "02",
    pages = "018",
    year = "2019"
}

@article{Ando:2018nge,
    author = "Ando, Kenta and Kawasaki, Masahiro and Nakatsuka, Hiromasa",
    title = "{Formation of primordial black holes in an axionlike curvaton model}",
    eprint = "1805.07757",
    archivePrefix = "arXiv",
    primaryClass = "astro-ph.CO",
    reportNumber = "IPMU18-0087",
    doi = "10.1103/PhysRevD.98.083508",
    journal = "Phys. Rev. D",
    volume = "98",
    number = "8",
    pages = "083508",
    year = "2018"
}

@article{Gow:2022jfb,
    author = "Gow, Andrew D. and Assadullahi, Hooshyar and Jackson, Joseph H. P. and Koyama, Kazuya and Vennin, Vincent and Wands, David",
    title = "{Non-perturbative non-Gaussianity and primordial black holes}",
    eprint = "2211.08348",
    archivePrefix = "arXiv",
    primaryClass = "astro-ph.CO",
    doi = "10.1209/0295-5075/acd417",
    journal = "EPL",
    volume = "142",
    number = "4",
    pages = "49001",
    year = "2023"
}

@article{Meng:2022ixx,
    author = "Meng, De-Shuang and Yuan, Chen and Huang, Qing-guo",
    title = "{One-loop correction to the enhanced curvature perturbation with local-type non-Gaussianity for the formation of primordial black holes}",
    eprint = "2207.07668",
    archivePrefix = "arXiv",
    primaryClass = "astro-ph.CO",
    doi = "10.1103/PhysRevD.106.063508",
    journal = "Phys. Rev. D",
    volume = "106",
    number = "6",
    pages = "063508",
    year = "2022"
}

@article{Sasaki:1995aw,
    author = "Sasaki, Misao and Stewart, Ewan D.",
    title = "{A General analytic formula for the spectral index of the density perturbations produced during inflation}",
    eprint = "astro-ph/9507001",
    archivePrefix = "arXiv",
    reportNumber = "LANCS-TH-9504, OU-TAP-22",
    doi = "10.1143/PTP.95.71",
    journal = "Prog. Theor. Phys.",
    volume = "95",
    pages = "71--78",
    year = "1996"
}

@article{Starobinsky:1982ee,
    author = "Starobinsky, Alexei A.",
    title = "{Dynamics of Phase Transition in the New Inflationary Universe Scenario and Generation of Perturbations}",
    doi = "10.1016/0370-2693(82)90541-X",
    journal = "Phys. Lett. B",
    volume = "117",
    pages = "175--178",
    year = "1982"
}

@article{Marchesano:2014mla,
    author = "Marchesano, Fernando and Shiu, Gary and Uranga, Angel M.",
    title = "{F-term Axion Monodromy Inflation}",
    eprint = "1404.3040",
    archivePrefix = "arXiv",
    primaryClass = "hep-th",
    reportNumber = "IFT-UAM-CSIC-14-032, MAD-TH-04-01",
    doi = "10.1007/JHEP09(2014)184",
    journal = "JHEP",
    volume = "09",
    pages = "184",
    year = "2014"
}

@article{McAllister:2008hb,
    author = "McAllister, Liam and Silverstein, Eva and Westphal, Alexander",
    title = "{Gravity Waves and Linear Inflation from Axion Monodromy}",
    eprint = "0808.0706",
    archivePrefix = "arXiv",
    primaryClass = "hep-th",
    reportNumber = "SLAC-PUB-13357, SU-ITP-08-15",
    doi = "10.1103/PhysRevD.82.046003",
    journal = "Phys. Rev. D",
    volume = "82",
    pages = "046003",
    year = "2010"
}

@article{Berg:2009tg,
    author = "Berg, Marcus and Pajer, Enrico and Sjors, Stefan",
    title = "{Dante's Inferno}",
    eprint = "0912.1341",
    archivePrefix = "arXiv",
    primaryClass = "hep-th",
    doi = "10.1103/PhysRevD.81.103535",
    journal = "Phys. Rev. D",
    volume = "81",
    pages = "103535",
    year = "2010"
}

@article{Salopek:1990jq,
    author = "Salopek, D. S. and Bond, J. R.",
    title = "{Nonlinear Evolution of Long Wavelength Metric Fluctuations in Inflationary Models}",
    reportNumber = "FERMILAB-PUB-90-131-A",
    doi = "10.1103/PhysRevD.42.3936",
    journal = "Phys. Rev. D",
    volume = "42",
    pages = "3936--3962",
    year = "1990"
}

@article{Turner:1983he,
    author = "Turner, Michael S.",
    title = "{Coherent Scalar Field Oscillations in an Expanding Universe}",
    reportNumber = "EFI-83-29-CHICAGO",
    doi = "10.1103/PhysRevD.28.1243",
    journal = "Phys. Rev. D",
    volume = "28",
    pages = "1243",
    year = "1983"
}

@article{Tomberg:2021bll,
    author = {Tomberg, Eemeli and Veerm{\"a}e, Hardi},
    title = "{Tachyonic preheating in plateau inflation}",
    eprint = "2108.10767",
    archivePrefix = "arXiv",
    primaryClass = "astro-ph.CO",
    doi = "10.1088/1475-7516/2021/12/035",
    journal = "JCAP",
    volume = "12",
    number = "12",
    pages = "035",
    year = "2021"
}

@article{Lozanov:2019jxc,
    author = "Lozanov, Kaloian D.",
    title = "{Lectures on Reheating after Inflation}",
    eprint = "1907.04402",
    archivePrefix = "arXiv",
    primaryClass = "astro-ph.CO",
    month = "7",
    year = "2019"
}

@article{Mollerach:1989hu,
    author = "Mollerach, Silvia",
    title = "{Isocurvature Baryon Perturbations and Inflation}",
    reportNumber = "SISSA-123-89-EP",
    doi = "10.1103/PhysRevD.42.313",
    journal = "Phys. Rev. D",
    volume = "42",
    pages = "313--325",
    year = "1990"
}

@article{Bardeen:1983qw,
    author = "Bardeen, James M. and Steinhardt, Paul J. and Turner, Michael S.",
    title = "{Spontaneous Creation of Almost Scale - Free Density Perturbations in an Inflationary Universe}",
    reportNumber = "UPR-0202T, EFI-83-13-CHICAGO",
    doi = "10.1103/PhysRevD.28.679",
    journal = "Phys. Rev. D",
    volume = "28",
    pages = "679",
    year = "1983"
}

@article{Lyth:2004gb,
    author = "Lyth, David H. and Malik, Karim A. and Sasaki, Misao",
    title = "{A General proof of the conservation of the curvature perturbation}",
    eprint = "astro-ph/0411220",
    archivePrefix = "arXiv",
    reportNumber = "YITP-04-67",
    doi = "10.1088/1475-7516/2005/05/004",
    journal = "JCAP",
    volume = "05",
    pages = "004",
    year = "2005"
}

@article{Kawasaki:2012wr,
    author = "Kawasaki, Masahiro and Kitajima, Naoya and Yanagida, Tsutomu T.",
    title = "{Primordial black hole formation from an axionlike curvaton model}",
    eprint = "1207.2550",
    archivePrefix = "arXiv",
    primaryClass = "hep-ph",
    reportNumber = "ICRR-REPORT-616-2012-5, IPMU-12-0116",
    doi = "10.1103/PhysRevD.87.063519",
    journal = "Phys. Rev. D",
    volume = "87",
    number = "6",
    pages = "063519",
    year = "2013"
}

@article{Enqvist:2001zp,
    author = "Enqvist, Kari and Sloth, Martin S.",
    title = "{Adiabatic CMB perturbations in pre - big bang string cosmology}",
    eprint = "hep-ph/0109214",
    archivePrefix = "arXiv",
    reportNumber = "HIP-2001-51-TH",
    doi = "10.1016/S0550-3213(02)00043-3",
    journal = "Nucl. Phys. B",
    volume = "626",
    pages = "395--409",
    year = "2002"
}

@article{Miller:2024fpo,
    author = "Miller, Andrew L. and Aggarwal, Nancy and Clesse, S{\'e}bastien and De Lillo, Federico and Sachdev, Surabhi and Astone, Pia and Palomba, Cristiano and Piccinni, Ornella J. and Pierini, Lorenzo",
    title = "{Gravitational Wave Constraints on Planetary-Mass Primordial Black Holes Using LIGO O3a Data}",
    eprint = "2402.19468",
    archivePrefix = "arXiv",
    primaryClass = "gr-qc",
    doi = "10.1103/PhysRevLett.133.111401",
    journal = "Phys. Rev. Lett.",
    volume = "133",
    number = "11",
    pages = "111401",
    year = "2024"
}

@article{Jedamzik:1996mr,
    author = "Jedamzik, Karsten",
    title = "{Primordial black hole formation during the QCD epoch}",
    eprint = "astro-ph/9605152",
    archivePrefix = "arXiv",
    doi = "10.1103/PhysRevD.55.R5871",
    journal = "Phys. Rev. D",
    volume = "55",
    pages = "5871--5875",
    year = "1997"
}

@article{Ashoorioon:2008pj,
    author = "Ashoorioon, Amjad and Freese, Katherine and Liu, James T.",
    title = "{Slow nucleation rates in Chain Inflation with QCD Axions or Monodromy}",
    eprint = "0810.0228",
    archivePrefix = "arXiv",
    primaryClass = "hep-ph",
    doi = "10.1103/PhysRevD.79.067302",
    journal = "Phys. Rev. D",
    volume = "79",
    pages = "067302",
    year = "2009"
}

@article{Brandenberger:2008kn,
    author = "Brandenberger, Robert H. and Knauf, Anke and Lorenz, Larissa C.",
    title = "{Reheating in a Brane Monodromy Inflation Model}",
    eprint = "0808.3936",
    archivePrefix = "arXiv",
    primaryClass = "hep-th",
    doi = "10.1088/1126-6708/2008/10/110",
    journal = "JHEP",
    volume = "10",
    pages = "110",
    year = "2008"
}

@article{Silverstein:2008sg,
    author = "Silverstein, Eva and Westphal, Alexander",
    title = "{Monodromy in the CMB: Gravity Waves and String Inflation}",
    eprint = "0803.3085",
    archivePrefix = "arXiv",
    primaryClass = "hep-th",
    reportNumber = "SU-ITP-08-07, SLAC-PUB-13183",
    doi = "10.1103/PhysRevD.78.106003",
    journal = "Phys. Rev. D",
    volume = "78",
    pages = "106003",
    year = "2008"
}

@article{Hannestad:2009yx,
    author = "Hannestad, Steen and Haugbolle, Troels and Jarnhus, Philip R. and Sloth, Martin S.",
    title = "{Non-Gaussianity from Axion Monodromy Inflation}",
    eprint = "0912.3527",
    archivePrefix = "arXiv",
    primaryClass = "hep-ph",
    reportNumber = "CERN-PH-TH-2009-253",
    doi = "10.1088/1475-7516/2010/06/001",
    journal = "JCAP",
    volume = "06",
    pages = "001",
    year = "2010"
}

@article{Serpico:2020ehh,
    author = "Serpico, Pasquale D. and Poulin, Vivian and Inman, Derek and Kohri, Kazunori",
    title = "{Cosmic microwave background bounds on primordial black holes including dark matter halo accretion}",
    eprint = "2002.10771",
    archivePrefix = "arXiv",
    primaryClass = "astro-ph.CO",
    reportNumber = "LAPTH-005/20, KEK-Cosmo-248, KEK-TH-2198, IPMU20-0021",
    doi = "10.1103/PhysRevResearch.2.023204",
    journal = "Phys. Rev. Res.",
    volume = "2",
    number = "2",
    pages = "023204",
    year = "2020"
}

@article{Inomata:2023drn,
    author = "Inomata, Keisuke and Kawasaki, Masahiro and Mukaida, Kyohei and Yanagida, Tsutomu T.",
    title = "{Axion curvaton model for the gravitational waves observed by pulsar timing arrays}",
    eprint = "2309.11398",
    archivePrefix = "arXiv",
    primaryClass = "astro-ph.CO",
    reportNumber = "KEK-TH-2554, KEK-Cosmo-0324",
    doi = "10.1103/PhysRevD.109.043508",
    journal = "Phys. Rev. D",
    volume = "109",
    number = "4",
    pages = "043508",
    year = "2024"
}

@article{Ando:2017veq,
    author = "Ando, Kenta and Inomata, Keisuke and Kawasaki, Masahiro and Mukaida, Kyohei and Yanagida, Tsutomu T.",
    title = "{Primordial black holes for the LIGO events in the axionlike curvaton model}",
    eprint = "1711.08956",
    archivePrefix = "arXiv",
    primaryClass = "astro-ph.CO",
    reportNumber = "IPMU-17-0165, DESY-17-209",
    doi = "10.1103/PhysRevD.97.123512",
    journal = "Phys. Rev. D",
    volume = "97",
    number = "12",
    pages = "123512",
    year = "2018"
}

@article{Kawasaki:2011pd,
    author = "Kawasaki, Masahiro and Kobayashi, Takeshi and Takahashi, Fuminobu",
    title = "{Non-Gaussianity from Curvatons Revisited}",
    eprint = "1107.6011",
    archivePrefix = "arXiv",
    primaryClass = "astro-ph.CO",
    reportNumber = "RESCEU-28-11, ICRR-REPORT-592-2011-9, IPMU11-0122, TU-885",
    doi = "10.1103/PhysRevD.84.123506",
    journal = "Phys. Rev. D",
    volume = "84",
    pages = "123506",
    year = "2011"
}

@article{DeLuca:2021wjr,
    author = "De Luca, V. and Franciolini, G. and Pani, P. and Riotto, A.",
    title = "{Bayesian Evidence for Both Astrophysical and Primordial Black Holes: Mapping the GWTC-2 Catalog to Third-Generation Detectors}",
    eprint = "2102.03809",
    archivePrefix = "arXiv",
    primaryClass = "astro-ph.CO",
    doi = "10.1088/1475-7516/2021/05/003",
    journal = "JCAP",
    volume = "05",
    pages = "003",
    year = "2021"
}

@article{Nakama:2016gzw,
    author = "Nakama, Tomohiro and Silk, Joseph and Kamionkowski, Marc",
    title = "{Stochastic gravitational waves associated with the formation of primordial black holes}",
    eprint = "1612.06264",
    archivePrefix = "arXiv",
    primaryClass = "astro-ph.CO",
    doi = "10.1103/PhysRevD.95.043511",
    journal = "Phys. Rev. D",
    volume = "95",
    number = "4",
    pages = "043511",
    year = "2017"
}

@article{Hooper:2023nnl,
    author = "Hooper, Dan and Ireland, Aurora and Krnjaic, Gordan and Stebbins, Albert",
    title = "{Supermassive primordial black holes from inflation}",
    eprint = "2308.00756",
    archivePrefix = "arXiv",
    primaryClass = "astro-ph.CO",
    reportNumber = "FERMILAB-PUB-23-390-T",
    doi = "10.1088/1475-7516/2024/04/021",
    journal = "JCAP",
    volume = "04",
    pages = "021",
    year = "2024"
}

@article{Wong:2020yig,
    author = "Wong, Kaze W. K. and Franciolini, Gabriele and De Luca, Valerio and Baibhav, Vishal and Berti, Emanuele and Pani, Paolo and Riotto, Antonio",
    title = "{Constraining the primordial black hole scenario with Bayesian inference and machine learning: the GWTC-2 gravitational wave catalog}",
    eprint = "2011.01865",
    archivePrefix = "arXiv",
    primaryClass = "gr-qc",
    doi = "10.1103/PhysRevD.103.023026",
    journal = "Phys. Rev. D",
    volume = "103",
    number = "2",
    pages = "023026",
    year = "2021"
}

@article{Kushwaha:2026msi,
    author = "Kushwaha, Ashu and Suyama, Teruaki",
    title = "{Revisiting Constraints on Primordial Curvature Power Spectrum from PBH Abundances}",
    eprint = "2603.23025",
    archivePrefix = "arXiv",
    primaryClass = "astro-ph.CO",
    month = "3",
    year = "2026"
}

@article{Inomata:2018epa,
    author = "Inomata, Keisuke and Nakama, Tomohiro",
    title = "{Gravitational waves induced by scalar perturbations as probes of the small-scale primordial spectrum}",
    eprint = "1812.00674",
    archivePrefix = "arXiv",
    primaryClass = "astro-ph.CO",
    reportNumber = "IPMU 18-0200",
    doi = "10.1103/PhysRevD.99.043511",
    journal = "Phys. Rev. D",
    volume = "99",
    number = "4",
    pages = "043511",
    year = "2019"
}

@article{Perna:2026szd,
    author = "Perna, Gabriele and Dom{\`e}nech, Guillem",
    title = "{Probing non-Gaussianity during reheating with SIGW in the LISA band}",
    eprint = "2604.08493",
    archivePrefix = "arXiv",
    primaryClass = "astro-ph.CO",
    month = "4",
    year = "2026"
}

@article{Domenech:2019quo,
    author = "Dom{\`e}nech, Guillem",
    title = "{Induced gravitational waves in a general cosmological background}",
    eprint = "1912.05583",
    archivePrefix = "arXiv",
    primaryClass = "gr-qc",
    doi = "10.1142/S0218271820500285",
    journal = "Int. J. Mod. Phys. D",
    volume = "29",
    number = "03",
    pages = "2050028",
    year = "2020"
}

@article{Ferrante:2022mui,
    author = "Ferrante, Giacomo and Franciolini, Gabriele and Iovino, Junior., Antonio and Urbano, Alfredo",
    title = "{Primordial non-Gaussianity up to all orders: Theoretical aspects and implications for primordial black hole models}",
    eprint = "2211.01728",
    archivePrefix = "arXiv",
    primaryClass = "astro-ph.CO",
    doi = "10.1103/PhysRevD.107.043520",
    journal = "Phys. Rev. D",
    volume = "107",
    number = "4",
    pages = "043520",
    year = "2023"
}

@article{Raidal:2017mfl,
    author = {Raidal, Martti and Vaskonen, Ville and Veerm\"ae, Hardi},
    title = "{Gravitational Waves from Primordial Black Hole Mergers}",
    eprint = "1707.01480",
    archivePrefix = "arXiv",
    primaryClass = "astro-ph.CO",
    doi = "10.1088/1475-7516/2017/09/037",
    journal = "JCAP",
    volume = "09",
    pages = "037",
    year = "2017"
}

@article{Yoo:2018kvb,
    author = "Yoo, Chul-Moon and Harada, Tomohiro and Garriga, Jaume and Kohri, Kazunori",
    title = "{Primordial black hole abundance from random Gaussian curvature perturbations and a local density threshold}",
    eprint = "1805.03946",
    archivePrefix = "arXiv",
    primaryClass = "astro-ph.CO",
    reportNumber = "RUP-18-15, KEK-Cosmo-225, KEK-TH-2052",
    doi = "10.1093/ptep/pty120",
    journal = "PTEP",
    volume = "2018",
    number = "12",
    pages = "123E01",
    year = "2018"
}

@article{Carr:2020gox,
    author = "Carr, Bernard and Kohri, Kazunori and Sendouda, Yuuiti and Yokoyama, Jun'ichi",
    title = "{Constraints on primordial black holes}",
    eprint = "2002.12778",
    archivePrefix = "arXiv",
    primaryClass = "astro-ph.CO",
    reportNumber = "RESCEU-03/20; KEK-Cosmo-249; KEK-TH-2199; IPMU20-0024",
    doi = "10.1088/1361-6633/ac1e31",
    journal = "Rept. Prog. Phys.",
    volume = "84",
    number = "11",
    pages = "116902",
    year = "2021"
}

@article{Dimopoulos:2003az,
    author = "Dimopoulos, Konstantinos and Lyth, D. H. and Notari, A. and Riotto, A.",
    title = "{The Curvaton as a pseudoNambu-Goldstone boson}",
    eprint = "hep-ph/0304050",
    archivePrefix = "arXiv",
    reportNumber = "DFPD-TH-03-12",
    doi = "10.1088/1126-6708/2003/07/053",
    journal = "JHEP",
    volume = "07",
    pages = "053",
    year = "2003"
}

@article{Dimopoulos:2003ss,
    author = "Dimopoulos, Konstantinos and Lazarides, George and Lyth, David and Ruiz de Austri, Roberto",
    title = "{Curvaton dynamics}",
    eprint = "hep-ph/0308015",
    archivePrefix = "arXiv",
    reportNumber = "UT-STPD-3-03, LANCS-16-07-03",
    doi = "10.1103/PhysRevD.68.123515",
    journal = "Phys. Rev. D",
    volume = "68",
    pages = "123515",
    year = "2003"
}

@article{Bartolo:2002vf,
    author = "Bartolo, N and Liddle, Andrew R",
    title = "{The Simplest curvaton model}",
    eprint = "astro-ph/0203076",
    archivePrefix = "arXiv",
    doi = "10.1103/PhysRevD.65.121301",
    journal = "Phys. Rev. D",
    volume = "65",
    pages = "121301",
    year = "2002"
}

@article{Gordon:2002gv,
    author = "Gordon, Christopher and Lewis, Antony",
    title = "{Observational constraints on the curvaton model of inflation}",
    eprint = "astro-ph/0212248",
    archivePrefix = "arXiv",
    doi = "10.1103/PhysRevD.67.123513",
    journal = "Phys. Rev. D",
    volume = "67",
    pages = "123513",
    year = "2003"
}

@article{Bartolo:2003jx,
    author = "Bartolo, N. and Matarrese, S. and Riotto, A.",
    title = "{On nonGaussianity in the curvaton scenario}",
    eprint = "hep-ph/0309033",
    archivePrefix = "arXiv",
    reportNumber = "DFPD-A-03-33",
    doi = "10.1103/PhysRevD.69.043503",
    journal = "Phys. Rev. D",
    volume = "69",
    pages = "043503",
    year = "2004"
}

@article{Wands:2002bn,
    author = "Wands, David and Bartolo, Nicola and Matarrese, Sabino and Riotto, Antonio",
    title = "{An Observational test of two-field inflation}",
    eprint = "astro-ph/0205253",
    archivePrefix = "arXiv",
    reportNumber = "PU-ICG-02-06",
    doi = "10.1103/PhysRevD.66.043520",
    journal = "Phys. Rev. D",
    volume = "66",
    pages = "043520",
    year = "2002"
}

@article{Musco:2020jjb,
    author = "Musco, Ilia and De Luca, Valerio and Franciolini, Gabriele and Riotto, Antonio",
    title = "{Threshold for primordial black holes. II. A simple analytic prescription}",
    eprint = "2011.03014",
    archivePrefix = "arXiv",
    primaryClass = "astro-ph.CO",
    doi = "10.1103/PhysRevD.103.063538",
    journal = "Phys. Rev. D",
    volume = "103",
    number = "6",
    pages = "063538",
    year = "2021"
}

@article{Cole:2023wyx,
    author = "Cole, Philippa S. and Gow, Andrew D. and Byrnes, Christian T. and Patil, Subodh P.",
    title = "{Primordial black holes from single-field inflation: a fine-tuning audit}",
    eprint = "2304.01997",
    archivePrefix = "arXiv",
    primaryClass = "astro-ph.CO",
    doi = "10.1088/1475-7516/2023/08/031",
    journal = "JCAP",
    volume = "08",
    pages = "031",
    year = "2023"
}

@article{Lyth:2001nq,
    author = "Lyth, David H. and Wands, David",
    title = "{Generating the curvature perturbation without an inflaton}",
    eprint = "hep-ph/0110002",
    archivePrefix = "arXiv",
    reportNumber = "PU-RCG-01-33",
    doi = "10.1016/S0370-2693(01)01366-1",
    journal = "Phys. Lett. B",
    volume = "524",
    pages = "5--14",
    year = "2002"
}

@article{Enqvist:2009ww,
    author = "Enqvist, Kari and Nurmi, Sami and Taanila, Olli and Takahashi, Tomo",
    title = "{Non-Gaussian Fingerprints of Self-Interacting Curvaton}",
    eprint = "0912.4657",
    archivePrefix = "arXiv",
    primaryClass = "astro-ph.CO",
    doi = "10.1088/1475-7516/2010/04/009",
    journal = "JCAP",
    volume = "04",
    pages = "009",
    year = "2010"
}

@article{Enqvist:2010dt,
    author = "Enqvist, Kari",
    editor = "Sasaki, Misao and Tanaka, Takahiro",
    title = "{The self-interacting curvaton}",
    eprint = "1012.1711",
    archivePrefix = "arXiv",
    primaryClass = "astro-ph.CO",
    reportNumber = "HIP-2010-37",
    doi = "10.1143/PTPS.190.62",
    journal = "Prog. Theor. Phys. Suppl.",
    volume = "190",
    pages = "62--74",
    year = "2011"
}

@article{Byrnes:2018txb,
    author = "Byrnes, Christian T. and Cole, Philippa S. and Patil, Subodh P.",
    title = "{Steepest growth of the power spectrum and primordial black holes}",
    eprint = "1811.11158",
    archivePrefix = "arXiv",
    primaryClass = "astro-ph.CO",
    doi = "10.1088/1475-7516/2019/06/028",
    journal = "JCAP",
    volume = "06",
    pages = "028",
    year = "2019"
}

@article{Ianniccari:2024bkh,
    author = "Ianniccari, A. and Iovino, A. J. and Kehagias, A. and Perrone, D. and Riotto, A.",
    title = "{The Primordial Black Hole Abundance: The Broader, the Better}",
    eprint = "2402.11033",
    archivePrefix = "arXiv",
    primaryClass = "astro-ph.CO",
    month = "2",
    year = "2024"
}

@article{Kawasaki:2019mbl,
    author = "Kawasaki, Masahiro and Nakatsuka, Hiromasa",
    title = "{Effect of nonlinearity between density and curvature perturbations on the primordial black hole formation}",
    eprint = "1903.02994",
    archivePrefix = "arXiv",
    primaryClass = "astro-ph.CO",
    reportNumber = "IPMU19-0029",
    doi = "10.1103/PhysRevD.99.123501",
    journal = "Phys. Rev. D",
    volume = "99",
    number = "12",
    pages = "123501",
    year = "2019"
}

@article{Carr:1975qj,
    author = "Carr, Bernard J.",
    title = "{The Primordial black hole mass spectrum}",
    doi = "10.1086/153853",
    journal = "Astrophys. J.",
    volume = "201",
    pages = "1--19",
    year = "1975"
}

@article{Pi:2022ysn,
    author = "Pi, Shi and Sasaki, Misao",
    title = "{Logarithmic Duality of the Curvature Perturbation}",
    eprint = "2211.13932",
    archivePrefix = "arXiv",
    primaryClass = "astro-ph.CO",
    reportNumber = "IPMU22-0060, YITP-22-144",
    doi = "10.1103/PhysRevLett.131.011002",
    journal = "Phys. Rev. Lett.",
    volume = "131",
    number = "1",
    pages = "011002",
    year = "2023"
}

@article{Musco:2008hv,
    author = "Musco, Ilia and Miller, John C. and Polnarev, Alexander G.",
    title = "{Primordial black hole formation in the radiative era: Investigation of the critical nature of the collapse}",
    eprint = "0811.1452",
    archivePrefix = "arXiv",
    primaryClass = "gr-qc",
    doi = "10.1088/0264-9381/26/23/235001",
    journal = "Class. Quant. Grav.",
    volume = "26",
    pages = "235001",
    year = "2009"
}

@article{Liu:2020zlr,
    author = "Liu, Lei-Hua and Liang, Bin and Zhou, Ya-Chen and Liu, Xiao-Dan and Xu, Wu-Long and Li, Ai-Chen",
    title = "{Revised $f_{NL}$ parameter in a curvaton scenario}",
    eprint = "2007.08278",
    archivePrefix = "arXiv",
    primaryClass = "astro-ph.CO",
    doi = "10.1103/PhysRevD.103.063515",
    journal = "Phys. Rev. D",
    volume = "103",
    number = "6",
    pages = "063515",
    year = "2021"
}

@article{Tomberg:2023kli,
    author = "Tomberg, Eemeli",
    title = "{Stochastic constant-roll inflation and primordial black holes}",
    eprint = "2304.10903",
    archivePrefix = "arXiv",
    primaryClass = "astro-ph.CO",
    doi = "10.1103/PhysRevD.108.043502",
    journal = "Phys. Rev. D",
    volume = "108",
    number = "4",
    pages = "043502",
    year = "2023"
}

@article{Ferrante:2023bgz,
    author = "Ferrante, Giacomo and Franciolini, Gabriele and Iovino, Junior., Antonio and Urbano, Alfredo",
    title = "{Primordial black holes in the curvaton model: possible connections to pulsar timing arrays and dark matter}",
    eprint = "2305.13382",
    archivePrefix = "arXiv",
    primaryClass = "astro-ph.CO",
    doi = "10.1088/1475-7516/2023/06/057",
    journal = "JCAP",
    volume = "06",
    pages = "057",
    year = "2023"
}

@article{Byrnes:2012yx,
    author = "Byrnes, Christian T. and Copeland, Edmund J. and Green, Anne M.",
    title = "{Primordial black holes as a tool for constraining non-Gaussianity}",
    eprint = "1206.4188",
    archivePrefix = "arXiv",
    primaryClass = "astro-ph.CO",
    reportNumber = "CERN-PH-TH-2012-167",
    doi = "10.1103/PhysRevD.86.043512",
    journal = "Phys. Rev. D",
    volume = "86",
    pages = "043512",
    year = "2012"
}

@article{Carr:1974nx,
    author = "Carr, Bernard J. and Hawking, S. W.",
    title = "{Black holes in the early Universe}",
    doi = "10.1093/mnras/168.2.399",
    journal = "Mon. Not. Roy. Astron. Soc.",
    volume = "168",
    pages = "399--415",
    year = "1974"
}

@article{Shibata:1999zs,
    author = "Shibata, Masaru and Sasaki, Misao",
    title = "{Black hole formation in the Friedmann universe: Formulation and computation in numerical relativity}",
    eprint = "gr-qc/9905064",
    archivePrefix = "arXiv",
    reportNumber = "OU-TAP-93",
    doi = "10.1103/PhysRevD.60.084002",
    journal = "Phys. Rev. D",
    volume = "60",
    pages = "084002",
    year = "1999"
}

@article{Musco:2023dak,
    author = "Musco, Ilia and Jedamzik, Karsten and Young, Sam",
    title = "{Primordial black hole formation during the QCD phase transition: Threshold, mass distribution, and abundance}",
    eprint = "2303.07980",
    archivePrefix = "arXiv",
    primaryClass = "astro-ph.CO",
    doi = "10.1103/PhysRevD.109.083506",
    journal = "Phys. Rev. D",
    volume = "109",
    number = "8",
    pages = "083506",
    year = "2024"
}

@article{Franciolini:2022tfm,
    author = "Franciolini, Gabriele and Musco, Ilia and Pani, Paolo and Urbano, Alfredo",
    title = "{From inflation to black hole mergers and back again: Gravitational-wave data-driven constraints on inflationary scenarios with a first-principle model of primordial black holes across the QCD epoch}",
    eprint = "2209.05959",
    archivePrefix = "arXiv",
    primaryClass = "astro-ph.CO",
    doi = "10.1103/PhysRevD.106.123526",
    journal = "Phys. Rev. D",
    volume = "106",
    number = "12",
    pages = "123526",
    year = "2022"
}

@article{Bird:2016dcv,
    author = {Bird, Simeon and Cholis, Ilias and Mu\~noz, Julian B. and Ali-Ha\"\i{}moud, Yacine and Kamionkowski, Marc and Kovetz, Ely D. and Raccanelli, Alvise and Riess, Adam G.},
    title = "{Did LIGO detect dark matter?}",
    eprint = "1603.00464",
    archivePrefix = "arXiv",
    primaryClass = "astro-ph.CO",
    doi = "10.1103/PhysRevLett.116.201301",
    journal = "Phys. Rev. Lett.",
    volume = "116",
    number = "20",
    pages = "201301",
    year = "2016"
}

@article{Young:2022phe,
    author = "Young, Sam",
    title = "{Peaks and primordial black holes: the~effect of non-Gaussianity}",
    eprint = "2201.13345",
    archivePrefix = "arXiv",
    primaryClass = "astro-ph.CO",
    doi = "10.1088/1475-7516/2022/05/037",
    journal = "JCAP",
    volume = "05",
    number = "05",
    pages = "037",
    year = "2022"
}

@article{Ianniccari:2024ltb,
    author = "Ianniccari, Andrea and Iovino, Antonio J. and Kehagias, Alex and Perrone, Davide and Riotto, Antonio",
    title = "{The Primordial Black Hole Formation -- Null Geodesic Correspondence}",
    eprint = "2404.02801",
    archivePrefix = "arXiv",
    primaryClass = "astro-ph.CO",
    month = "4",
    year = "2024"
}

@article{Raatikainen:2025gpd,
    author = "Raatikainen, Sami and Rasanen, Syksy and Tomberg, Eemeli",
    title = "{Effect of stochastic kicks on primordial black hole abundance and mass via the compaction function}",
    eprint = "2510.09303",
    archivePrefix = "arXiv",
    primaryClass = "astro-ph.CO",
    reportNumber = "HIP-2025-28/TH",
    doi = "10.1088/1475-7516/2026/03/063",
    journal = "JCAP",
    volume = "03",
    pages = "063",
    year = "2026"
}

@article{Byrnes:2011gh,
    author = "Byrnes, Christian T. and Enqvist, Kari and Nurmi, Sami and Takahashi, Tomo",
    title = "{Strongly scale-dependent polyspectra from curvaton self-interactions}",
    eprint = "1108.2708",
    archivePrefix = "arXiv",
    primaryClass = "astro-ph.CO",
    reportNumber = "BI-TP-2011-24, HIP-2011-22-TH, NORDITA-2011-66",
    doi = "10.1088/1475-7516/2011/11/011",
    journal = "JCAP",
    volume = "11",
    pages = "011",
    year = "2011"
}

@article{Enqvist:2013qba,
    author = "Enqvist, Kari and Lerner, Rose N. and Rusak, Stanislav",
    title = "{Reheating dynamics affects non-perturbative decay of spectator fields}",
    eprint = "1308.3321",
    archivePrefix = "arXiv",
    primaryClass = "astro-ph.CO",
    reportNumber = "HIP-2013-11-TH",
    doi = "10.1088/1475-7516/2013/11/034",
    journal = "JCAP",
    volume = "11",
    pages = "034",
    year = "2013"
}

@article{Enqvist:2012tc,
    author = "Enqvist, Kari and Figueroa, Daniel G. and Lerner, Rose N.",
    title = "{Curvaton Decay by Resonant Production of the Standard Model Higgs}",
    eprint = "1211.5028",
    archivePrefix = "arXiv",
    primaryClass = "astro-ph.CO",
    reportNumber = "HIP-2012-19-TH",
    doi = "10.1088/1475-7516/2013/01/040",
    journal = "JCAP",
    volume = "01",
    pages = "040",
    year = "2013"
}

@article{Enqvist:2009zf,
    author = "Enqvist, Kari and Nurmi, Sami and Rigopoulos, Gerasimos and Taanila, Olli and Takahashi, Tomo",
    title = "{The Subdominant Curvaton}",
    eprint = "0906.3126",
    archivePrefix = "arXiv",
    primaryClass = "astro-ph.CO",
    doi = "10.1088/1475-7516/2009/11/003",
    journal = "JCAP",
    volume = "11",
    pages = "003",
    year = "2009"
}

@article{Enqvist:2003mr,
    author = "Enqvist, Kari and Jokinen, Asko and Kasuya, Shinta and Mazumdar, Anupam",
    title = "{MSSM flat direction as a curvaton}",
    eprint = "hep-ph/0303165",
    archivePrefix = "arXiv",
    reportNumber = "HIP-2003-13-TH",
    doi = "10.1103/PhysRevD.68.103507",
    journal = "Phys. Rev. D",
    volume = "68",
    pages = "103507",
    year = "2003"
}

@article{Enqvist:2004gz,
    author = "Enqvist, Kari",
    title = "{Curvatons in the minimally supersymmetric standard model}",
    eprint = "hep-ph/0403273",
    archivePrefix = "arXiv",
    reportNumber = "HIP-2004-10-TH",
    doi = "10.1142/S0217732304013970",
    journal = "Mod. Phys. Lett. A",
    volume = "19",
    pages = "1421--1434",
    year = "2004"
}

@article{Enqvist:2005pg,
    author = "Enqvist, Kari and Nurmi, Sami",
    title = "{Non-gaussianity in curvaton models with nearly quadratic potential}",
    eprint = "astro-ph/0508573",
    archivePrefix = "arXiv",
    reportNumber = "HIP-2005-33-TH",
    doi = "10.1088/1475-7516/2005/10/013",
    journal = "JCAP",
    volume = "10",
    pages = "013",
    year = "2005"
}

@article{Allahverdi:2006dr,
    author = "Allahverdi, Rouzbeh and Enqvist, Kari and Jokinen, Asko and Mazumdar, Anupam",
    title = "{Identifying the curvaton within MSSM}",
    eprint = "hep-ph/0603255",
    archivePrefix = "arXiv",
    doi = "10.1088/1475-7516/2006/10/007",
    journal = "JCAP",
    volume = "10",
    pages = "007",
    year = "2006"
}

@article{Enqvist:2008gk,
    author = "Enqvist, Kari and Takahashi, Tomo",
    title = "{Signatures of Non-Gaussianity in the Curvaton Model}",
    eprint = "0807.3069",
    archivePrefix = "arXiv",
    primaryClass = "astro-ph",
    doi = "10.1088/1475-7516/2008/09/012",
    journal = "JCAP",
    volume = "09",
    pages = "012",
    year = "2008"
}

@article{Enqvist:2008be,
    author = "Enqvist, K. and Nurmi, S. and Rigopoulos, G. I.",
    title = "{Parametric Decay of the Curvaton}",
    eprint = "0807.0382",
    archivePrefix = "arXiv",
    primaryClass = "astro-ph",
    reportNumber = "HIP-2008-23-TH",
    doi = "10.1088/1475-7516/2008/10/013",
    journal = "JCAP",
    volume = "10",
    pages = "013",
    year = "2008"
}

@article{Enqvist:2009eq,
    author = "Enqvist, Kari and Takahashi, Tomo",
    title = "{Effect of Background Evolution on the Curvaton Non-Gaussianity}",
    eprint = "0909.5362",
    archivePrefix = "arXiv",
    primaryClass = "astro-ph.CO",
    doi = "10.1088/1475-7516/2009/12/001",
    journal = "JCAP",
    volume = "12",
    pages = "001",
    year = "2009"
}

@article{Byrnes:2010xd,
    author = "Byrnes, Christian T. and Enqvist, Kari and Takahashi, Tomo",
    title = "{Scale-dependence of Non-Gaussianity in the Curvaton Model}",
    eprint = "1007.5148",
    archivePrefix = "arXiv",
    primaryClass = "astro-ph.CO",
    doi = "10.1088/1475-7516/2010/09/026",
    journal = "JCAP",
    volume = "09",
    pages = "026",
    year = "2010"
}

@article{Enqvist:2011jf,
    author = "Enqvist, Kari and Lerner, Rose N. and Taanila, Olli",
    title = "{Curvaton model completed}",
    eprint = "1105.0498",
    archivePrefix = "arXiv",
    primaryClass = "astro-ph.CO",
    reportNumber = "HIP-2011-13-TH",
    doi = "10.1088/1475-7516/2011/12/016",
    journal = "JCAP",
    volume = "12",
    pages = "016",
    year = "2011"
}

@article{DeLuca:2025nao,
    author = "De Luca, Valerio and Del Grosso, Loris and Franciolini, Gabriele and Kritos, Konstantinos and Berti, Emanuele and D'Orazio, Daniel J. and Silk, Joseph",
    title = "{Primordial-Black-Hole-Based Pathways to Little Red Dots}",
    eprint = "2512.19666",
    archivePrefix = "arXiv",
    primaryClass = "astro-ph.CO",
    reportNumber = "CERN-TH-2025-261",
    doi = "10.1103/6y1w-87pd",
    journal = "Phys. Rev. Lett.",
    volume = "136",
    number = "23",
    pages = "231402",
    year = "2026"
}

@article{Moroi:2001ct,
    author = "Moroi, Takeo and Takahashi, Tomo",
    title = "{Effects of cosmological moduli fields on cosmic microwave background}",
    eprint = "hep-ph/0110096",
    archivePrefix = "arXiv",
    reportNumber = "TU-632",
    doi = "10.1016/S0370-2693(01)01295-3",
    journal = "Phys. Lett. B",
    volume = "522",
    pages = "215--221",
    year = "2001",
    note = "[Erratum: Phys.Lett.B 539, 303--303 (2002)]"
}

@article{Perna:2024ehx,
    author = "Perna, Gabriele and Testini, Chiara and Ricciardone, Angelo and Matarrese, Sabino",
    title = "{Fully non-Gaussian Scalar-Induced Gravitational Waves}",
    eprint = "2403.06962",
    archivePrefix = "arXiv",
    primaryClass = "astro-ph.CO",
    month = "3",
    year = "2024"
}

@article{Tomita:1975kj,
    author = "Tomita, Kenji",
    title = "{Evolution of Irregularities in a Chaotic Early Universe}",
    reportNumber = "RRK 75-3",
    doi = "10.1143/PTP.54.730",
    journal = "Prog. Theor. Phys.",
    volume = "54",
    pages = "730",
    year = "1975"
}

@article{Matarrese:1993zf,
    author = "Matarrese, Sabino and Pantano, Ornella and Saez, Diego",
    title = "{General relativistic dynamics of irrotational dust: Cosmological implications}",
    eprint = "astro-ph/9310036",
    archivePrefix = "arXiv",
    reportNumber = "DFPD-93-A-67",
    doi = "10.1103/PhysRevLett.72.320",
    journal = "Phys. Rev. Lett.",
    volume = "72",
    pages = "320--323",
    year = "1994"
}

@article{Acquaviva:2002ud,
    author = "Acquaviva, Viviana and Bartolo, Nicola and Matarrese, Sabino and Riotto, Antonio",
    title = "{Second order cosmological perturbations from inflation}",
    eprint = "astro-ph/0209156",
    archivePrefix = "arXiv",
    reportNumber = "DFPD-A-02-21",
    doi = "10.1016/S0550-3213(03)00550-9",
    journal = "Nucl. Phys. B",
    volume = "667",
    pages = "119--148",
    year = "2003"
}

@article{Mollerach:2003nq,
    author = "Mollerach, Silvia and Harari, Diego and Matarrese, Sabino",
    title = "{CMB polarization from secondary vector and tensor modes}",
    eprint = "astro-ph/0310711",
    archivePrefix = "arXiv",
    doi = "10.1103/PhysRevD.69.063002",
    journal = "Phys. Rev. D",
    volume = "69",
    pages = "063002",
    year = "2004"
}

@article{Ananda:2006af,
    author = "Ananda, Kishore N. and Clarkson, Chris and Wands, David",
    title = "{The Cosmological gravitational wave background from primordial density perturbations}",
    eprint = "gr-qc/0612013",
    archivePrefix = "arXiv",
    doi = "10.1103/PhysRevD.75.123518",
    journal = "Phys. Rev. D",
    volume = "75",
    pages = "123518",
    year = "2007"
}

@article{Baumann:2007zm,
    author = "Baumann, Daniel and Steinhardt, Paul J. and Takahashi, Keitaro and Ichiki, Kiyotomo",
    title = "{Gravitational Wave Spectrum Induced by Primordial Scalar Perturbations}",
    eprint = "hep-th/0703290",
    archivePrefix = "arXiv",
    doi = "10.1103/PhysRevD.76.084019",
    journal = "Phys. Rev. D",
    volume = "76",
    pages = "084019",
    year = "2007"
}

@article{Domenech:2021ztg,
    author = "Dom\`enech, Guillem",
    title = "{Scalar Induced Gravitational Waves Review}",
    eprint = "2109.01398",
    archivePrefix = "arXiv",
    primaryClass = "gr-qc",
    doi = "10.3390/universe7110398",
    journal = "Universe",
    volume = "7",
    number = "11",
    pages = "398",
    year = "2021"
}

@article{Kohri:2018awv,
    author = "Kohri, Kazunori and Terada, Takahiro",
    title = "{Semianalytic calculation of gravitational wave spectrum nonlinearly induced from primordial curvature perturbations}",
    eprint = "1804.08577",
    archivePrefix = "arXiv",
    primaryClass = "gr-qc",
    reportNumber = "KEK-TH-2046, KEK-COSMO-223",
    doi = "10.1103/PhysRevD.97.123532",
    journal = "Phys. Rev. D",
    volume = "97",
    number = "12",
    pages = "123532",
    year = "2018"
}

@article{Adshead:2021hnm,
    author = "Adshead, Peter and Lozanov, Kaloian D. and Weiner, Zachary J.",
    title = "{Non-Gaussianity and the induced gravitational wave background}",
    eprint = "2105.01659",
    archivePrefix = "arXiv",
    primaryClass = "astro-ph.CO",
    doi = "10.1088/1475-7516/2021/10/080",
    journal = "JCAP",
    volume = "10",
    pages = "080",
    year = "2021"
}

@article{Garcia-Saenz:2022tzu,
    author = "Garcia-Saenz, Sebastian and Pinol, Lucas and Renaux-Petel, S\'ebastien and Werth, Denis",
    title = "{No-go theorem for scalar-trispectrum-induced gravitational waves}",
    eprint = "2207.14267",
    archivePrefix = "arXiv",
    primaryClass = "astro-ph.CO",
    doi = "10.1088/1475-7516/2023/03/057",
    journal = "JCAP",
    volume = "03",
    pages = "057",
    year = "2023"
}

@article{Cai:2018dig,
    author = "Cai, Rong-gen and Pi, Shi and Sasaki, Misao",
    title = "{Gravitational Waves Induced by non-Gaussian Scalar Perturbations}",
    eprint = "1810.11000",
    archivePrefix = "arXiv",
    primaryClass = "astro-ph.CO",
    reportNumber = "IPMU18-0172, YITP-18-114",
    doi = "10.1103/PhysRevLett.122.201101",
    journal = "Phys. Rev. Lett.",
    volume = "122",
    number = "20",
    pages = "201101",
    year = "2019"
}

@article{Unal:2018yaa,
    author = "Unal, Caner",
    title = "{Imprints of Primordial Non-Gaussianity on Gravitational Wave Spectrum}",
    eprint = "1811.09151",
    archivePrefix = "arXiv",
    primaryClass = "astro-ph.CO",
    doi = "10.1103/PhysRevD.99.041301",
    journal = "Phys. Rev. D",
    volume = "99",
    number = "4",
    pages = "041301",
    year = "2019"
}

@article{Atal:2021jyo,
    author = "Atal, Vicente and Dom\`enech, Guillem",
    title = "{Probing non-Gaussianities with the high frequency tail of induced gravitational waves}",
    eprint = "2103.01056",
    archivePrefix = "arXiv",
    primaryClass = "astro-ph.CO",
    doi = "10.1088/1475-7516/2021/06/001",
    journal = "JCAP",
    volume = "06",
    pages = "001",
    year = "2021",
    note = "[Erratum: JCAP 10, E01 (2023)]"
}

@article{Yuan:2020iwf,
    author = "Yuan, Chen and Huang, Qing-Guo",
    title = "{Gravitational waves induced by the local-type non-Gaussian curvature perturbations}",
    eprint = "2007.10686",
    archivePrefix = "arXiv",
    primaryClass = "astro-ph.CO",
    doi = "10.1016/j.physletb.2021.136606",
    journal = "Phys. Lett. B",
    volume = "821",
    pages = "136606",
    year = "2021"
}

@article{Abe:2022xur,
    author = "Abe, Katsuya T. and Inui, Ryoto and Tada, Yuichiro and Yokoyama, Shuichiro",
    title = "{Primordial black holes and gravitational waves induced by exponential-tailed perturbations}",
    eprint = "2209.13891",
    archivePrefix = "arXiv",
    primaryClass = "astro-ph.CO",
    doi = "10.1088/1475-7516/2023/05/044",
    journal = "JCAP",
    volume = "05",
    pages = "044",
    year = "2023"
}

@article{Felder:2000hj,
    author = "Felder, Gary N. and Garcia-Bellido, Juan and Greene, Patrick B. and Kofman, Lev and Linde, Andrei D. and Tkachev, Igor",
    title = "{Dynamics of symmetry breaking and tachyonic preheating}",
    eprint = "hep-ph/0012142",
    archivePrefix = "arXiv",
    reportNumber = "CITA-2000-60, SU-ITP-00-35, IFT-UAM-CSIC-00-40, FT-UAM-00-26, CERN-TH-2000-365",
    doi = "10.1103/PhysRevLett.87.011601",
    journal = "Phys. Rev. Lett.",
    volume = "87",
    pages = "011601",
    year = "2001"
}

@article{Felder:2001kt,
    author = "Felder, Gary N. and Kofman, Lev and Linde, Andrei D.",
    title = "{Tachyonic instability and dynamics of spontaneous symmetry breaking}",
    eprint = "hep-th/0106179",
    archivePrefix = "arXiv",
    reportNumber = "CITA-2001-12, SU-ITP-01-26",
    doi = "10.1103/PhysRevD.64.123517",
    journal = "Phys. Rev. D",
    volume = "64",
    pages = "123517",
    year = "2001"
}

@article{Franciolini:2023pbf,
    author = "Franciolini, Gabriele and Iovino, Junior., Antonio and Vaskonen, Ville and Veermae, Hardi",
    title = "{Recent Gravitational Wave Observation by Pulsar Timing Arrays and Primordial Black Holes: The Importance of Non-Gaussianities}",
    eprint = "2306.17149",
    archivePrefix = "arXiv",
    primaryClass = "astro-ph.CO",
    doi = "10.1103/PhysRevLett.131.201401",
    journal = "Phys. Rev. Lett.",
    volume = "131",
    number = "20",
    pages = "201401",
    year = "2023"
}

@article{Planck:2018vyg,
    author = "Aghanim, N. and others",
    collaboration = "Planck",
    title = "{Planck 2018 results. VI. Cosmological parameters}",
    eprint = "1807.06209",
    archivePrefix = "arXiv",
    primaryClass = "astro-ph.CO",
    doi = "10.1051/0004-6361/201833910",
    journal = "Astron. Astrophys.",
    volume = "641",
    pages = "A6",
    year = "2020",
    note = "[Erratum: Astron.Astrophys. 652, C4 (2021)]"
}

@article{Hawking:1971ei,
    author = "Hawking, Stephen",
    title = "{Gravitationally collapsed objects of very low mass}",
    doi = "10.1093/mnras/152.1.75",
    journal = "Mon. Not. Roy. Astron. Soc.",
    volume = "152",
    pages = "75",
    year = "1971"
}

@article{Lyth:2005fi,
    author = "Lyth, David H. and Rodriguez, Yeinzon",
    title = "{The Inflationary prediction for primordial non-Gaussianity}",
    eprint = "astro-ph/0504045",
    archivePrefix = "arXiv",
    doi = "10.1103/PhysRevLett.95.121302",
    journal = "Phys. Rev. Lett.",
    volume = "95",
    pages = "121302",
    year = "2005"
}

@article{Zeldovich:1967lct,
    author = "Zel'dovich, Ya. B. and Novikov, I. D.",
    title = "{The Hypothesis of Cores Retarded during Expansion and the Hot Cosmological Model}",
    journal = "Sov. Astron.",
    volume = "10",
    pages = "602",
    year = "1967"
}

@article{Green:2020jor,
    author = "Green, Anne M. and Kavanagh, Bradley J.",
    title = "{Primordial Black Holes as a dark matter candidate}",
    eprint = "2007.10722",
    archivePrefix = "arXiv",
    primaryClass = "astro-ph.CO",
    doi = "10.1088/1361-6471/abc534",
    journal = "J. Phys. G",
    volume = "48",
    number = "4",
    pages = "043001",
    year = "2021"
}

@article{Kehagias:2019eil,
    author = "Kehagias, Alex and Musco, Ilia and Riotto, Antonio",
    title = "{Non-Gaussian Formation of Primordial Black Holes: Effects on the Threshold}",
    eprint = "1906.07135",
    archivePrefix = "arXiv",
    primaryClass = "astro-ph.CO",
    doi = "10.1088/1475-7516/2019/12/029",
    journal = "JCAP",
    volume = "12",
    pages = "029",
    year = "2019"
}

@article{Cai:2019elf,
    author = "Cai, Rong-Gen and Pi, Shi and Wang, Shao-Jiang and Yang, Xing-Yu",
    title = "{Pulsar Timing Array Constraints on the Induced Gravitational Waves}",
    eprint = "1907.06372",
    archivePrefix = "arXiv",
    primaryClass = "astro-ph.CO",
    doi = "10.1088/1475-7516/2019/10/059",
    journal = "JCAP",
    volume = "10",
    pages = "059",
    year = "2019"
}

@article{Domenech:2021and,
    author = "Dom{\`e}nech, Guillem and Passaglia, Samuel and Renaux-Petel, S{\'e}bastien",
    title = "{Gravitational waves from dark matter isocurvature}",
    eprint = "2112.10163",
    archivePrefix = "arXiv",
    primaryClass = "astro-ph.CO",
    reportNumber = "ET-0466A-21",
    doi = "10.1088/1475-7516/2022/03/023",
    journal = "JCAP",
    volume = "03",
    number = "03",
    pages = "023",
    year = "2022"
}

@article{Yuan:2023ofl,
    author = "Yuan, Chen and Meng, De-Shuang and Huang, Qing-Guo",
    title = "{Full analysis of the scalar-induced gravitational waves for the curvature perturbation with local-type non-Gaussianities}",
    eprint = "2308.07155",
    archivePrefix = "arXiv",
    primaryClass = "astro-ph.CO",
    doi = "10.1088/1475-7516/2023/12/036",
    journal = "JCAP",
    volume = "12",
    pages = "036",
    year = "2023"
}

@article{Cyr:2023pgw,
    author = "Cyr, Bryce and Kite, Thomas and Chluba, Jens and Hill, J. Colin and Jeong, Donghui and Acharya, Sandeep Kumar and Bolliet, Boris and Patil, Subodh P.",
    title = "{Disentangling the primordial nature of stochastic gravitational wave backgrounds with CMB spectral distortions}",
    eprint = "2309.02366",
    archivePrefix = "arXiv",
    primaryClass = "astro-ph.CO",
    doi = "10.1093/mnras/stad3861",
    journal = "Mon. Not. Roy. Astron. Soc.",
    volume = "528",
    number = "1",
    pages = "883--897",
    year = "2024"
}

@article{Andres-Carcasona:2026avd,
    author = {Andr{\'e}s-Carcasona, M. and Iovino, A. J. and Vallejo-Pag{\`e}s, E. and Vaskonen, V. and Veerm{\"a}e, H. and Mart{\'\i}nez, M. and Mir, Ll. M.},
    title = "{Constraints on primordial black holes from the first part of LIGO-Virgo-KAGRA fourth observing run}",
    eprint = "2605.15749",
    archivePrefix = "arXiv",
    primaryClass = "astro-ph.CO",
    month = "5",
    year = "2026"
}

@article{Li:2023xtl,
    author = "Li, Jun-Peng and Wang, Sai and Zhao, Zhi-Chao and Kohri, Kazunori",
    title = "{Complete analysis of the background and anisotropies of scalar-induced gravitational waves: primordial non-Gaussianity f $_{NL}$ and g $_{NL}$ considered}",
    eprint = "2309.07792",
    archivePrefix = "arXiv",
    primaryClass = "astro-ph.CO",
    reportNumber = "KEK-Cosmo-0326, KEK-TH-2556, KEK-QUP-2023-0024, KEK-QUP-2023-0024,
  https://github.com/Zhi-ChaoZhao/sigw{\_}class",
    doi = "10.1088/1475-7516/2024/06/039",
    journal = "JCAP",
    volume = "06",
    pages = "039",
    year = "2024"
}

@article{Franciolini:2018vbk,
    author = "Franciolini, G. and Kehagias, A. and Matarrese, S. and Riotto, A.",
    title = "{Primordial Black Holes from Inflation and non-Gaussianity}",
    eprint = "1801.09415",
    archivePrefix = "arXiv",
    primaryClass = "astro-ph.CO",
    doi = "10.1088/1475-7516/2018/03/016",
    journal = "JCAP",
    volume = "03",
    pages = "016",
    year = "2018"
}

@article{Biagetti:2021eep,
    author = "Biagetti, Matteo and De Luca, Valerio and Franciolini, Gabriele and Kehagias, Alex and Riotto, Antonio",
    title = "{The formation probability of primordial black holes}",
    eprint = "2105.07810",
    archivePrefix = "arXiv",
    primaryClass = "astro-ph.CO",
    doi = "10.1016/j.physletb.2021.136602",
    journal = "Phys. Lett. B",
    volume = "820",
    pages = "136602",
    year = "2021"
}

@article{Kitajima:2021fpq,
    author = "Kitajima, Naoya and Tada, Yuichiro and Yokoyama, Shuichiro and Yoo, Chul-Moon",
    title = "{Primordial black holes in peak theory with a non-Gaussian tail}",
    eprint = "2109.00791",
    archivePrefix = "arXiv",
    primaryClass = "astro-ph.CO",
    reportNumber = "TU-1130",
    doi = "10.1088/1475-7516/2021/10/053",
    journal = "JCAP",
    volume = "10",
    pages = "053",
    year = "2021"
}

@article{Hooshangi:2021ubn,
    author = "Hooshangi, Sina and Namjoo, Mohammad Hossein and Noorbala, Mahdiyar",
    title = "{Rare events are nonperturbative: Primordial black holes from heavy-tailed distributions}",
    eprint = "2112.04520",
    archivePrefix = "arXiv",
    primaryClass = "astro-ph.CO",
    doi = "10.1016/j.physletb.2022.137400",
    journal = "Phys. Lett. B",
    volume = "834",
    pages = "137400",
    year = "2022"
}

@article{Atal:2018neu,
    author = "Atal, Vicente and Germani, Cristiano",
    title = "{The role of non-gaussianities in Primordial Black Hole formation}",
    eprint = "1811.07857",
    archivePrefix = "arXiv",
    primaryClass = "astro-ph.CO",
    reportNumber = "ICCUB-18-022",
    doi = "10.1016/j.dark.2019.100275",
    journal = "Phys. Dark Univ.",
    volume = "24",
    pages = "100275",
    year = "2019"
}

@article{Veermae:2026yzz,
    author = {Veerm{\"a}e, Hardi},
    title = "{A non-perturbative framework for N-point functions of locally non-Gaussian fields}",
    eprint = "2602.10151",
    archivePrefix = "arXiv",
    primaryClass = "astro-ph.CO",
    month = "2",
    year = "2026"
}

@article{Choptuik:1993,
  title = {Universality and scaling in gravitational collapse of a massless scalar field},
  author = {Choptuik, Matthew W.},
  journal = {Phys. Rev. Lett.},
  volume = {70},
  issue = {1},
  pages = {9--12},
  numpages = {0},
  year = {1993},
  month = {Jan},
  publisher = {American Physical Society},
  doi = {10.1103/PhysRevLett.70.9},
  url = {https://link.aps.org/doi/10.1103/PhysRevLett.70.9}
}

@article{Chluba:2012we,
    author = "Chluba, Jens and Erickcek, Adrienne L. and Ben-Dayan, Ido",
    title = "{Probing the inflaton: Small-scale power spectrum constraints from measurements of the CMB energy spectrum}",
    eprint = "1203.2681",
    archivePrefix = "arXiv",
    primaryClass = "astro-ph.CO",
    doi = "10.1088/0004-637X/758/2/76",
    journal = "Astrophys. J.",
    volume = "758",
    pages = "76",
    year = "2012"
}

@article{Chluba:2013dna,
    author = "Chluba, Jens and Grin, Daniel",
    title = "{CMB spectral distortions from small-scale isocurvature fluctuations}",
    eprint = "1304.4596",
    archivePrefix = "arXiv",
    primaryClass = "astro-ph.CO",
    doi = "10.1093/mnras/stt1129",
    journal = "Mon. Not. Roy. Astron. Soc.",
    volume = "434",
    pages = "1619--1635",
    year = "2013"
}

@article{Bianchini:2022dqh,
    author = "Bianchini, Federico and Fabbian, Giulio",
    title = "{CMB spectral distortions revisited: A new take on {\ensuremath{\mu}} distortions and primordial non-Gaussianities from FIRAS data}",
    eprint = "2206.02762",
    archivePrefix = "arXiv",
    primaryClass = "astro-ph.CO",
    doi = "10.1103/PhysRevD.106.063527",
    journal = "Phys. Rev. D",
    volume = "106",
    number = "6",
    pages = "063527",
    year = "2022"
}

@article{Kormendy:1995er,
    author = "Kormendy, John and Richstone, Douglas",
    title = "{Inward bound: The Search for supermassive black holes in galactic nuclei}",
    doi = "10.1146/annurev.aa.33.090195.003053",
    journal = "Ann. Rev. Astron. Astrophys.",
    volume = "33",
    pages = "581",
    year = "1995"
}

@article{Kohri:2014lza,
    author = "Kohri, Kazunori and Nakama, Tomohiro and Suyama, Teruaki",
    title = "{Testing scenarios of primordial black holes being the seeds of supermassive black holes by ultracompact minihalos and CMB $\mu$-distortions}",
    eprint = "1405.5999",
    archivePrefix = "arXiv",
    primaryClass = "astro-ph.CO",
    reportNumber = "KEK-TH-1736, KEK-COSMO-146, RESCEU-9-14",
    doi = "10.1103/PhysRevD.90.083514",
    journal = "Phys. Rev. D",
    volume = "90",
    number = "8",
    pages = "083514",
    year = "2014"
}

@article{2023ApJ...959...39H,
       author = {{Harikane}, Yuichi and {Zhang}, Yechi and {Nakajima}, Kimihiko and {Ouchi}, Masami and {Isobe}, Yuki and {Ono}, Yoshiaki and {Hatano}, Shun and {Xu}, Yi and {Umeda}, Hiroya},
        title = "{A JWST/NIRSpec First Census of Broad-line AGNs at z = 4-7: Detection of 10 Faint AGNs with M $_{BH}$ {}10$^{6}$-{}10$^{8}$ M $_{{\ensuremath{\odot}}}$ and Their Host Galaxy Properties}",
      journal = "Astrophys. J.",
         year = 2023,
        month = dec,
       volume = {959},
       number = {1},
          eid = {39},
        pages = {39},
          doi = {10.3847/1538-4357/ad029e},
archivePrefix = {arXiv},
       eprint = {2303.11946},
 primaryClass = {astro-ph.GA},
       adsurl = {https://ui.adsabs.harvard.edu/abs/2023ApJ...959...39H}
}

@article{Pacucci:2024tws,
    author = "Pacucci, Fabio and Narayan, Ramesh",
    title = "{Mildly Super-Eddington Accretion onto Slowly Spinning Black Holes Explains the X-Ray Weakness of the Little Red Dots}",
    eprint = "2407.15915",
    archivePrefix = "arXiv",
    primaryClass = "astro-ph.HE",
    doi = "10.3847/1538-4357/ad84f7",
    journal = "Astrophys. J.",
    volume = "976",
    number = "1",
    pages = "96",
    year = "2024"
}

@article{Madau:2024fdv,
    author = "Madau, Piero and Haardt, Francesco",
    title = "{X-Ray Weak Active Galactic Nuclei from Super-Eddington Accretion onto Infant Black Holes}",
    eprint = "2410.00417",
    archivePrefix = "arXiv",
    primaryClass = "astro-ph.GA",
    doi = "10.3847/2041-8213/ad90e1",
    journal = "Astrophys. J. Lett.",
    volume = "976",
    number = "2",
    pages = "L24",
    year = "2024"
}

@article{Inayoshi:2025lsf,
    author = "Inayoshi, Kohei and Maiolino, Roberto",
    title = "{Extremely Dense Gas around Little Red Dots and High-redshift Active Galactic Nuclei: A Nonstellar Origin of the Balmer Break and Absorption Features}",
    doi = "10.3847/2041-8213/adaebd",
    journal = "Astrophys. J. Lett.",
    volume = "980",
    number = "2",
    pages = "L27",
    year = "2025"
}

@article{Liu:2025lrd,
    author = "Liu, Hanpu and Jiang, Yan-Fei and Quataert, Eliot and Greene, Jenny E. and Ma, Yilun",
    title = "{The Balmer Break and Optical Continuum of Little Red Dots from Super-Eddington Accretion}",
    eprint = "2507.07190",
    archivePrefix = "arXiv",
    primaryClass = "astro-ph.GA",
    doi = "10.3847/1538-4357/ae0c19",
    journal = "Astrophys. J.",
    volume = "994",
    number = "1",
    pages = "113",
    year = "2025"
}

@article{Kritos:2025aqo,
    author = "Kritos, Konstantinos and Silk, Joseph",
    title = "{From nuclear star clusters to Little Red Dots: black hole growth, mergers, and tidal disruptions}",
    eprint = "2510.21709",
    archivePrefix = "arXiv",
    primaryClass = "astro-ph.HE",
    month = "10",
    year = "2025"
}

@article{Maiolino:2024uon,
    author = "Maiolino, Roberto and others",
    title = "{JWST meets Chandra: a large population of Compton thick, feedback-free, and intrinsically X-ray weak AGN, with a sprinkle of SNe}",
    eprint = "2405.00504",
    archivePrefix = "arXiv",
    primaryClass = "astro-ph.GA",
    doi = "10.1093/mnras/staf359",
    journal = "Mon. Not. Roy. Astron. Soc.",
    volume = "538",
    number = "3",
    pages = "1921--1943",
    year = "2025"
}

@article{Zhang:2026vjk,
    author = "Zhang, Borui and Feng, Wei-Xiang and An, Haipeng",
    title = "{Smoluchowski Coagulation Equation and the Evolution of Primordial Black Hole Clusters}",
    eprint = "2604.01684",
    archivePrefix = "arXiv",
    primaryClass = "astro-ph.CO",
    month = "4",
    year = "2026"
}

@article{Zhang:2025tgm,
    author = "Zhang, Borui and Feng, Wei-Xiang and An, Haipeng",
    title = "{Little Red Dots from Small-Scale Primordial Black Hole Clustering}",
    eprint = "2507.07171",
    archivePrefix = "arXiv",
    primaryClass = "astro-ph.CO",
    month = "7",
    year = "2025"
}

@article{Guo:2026cuv,
    author = "Guo, Jinhui and Liu, Jia and Tanaka, Masanori and Wang, Xiao-Ping and Xiao, Huangyu",
    title = "{Supermassive Primordial Black Holes from a Catalyzed Dark Phase Transition for Little Red Dots}",
    eprint = "2604.01304",
    archivePrefix = "arXiv",
    primaryClass = "hep-ph",
    month = "4",
    year = "2026"
}

@article{Lambrides:2024ugh,
    author = "Lambrides, Erini and others",
    title = "{The Case for Super-Eddington Accretion: Connecting Weak X-ray and UV Line Emission in JWST Broad-Line AGN During the First Gyr of Cosmic Time}",
    eprint = "2409.13047",
    archivePrefix = "arXiv",
    primaryClass = "astro-ph.HE",
    month = "9",
    year = "2024"
}

@article{Inayoshi:2025crit,
    author = "Inayoshi, Kohei and Ho, Luis C.",
    title = "{A Critical Evaluation of the Physical Nature of the Little Red Dots}",
    eprint = "2512.03130",
    archivePrefix = "arXiv",
    primaryClass = "astro-ph.GA",
    doi = "10.48550/arXiv.2512.03130",
    journal = "arXiv e-prints",
    year = "2025"
}

@article{Umeda:2025bha,
    author = "Umeda, Hiroya and Inayoshi, Kohei and Harikane, Yuichi and Murase, Kohta",
    title = "{A Black Hole Envelope Interpretation for Cosmological Demographics of Little Red Dots}",
    eprint = "2512.04208",
    archivePrefix = "arXiv",
    primaryClass = "astro-ph.GA",
    doi = "10.3847/1538-4357/ae4101",
    journal = "Astrophys. J.",
    volume = "999",
    number = "2",
    pages = "183",
    year = "2026"
}

@article{Akins:2025iqo,
    author = "Akins, Hollis B. and others",
    title = "{COSMOS-Web: The Overabundance and Physical Nature of {\textquotedblleft}Little Red Dots{\textquotedblright}{\textemdash}Implications for Early Galaxy and SMBH Assembly}",
    doi = "10.3847/1538-4357/ade984",
    journal = "Astrophys. J.",
    volume = "991",
    number = "1",
    pages = "37",
    year = "2025"
}

@article{2023ApJ...954L...4K,
author = "Kocevski, D.D. and others",
        title = "{Hidden Little Monsters: Spectroscopic Identification of Low-mass, Broad-line AGNs at z > 5 with CEERS}",
      journal = {Astrophys. J. Lett.},
         year = 2023,
        month = sep,
       volume = {954},
       number = {1},
          eid = {L4},
        pages = {L4},
          doi = {10.3847/2041-8213/ace5a0},
archivePrefix = {arXiv},
       eprint = {2302.00012},
 primaryClass = {astro-ph.GA},
       adsurl = {https://ui.adsabs.harvard.edu/abs/2023ApJ...954L...4K}
}

@article{Greene:2024phl,
    author = "Greene, Jenny E. and others",
    title = "{UNCOVER Spectroscopy Confirms the Surprising Ubiquity of Active Galactic Nuclei in Red Sources at z {\ensuremath{>}} 5}",
    doi = "10.3847/1538-4357/ad1e5f",
    journal = "Astrophys. J.",
    volume = "964",
    number = "1",
    pages = "39",
    year = "2024"
}

@article{Matthee:2023utn,
    author = "Matthee, Jorryt and others",
    title = "{Little Red Dots: An Abundant Population of Faint Active Galactic Nuclei at z {\ensuremath{\sim}} 5 Revealed by the EIGER and FRESCO JWST Surveys}",
    eprint = "2306.05448",
    archivePrefix = "arXiv",
    primaryClass = "astro-ph.GA",
    doi = "10.3847/1538-4357/ad2345",
    journal = "Astrophys. J.",
    volume = "963",
    number = "2",
    pages = "129",
    year = "2024"
}

@article{Bernal:2017nec,
    author = "Bernal, Jos{\'e} Luis and Raccanelli, Alvise and Verde, Licia and Silk, Joseph",
    title = "{Signatures of primordial black holes as seeds of supermassive black holes}",
    eprint = "1712.01311",
    archivePrefix = "arXiv",
    primaryClass = "astro-ph.CO",
    doi = "10.1088/1475-7516/2018/05/017",
    journal = "JCAP",
    volume = "05",
    pages = "017",
    year = "2018",
    note = "[Erratum: JCAP 01, E01 (2020)]"
}

@article{Magorrian:1997hw,
    author = "Magorrian, John and others",
    title = "{The Demography of massive dark objects in galaxy centers}",
    eprint = "astro-ph/9708072",
    archivePrefix = "arXiv",
    reportNumber = "CITA-97-28",
    doi = "10.1086/300353",
    journal = "Astron. J.",
    volume = "115",
    pages = "2285",
    year = "1998"
}

@article{Duechting:2004dk,
    author = "Duechting, Norbert",
    title = "{Supermassive black holes from primordial black hole seeds}",
    eprint = "astro-ph/0406260",
    archivePrefix = "arXiv",
    doi = "10.1103/PhysRevD.70.064015",
    journal = "Phys. Rev. D",
    volume = "70",
    pages = "064015",
    year = "2004"
}

@article{Richstone:1998ky,
    author = "Richstone, D. and others",
    title = "{Supermassive black holes and the evolution of galaxies}",
    eprint = "astro-ph/9810378",
    archivePrefix = "arXiv",
    journal = "Nature",
    volume = "395",
    pages = "A14--A19",
    year = "1998"
}

@article{Byrnes:2024vjt,
    author = "Byrnes, Christian T. and Lesgourgues, Julien and Sharma, Devanshu",
    title = "{Robust {\ensuremath{\mu}}-distortion constraints on primordial supermassive black holes from non-Gaussian perturbations}",
    eprint = "2404.18475",
    archivePrefix = "arXiv",
    primaryClass = "astro-ph.CO",
    reportNumber = "TTK-23-16",
    doi = "10.1088/1475-7516/2024/09/012",
    journal = "JCAP",
    volume = "09",
    pages = "012",
    year = "2024"
}

@article{Pritchard:2025yda,
    author = "Pritchard, Xavier and Byrnes, Christian T. and Lesgourgues, Julien and Sharma, Devanshu",
    title = "{Robust {\ensuremath{\mu}}-distortion constraints on primordial supermassive black holes from cubic (gNL) non-Gaussian perturbations}",
    eprint = "2505.08442",
    archivePrefix = "arXiv",
    primaryClass = "astro-ph.CO",
    doi = "10.1088/1475-7516/2025/07/079",
    journal = "JCAP",
    volume = "07",
    pages = "079",
    year = "2025"
}

@article{Fixsen:1996nj,
    author = "Fixsen, D. J. and Cheng, E. S. and Gales, J. M. and Mather, John C. and Shafer, R. A. and Wright, E. L.",
    title = "{The Cosmic Microwave Background spectrum from the full COBE FIRAS data set}",
    eprint = "astro-ph/9605054",
    archivePrefix = "arXiv",
    doi = "10.1086/178173",
    journal = "Astrophys. J.",
    volume = "473",
    pages = "576",
    year = "1996"
}

@article{Fabbian:2025txv,
    author = "Fabbian, Giulio and Bianchini, Federico and Sabyr, Alina and Hill, J. Colin and Lovell, Christopher C. and Thiele, Leander and Spergel, David N.",
    title = "{A new constraint on the $y$-distortion with FIRAS: implications for feedback models in galaxy formation and cosmic shear measurements}",
    eprint = "2512.03038",
    archivePrefix = "arXiv",
    primaryClass = "astro-ph.CO",
    month = "12",
    year = "2025"
}

@article{Evans:1994,
  title = {Critical phenomena and self-similarity in the gravitational collapse of radiation fluid},
  author = {Evans, Charles R. and Coleman, Jason S.},
  journal = {Phys. Rev. Lett.},
  volume = {72},
  issue = {12},
  pages = {1782--1785},
  numpages = {0},
  year = {1994},
  month = {Mar},
  publisher = {American Physical Society},
  doi = {10.1103/PhysRevLett.72.1782},
  url = {https://link.aps.org/doi/10.1103/PhysRevLett.72.1782}
}

@article{Hooshangi:2023kss,
    author = "Hooshangi, Sina and Namjoo, Mohammad Hossein and Noorbala, Mahdiyar",
    title = "{Tail diversity from inflation}",
    eprint = "2305.19257",
    archivePrefix = "arXiv",
    primaryClass = "astro-ph.CO",
    doi = "10.1088/1475-7516/2023/09/023",
    journal = "JCAP",
    volume = "09",
    pages = "023",
    year = "2023"
}

@article{Germani:2023ojx,
    author = "Germani, Cristiano and Sheth, Ravi K.",
    title = "{The Statistics of Primordial Black Holes in a Radiation-Dominated Universe: Recent and New Results}",
    eprint = "2308.02971",
    archivePrefix = "arXiv",
    primaryClass = "astro-ph.CO",
    doi = "10.3390/universe9090421",
    journal = "Universe",
    volume = "9",
    number = "9",
    pages = "421",
    year = "2023"
}

@article{Germani:2019zez,
    author = "Germani, Cristiano and Sheth, Ravi K.",
    title = "{Nonlinear statistics of primordial black holes from Gaussian curvature perturbations}",
    eprint = "1912.07072",
    archivePrefix = "arXiv",
    primaryClass = "astro-ph.CO",
    reportNumber = "ICCUB-19-021",
    doi = "10.1103/PhysRevD.101.063520",
    journal = "Phys. Rev. D",
    volume = "101",
    number = "6",
    pages = "063520",
    year = "2020"
}

@article{Sasaki:2006kq,
    author = "Sasaki, Misao and Valiviita, Jussi and Wands, David",
    title = "{Non-Gaussianity of the primordial perturbation in the curvaton model}",
    eprint = "astro-ph/0607627",
    archivePrefix = "arXiv",
    reportNumber = "YITP-06-33",
    doi = "10.1103/PhysRevD.74.103003",
    journal = "Phys. Rev. D",
    volume = "74",
    pages = "103003",
    year = "2006"
}

@article{Cirelli:2024ssz,
    author = "Cirelli, Marco and Strumia, Alessandro and Zupan, Jure",
    title = "{Dark Matter}",
    eprint = "2406.01705",
    archivePrefix = "arXiv",
    primaryClass = "hep-ph",
    month = "6",
    year = "2024"
}

@article{Zhang:2026hqg,
    author = "Zhang, He-Xu and Huang, Mei",
    title = "{Purely Quadratic Non-Gaussianity from Tachyonic Instability: Primordial Black Holes and Scalar-Induced Gravitational Waves}",
    eprint = "2604.20063",
    archivePrefix = "arXiv",
    primaryClass = "astro-ph.CO",
    month = "4",
    year = "2026"
}

@article{Iovino:2025cdy,
    author = {Iovino, Junior., Antonio and Perna, Gabriele and Veerm{\"a}e, Hardi},
    title = "{The impact of non-Gaussianity when searching for Primordial Black Holes with LISA}",
    eprint = "2512.13648",
    archivePrefix = "arXiv",
    primaryClass = "astro-ph.CO",
    month = "12",
    year = "2025"
}

@article{Carr:2018rid,
    author = "Carr, Bernard and Silk, Joseph",
    title = "{Primordial Black Holes as Generators of Cosmic Structures}",
    eprint = "1801.00672",
    archivePrefix = "arXiv",
    primaryClass = "astro-ph.CO",
    doi = "10.1093/mnras/sty1204",
    journal = "Mon. Not. Roy. Astron. Soc.",
    volume = "478",
    number = "3",
    pages = "3756--3775",
    year = "2018"
}

@article{Carr:2026hot,
    author = {Carr, Bernard and Iovino, Antonio J. and Perna, Gabriele and Vaskonen, Ville and Veerm{\"a}e, Hardi},
    title = "{Primordial black holes: constraints, potential evidence and prospects}",
    eprint = "2601.06024",
    archivePrefix = "arXiv",
    primaryClass = "astro-ph.CO",
    doi = "10.1007/s40766-026-00080-z",
    month = "1",
    year = "2026"
}

@article{Atal:2019cdz,
    author = "Atal, Vicente and Garriga, Jaume and Marcos-Caballero, Airam",
    title = "{Primordial black hole formation with non-Gaussian curvature perturbations}",
    eprint = "1905.13202",
    archivePrefix = "arXiv",
    primaryClass = "astro-ph.CO",
    doi = "10.1088/1475-7516/2019/09/073",
    journal = "JCAP",
    volume = "09",
    pages = "073",
    year = "2019"
}

@article{Figueroa:2020jkf,
    author = "Figueroa, Daniel G. and Raatikainen, Sami and Rasanen, Syksy and Tomberg, Eemeli",
    title = "{Non-Gaussian Tail of the Curvature Perturbation in Stochastic Ultraslow-Roll Inflation: Implications for Primordial Black Hole Production}",
    eprint = "2012.06551",
    archivePrefix = "arXiv",
    primaryClass = "astro-ph.CO",
    reportNumber = "HIP-2020-32/TH",
    doi = "10.1103/PhysRevLett.127.101302",
    journal = "Phys. Rev. Lett.",
    volume = "127",
    number = "10",
    pages = "101302",
    year = "2021"
}

@article{Bartolo:2018rku,
    author = "Bartolo, N. and De Luca, V. and Franciolini, G. and Peloso, M. and Racco, D. and Riotto, A.",
    title = "{Testing primordial black holes as dark matter with LISA}",
    eprint = "1810.12224",
    archivePrefix = "arXiv",
    primaryClass = "astro-ph.CO",
    doi = "10.1103/PhysRevD.99.103521",
    journal = "Phys. Rev. D",
    volume = "99",
    number = "10",
    pages = "103521",
    year = "2019"
}

@article{Jedamzik:1998hc,
    author = "Jedamzik, Karsten",
    editor = "Cline, D. B.",
    title = "{Could MACHOS be primordial black holes formed during the QCD epoch?}",
    eprint = "astro-ph/9805147",
    archivePrefix = "arXiv",
    doi = "10.1016/S0370-1573(98)00067-2",
    journal = "Phys. Rept.",
    volume = "307",
    pages = "155--162",
    year = "1998"
}

@article{Bartolo:2018evs,
    author = "Bartolo, N. and De Luca, V. and Franciolini, G. and Lewis, A. and Peloso, M. and Riotto, A.",
    title = "{Primordial Black Hole Dark Matter: LISA Serendipity}",
    eprint = "1810.12218",
    archivePrefix = "arXiv",
    primaryClass = "astro-ph.CO",
    doi = "10.1103/PhysRevLett.122.211301",
    journal = "Phys. Rev. Lett.",
    volume = "122",
    number = "21",
    pages = "211301",
    year = "2019"
}

@article{Iovino:2025xkq,
    author = "Iovino, A. J. and Perna, G. and Perrone, D. and Racco, D. and Riotto, A.",
    title = "{Understanding the Nature of Scalar-Induced Gravitational Waves}",
    eprint = "2509.24774",
    archivePrefix = "arXiv",
    primaryClass = "gr-qc",
    month = "9",
    year = "2025"
}

@article{Bugaev:2013vba,
    author = "Bugaev, E. V. and Klimai, P. A.",
    title = "{Primordial black hole constraints for curvaton models with predicted large non-Gaussianity}",
    eprint = "1303.3146",
    archivePrefix = "arXiv",
    primaryClass = "astro-ph.CO",
    doi = "10.1142/S021827181350034X",
    journal = "Int. J. Mod. Phys. D",
    volume = "22",
    pages = "1350034",
    year = "2013"
}

@article{Kohri:2012yw,
    author = "Kohri, Kazunori and Lin, Chia-Min and Matsuda, Tomohiro",
    title = "{Primordial black holes from the inflating curvaton}",
    eprint = "1211.2371",
    archivePrefix = "arXiv",
    primaryClass = "hep-ph",
    reportNumber = "KEK-TH-1590, KEK-COSMO-107",
    doi = "10.1103/PhysRevD.87.103527",
    journal = "Phys. Rev. D",
    volume = "87",
    number = "10",
    pages = "103527",
    year = "2013"
}

@article{Chen:2019zza,
    author = "Chen, Chao and Cai, Yi-Fu",
    title = "{Primordial black holes from sound speed resonance in the inflaton-curvaton mixed scenario}",
    eprint = "1908.03942",
    archivePrefix = "arXiv",
    primaryClass = "astro-ph.CO",
    doi = "10.1088/1475-7516/2019/10/068",
    journal = "JCAP",
    volume = "10",
    pages = "068",
    year = "2019"
}

@article{Inomata:2020xad,
    author = "Inomata, Keisuke and Kawasaki, Masahiro and Mukaida, Kyohei and Yanagida, Tsutomu T.",
    title = "{NANOGrav Results and LIGO-Virgo Primordial Black Holes in Axionlike Curvaton Models}",
    eprint = "2011.01270",
    archivePrefix = "arXiv",
    primaryClass = "astro-ph.CO",
    reportNumber = "CERN-TH-2020-182, DESY-20-188, DESY 20-188",
    doi = "10.1103/PhysRevLett.126.131301",
    journal = "Phys. Rev. Lett.",
    volume = "126",
    number = "13",
    pages = "131301",
    year = "2021"
}

@article{Kawasaki:2021ycf,
    author = "Kawasaki, Masahiro and Nakatsuka, Hiromasa",
    title = "{Gravitational waves from type II axion-like curvaton model and its implication for NANOGrav result}",
    eprint = "2101.11244",
    archivePrefix = "arXiv",
    primaryClass = "astro-ph.CO",
    doi = "10.1088/1475-7516/2021/05/023",
    journal = "JCAP",
    volume = "05",
    pages = "023",
    year = "2021"
}

@article{Pi:2021dft,
    author = "Pi, Shi and Sasaki, Misao",
    title = "{Primordial black hole formation in nonminimal curvaton scenarios}",
    eprint = "2112.12680",
    archivePrefix = "arXiv",
    primaryClass = "astro-ph.CO",
    reportNumber = "IPMU21-0075, YITP-21-131",
    doi = "10.1103/PhysRevD.108.L101301",
    journal = "Phys. Rev. D",
    volume = "108",
    number = "10",
    pages = "L101301",
    year = "2023"
}

@article{Chen:2023lou,
    author = "Chen, Chao and Ghoshal, Anish and Lalak, Zygmunt and Luo, Yudong and Naskar, Abhishek",
    title = "{Growth of curvature perturbations for PBH formation {\&} detectable GWs in non-minimal curvaton scenario revisited}",
    eprint = "2305.12325",
    archivePrefix = "arXiv",
    primaryClass = "astro-ph.CO",
    doi = "10.1088/1475-7516/2023/08/041",
    journal = "JCAP",
    volume = "08",
    pages = "041",
    year = "2023"
}

@article{Gow:2023zzp,
    author = "Gow, Andrew D. and Miranda, Tays and Nurmi, Sami",
    title = "{Primordial black holes from a curvaton scenario with strongly non-Gaussian perturbations}",
    eprint = "2307.03078",
    archivePrefix = "arXiv",
    primaryClass = "astro-ph.CO",
    doi = "10.1088/1475-7516/2023/11/006",
    journal = "JCAP",
    volume = "11",
    pages = "006",
    year = "2023"
}

@article{Chen:2024pge,
    author = "Chen, Chao and Ghoshal, Anish and Tasinato, Gianmassimo and Tomberg, Eemeli",
    title = "{Stochastic axionlike curvaton: Non-Gaussianity and primordial black holes without a large power spectrum}",
    eprint = "2409.12950",
    archivePrefix = "arXiv",
    primaryClass = "astro-ph.CO",
    doi = "10.1103/PhysRevD.111.063539",
    journal = "Phys. Rev. D",
    volume = "111",
    number = "6",
    pages = "063539",
    year = "2025"
}

@article{Kuroda:2025coa,
    author = "Kuroda, Tomotaka and Naruko, Atsushi and Vennin, Vincent and Yamaguchi, Masahide",
    title = "{Primordial black holes from a curvaton: the role of bimodal distributions}",
    eprint = "2504.09548",
    archivePrefix = "arXiv",
    primaryClass = "astro-ph.CO",
    doi = "10.1088/1475-7516/2025/07/052",
    journal = "JCAP",
    volume = "07",
    pages = "052",
    year = "2025"
}

@article{Kasai:2026yna,
    author = "Kasai, Kentaro and Kawasaki, Masahiro and Murai, Kai and Neda, Shunsuke",
    title = "{Microlensing events and primordial black holes in the axionlike curvaton model}",
    eprint = "2602.09558",
    archivePrefix = "arXiv",
    primaryClass = "astro-ph.CO",
    reportNumber = "TU-1298",
    month = "2",
    year = "2026"
}

@article{Kusenko:1997si,
    author = "Kusenko, Alexander and Shaposhnikov, Mikhail E.",
    title = "{Supersymmetric Q balls as dark matter}",
    eprint = "hep-ph/9709492",
    archivePrefix = "arXiv",
    reportNumber = "CERN-TH-97-259",
    doi = "10.1016/S0370-2693(97)01375-0",
    journal = "Phys. Lett. B",
    volume = "418",
    pages = "46--54",
    year = "1998"
}

@article{Dine:2003ax,
    author = "Dine, Michael and Kusenko, Alexander",
    title = "{The Origin of the matter - antimatter asymmetry}",
    eprint = "hep-ph/0303065",
    archivePrefix = "arXiv",
    reportNumber = "SCIPP-2003-08, UCLA-03-TEP-08",
    doi = "10.1103/RevModPhys.76.1",
    journal = "Rev. Mod. Phys.",
    volume = "76",
    pages = "1",
    year = "2003"
}

@article{Cotner:2017tir,
    author = "Cotner, Eric and Kusenko, Alexander",
    title = "{Primordial black holes from scalar field evolution in the early universe}",
    eprint = "1706.09003",
    archivePrefix = "arXiv",
    primaryClass = "astro-ph.CO",
    doi = "10.1103/PhysRevD.96.103002",
    journal = "Phys. Rev. D",
    volume = "96",
    number = "10",
    pages = "103002",
    year = "2017"
}

@article{Cotner:2018vug,
    author = "Cotner, Eric and Kusenko, Alexander and Takhistov, Volodymyr",
    title = "{Primordial Black Holes from Inflaton Fragmentation into Oscillons}",
    eprint = "1801.03321",
    archivePrefix = "arXiv",
    primaryClass = "astro-ph.CO",
    reportNumber = "IPMU18-0008",
    doi = "10.1103/PhysRevD.98.083513",
    journal = "Phys. Rev. D",
    volume = "98",
    number = "8",
    pages = "083513",
    year = "2018"
}

@article{Cotner:2019ykd,
    author = "Cotner, Eric and Kusenko, Alexander and Sasaki, Misao and Takhistov, Volodymyr",
    title = "{Analytic Description of Primordial Black Hole Formation from Scalar Field Fragmentation}",
    eprint = "1907.10613",
    archivePrefix = "arXiv",
    primaryClass = "astro-ph.CO",
    reportNumber = "IPMU19-0063, YITP-19-31",
    doi = "10.1088/1475-7516/2019/10/077",
    journal = "JCAP",
    volume = "10",
    pages = "077",
    year = "2019"
}

@article{Koivunen:2022mem,
    author = {Koivunen, Niko and Tomberg, Eemeli and Veerm{\"a}e, Hardi},
    title = "{The linear regime of tachyonic preheating}",
    eprint = "2201.04145",
    archivePrefix = "arXiv",
    primaryClass = "astro-ph.CO",
    doi = "10.1088/1475-7516/2022/07/028",
    journal = "JCAP",
    volume = "07",
    number = "07",
    pages = "028",
    year = "2022"
}

@article{Kim:2017duj,
    author = "Kim, Jinsu and McDonald, John",
    title = "{Inflaton Condensate Fragmentation: Analytical Conditions and Application to $\alpha$-Attractor Models}",
    eprint = "1702.08777",
    archivePrefix = "arXiv",
    primaryClass = "astro-ph.CO",
    doi = "10.1103/PhysRevD.95.123537",
    journal = "Phys. Rev. D",
    volume = "95",
    number = "12",
    pages = "123537",
    year = "2017"
}

@article{Kallosh:2013hoa,
    author = "Kallosh, Renata and Linde, Andrei",
    title = "{Universality Class in Conformal Inflation}",
    eprint = "1306.5220",
    archivePrefix = "arXiv",
    primaryClass = "hep-th",
    doi = "10.1088/1475-7516/2013/07/002",
    journal = "JCAP",
    volume = "07",
    pages = "002",
    year = "2013"
}

\end{document}